\documentclass[twocolumn]{aastex62} %{amsart}

\usepackage{xspace}
\usepackage{color}
\usepackage{rotating}
\usepackage{comment}
\usepackage{amsmath}
\usepackage{subfigure}
\usepackage[ampersand]{easylist}
\usepackage[version=4]{mhchem} % Molecular species! e.g. \ce{H2O}
\usepackage[normalem]{ulem}    % for strikeout text use: \sout{removed text}

\graphicspath{{./}{figures/}}

\makeatletter
\newcommand*{\linktocite}[2]{%
  \hyper@natlinkstart{#1}#2\hyper@natlinkend}
\makeatother

\PassOptionsToPackage{hyphens}{url}\usepackage{hyperref}

\newcommand{\smarter}{\textsc{smarter}\xspace}
\newcommand{\smart}{\textsc{smart}\xspace}
\newcommand{\lblabc}{\textsc{lblabc}\xspace}
\newcommand{\chimera}{\textsc{chimera}\xspace}
\newcommand{\multinest}{\textsc{MultiNest}\xspace}
\newcommand{\pymultinest}{\textsc{PyMultiNest}\xspace}
\newcommand{\dynesty}{\textsc{Dynesty}\xspace}
\newcommand{\pandexo}{\textsc{PandExo}\xspace}

\usepackage{textgreek}
\usepackage{upgreek}

\def\gsim{~\rlap{$>$}{\lower 1.0ex\hbox{$\sim$}}}
\def\lsim{~\rlap{$<$}{\lower 1.0ex\hbox{$\sim$}}}

\newcommand{\emcee}[0]{\textsc{emcee}\xspace}

\newcommand{\jwst}[0]{JWST\xspace}

\newcommand{\um}[0]{$\upmu$m\xspace}

\defcitealias{Macdonald2019}{M\&C19}

\accepted{August 22, 2023}
\submitjournal{PSJ}

\begin{document}

%% ApJ style
\bibliographystyle{apj}

%% Slugcomment
%\slugcomment{Draft version \today}

\title{Earth as a Transiting Exoplanet: A Validation of Transmission Spectroscopy \& Atmospheric Retrieval Methodologies for Terrestrial Exoplanets}

%% Short title, authors
\shorttitle{Earth as a Transiting Exoplanet}
\shortauthors{Lustig-Yaeger et al.}

\correspondingauthor{Jacob Lustig-Yaeger}
\email{Jacob.Lustig-Yaeger@jhuapl.edu}

\author[0000-0002-0746-1980]{Jacob Lustig-Yaeger}
\affiliation{Johns Hopkins University Applied Physics Laboratory, Laurel, MD 20723, USA}
\affiliation{NASA NExSS Virtual Planetary Laboratory, Box 351580, University of Washington, Seattle, Washington 98195, USA}

\author[0000-0002-1386-1710]{Victoria S. Meadows}
\affiliation{Department of Astronomy and Astrobiology Program, University of Washington, Box 351580, Seattle, Washington 98195, USA}
\affiliation{NASA NExSS Virtual Planetary Laboratory, Box 351580, University of Washington, Seattle, Washington 98195, USA}

\author[0000-0002-4573-9998]{David Crisp}
\affiliation{Jet Propulsion Laboratory, California Institute of Technology, Earth and Space Sciences Division, Pasadena, CA 91011, USA}
\affiliation{NASA NExSS Virtual Planetary Laboratory, Box 351580, University of Washington, Seattle, Washington 98195, USA}

\author[0000-0002-2338-476X]{Michael R. Line}
\affiliation{School of Earth and Space Exploration, Arizona State University, P.O. Box 871404, Tempe, AZ 85287-1404, USA}
\affiliation{NASA NExSS Virtual Planetary Laboratory, Box 351580, University of Washington, Seattle, Washington 98195, USA}

\author[0000-0002-3196-414X]{Tyler D. Robinson}
\affiliation{Lunar \& Planetary Laboratory, University of Arizona, Tucson, AZ 85721, USA}
\affiliation{Department of Astronomy and Planetary Science, Northern Arizona University, Flagstaff, AZ 86011, USA}
\affiliation{NASA NExSS Virtual Planetary Laboratory, Box 351580, University of Washington, Seattle, Washington 98195, USA}

%% Abstract %%

\begin{abstract}

The James Webb Space Telescope (JWST) will enable the search for and characterization of terrestrial exoplanet atmospheres in the habitable zone via transmission spectroscopy. 
However, relatively little work has been done to use solar system data, where  ground truth is known, to validate spectroscopic retrieval codes intended for exoplanet studies, particularly in the limit of high resolution and high signal-to-noise (S/N). 
In this work, we perform such a validation by analyzing a high S/N empirical transmission spectrum of Earth using a new terrestrial exoplanet atmospheric retrieval model with heritage in Solar System remote sensing and gaseous exoplanet retrievals. 
We fit the Earth's $2-14$ \um transmission spectrum in low resolution ($R=250$ at 5 \um) and high resolution ($R=100,000$ at 5 \um) under a variety of assumptions about the 1D vertical atmospheric structure. 
In the limit of noiseless transmission spectra, we find excellent agreement between model and data (deviations $< 10\%$) that enable the robust detection of \ce{H2O}, \ce{CO2}, \ce{O3}, \ce{CH4}, \ce{N2}, \ce{N2O}, \ce{NO2}, \ce{HNO3}, CFC-11, and CFC-12 thereby providing compelling support for the detection of habitability, biosignature, and technosignature gases in the atmosphere of the planet using an exoplanet-analog transmission spectrum. 
Our retrievals at high spectral resolution show a marked sensitivity to the thermal structure of the atmosphere, trace gas abundances, density-dependent effects, such as collision-induced absorption and refraction, and even hint at 3D spatial effects. 
However, we used synthetic observations of TRAPPIST-1e to verify that the use of simple 1D vertically homogeneous atmospheric models will likely suffice for JWST observations of terrestrial exoplanets transiting M dwarfs. 

\end{abstract}

%%% Keywords %%%

\section{Introduction} \label{sec:intro}

% Overarching intro 
% Solar system planets as exoplanets
Solar System planets provide a powerful opportunity to validate both the astronomical observing strategies and the theoretical modeling approaches required to characterize distant exoplanets. 
% Upcoming observations with JWST are expected to be high SNR for gaseous exoplanets and sufficient SNR for rocky planets to enable modest atmospheric inference 
Such ground-truth efforts are becoming increasingly important in the era of the James Webb Space Telescope (\jwst), which can collect relatively high signal-to-noise (S/N) spectra of gaseous exoplanets \citep[e.g.][]{Greene2016, ERS2022} and sufficient S/N spectra of rocky exoplanets to enable atmospheric detection and modest atmospheric characterization for the first time \citep[e.g.][]{Morley2017, Wunderlich2019, Lustig-Yaeger2019, Fauchez2019, Pidhorodetska2020, Suissa2020, Gialluca2021}. 
% Problem: Interpretation is hard
While these next-generation observations promise higher precision on exoplanetary atmospheres, they may present a greater challenge to interpret correctly in the presence of complicated physical, chemical, and three dimensional effects.  
% Solution provided
In this paper, we analyze the observed transmission spectrum of Earth, recently published by \citet{Macdonald2019}, using a state-of-the-art terrestrial exoplanet atmospheric retrieval code with extensive Solar System heritage. We examine common modeling assumptions used throughout the field, and then relate the observations to upcoming opportunities to study transiting Earth-like exoplanets on the path to constraining habitability and the presence of life beyond Earth. 

% Past Solar System as exoplanets studies have helped shape the outlook for future exoplanet observations
Shifting our perspective to envision Solar System planets as exoplanets has already helped to fundamentally shape the outlook for future exoplanet observations. 
% Carl Sagan's Galileo lookback study
At the dawn of the exoplanet era, \citet{Sagan1993} analyzed a combination of spectroscopic and imaging observations of Earth from the \textit{Galileo} spacecraft to argue for the remote detection of life on Earth based on detection of atmospheric oxygen and methane, ocean glint, and the vegetation red edge. 
% EPOXI observations of Earth
More recently time-dependent reflected-light observations of Earth using the Extrasolar Planet Observation and Deep Impact Extended Investigation (EPOXI) mission \citep{Livengood2011} were used to demonstrate a method of spatially mapping terrestrial exoplanets using rotation-induced photometric variability  \citep[e.g.][]{Cowan2009, Fujii2011, Cowan2013a, Fujii2017}. 
% DSCOVR
Observations of Earth using the Deep Space Climate Observatory (DSCOVR) have helped to mature these spatial mapping methods \citep{Jiang2018, Fan2019, Aizawa2020, Kawahara2020color, Kawahara2020bayes, Gu2021, Gu2022}. % Ty's Earth model validation
The EPOXI observations have also been used to validate high-fidelity, 3D models of Earth \citep{Robinson2010, Robinson2011} that have been used to predict the strength of the ocean glint signal for Earth-analog exoplanet observations \citep{Robinson2010, Lustig-Yaeger2018}. These predictions were further validated using Lunar CRater Observation and Sensing Satellite (LCROSS) observations of Earth, which successfully detected ocean glint \citep{Robinson2014}, opening a path for robust habitability assessments in the future.  
% Laura's phase curves
Measurements of Solar System bodies as a function of illumination, including Jupiter \citep{Mayorga2016}, Saturn \citep{Dyudina2005, Dyudina2016}, and the Galilean satellites \citep{Mayorga2020}, have brought to light the importance of phase-resolved observations for comparative planetology with future direct-imaging telescopes. 

% Transmission spectra within the solar system 
Of more direct relevance to JWST observations, Earth and Solar System planetary observations have been reanalyzed to produce exoplanet-analog transmission spectra that can be used to test our understanding and models of exoplanet transit phenomena. 
Measurements of Earthshine reflected off the moon during lunar eclipses have been used to study the characteristic spectroscopic observables of Earth as seen in a transit viewing geometry \citep{Palle2009, Vidal-Madjar2010, Ugolnikov2013, Arnold2014, Yan2015, Kawauchi2018, Youngblood2020}.   
In addition, solar occultation limb-sounding measurements have been used to construct exoplanet-analog transmission spectra for Saturn \citep{Dalba2015}, Titan \citep{Robinson2014titan, Tribbett2021}, and Earth \citep{Schreier2018, Macdonald2019}. These Solar System transmission spectra are particularly salient at present because transiting exoplanets will likely be the first terrestrial targets in the next few years to afford spectroscopic atmospheric characterization \citep{NAS2018}. 

% Upcoming terrestrial exoplanet characterization with JWST
Current and upcoming transmission spectroscopy observations of the TRAPPIST-1 system \citep{Gillon2016, Gillon2017, Luger2017b} with \jwst are predicted to probe the composition of terrestrial exoplanet atmospheres for the first time \citep{Barstow2016b, Morley2017, Batalha2018, Krissansen-Totton2018, Wunderlich2019, Lustig-Yaeger2019, Wunderlich2020, May2021, Lin2021, Fauchez2022, Mikal-Evans2022, Krissansen-Totton2022, Rotman2023}. These observations will be analyzed with atmospheric retrieval models that are designed to infer the physical and chemical state of the planetary atmosphere, including presence and abundance of gases, from spectroscopic measurements. For these rocky exoplanet observations, attempts will be made to look for signs of past evolutionary processes \citep[e.g.,][]{Lincowski2018, Lincowski2019, Lustig-Yaeger2019, Loftus2019} and assess the surface habitability and presence of biosignatures \citep[e.g.,][]{Krissansen-Totton2018, Wunderlich2019, Wunderlich2020,Meadows2023}. 

Over the past decade and a half, numerous atmospheric retrieval models have been developed to interpret the transmission and thermal emission spectra of gas giant exoplanets  \citep{Irwin2008, Madhusudhan2009, Line2013a, Benneke2013, Waldmann2015, Cubillos2016, Lavie2016, MacDonald2017, Howe2017, Kitzmann2019, Zhang2019, Damiano2020} and to validate different retrieval models against one another \citep{Kreidberg2015, Barstow2020b} (see \citet{MacDonald2023} for a more complete catalog of retrieval codes). Additionally, critical retrieval work has been done in preparation for data capable of characterizing rocky exoplanet atmospheres using, for example, JWST observations \citep{Barstow2016, Barstow2016b, Krissansen-Totton2018, Lin2021, Mikal-Evans2022} and next generation observing capabilities \citep{Feng2018, Quanz2019, Tremblay2020}. However, relatively few studies \citep[e.g.,][]{Robinson2023} have been performed to validate terrestrial retrieval models and their common underlying assumptions using exoplanet analog observations within the solar system, particularly over a broad wavelength range and using high precision observations.    
Thus, to fully realize the scientific value of upcoming terrestrial exoplanet observations it is critical to validate these retrieval codes on observations with known ground-truth, before (or in parallel with) attempts to understand the unexpected alien environments of exoplanets. 

% Data zoom!
\begin{figure*}[h!]
\centering
\includegraphics[width=0.85\textwidth]{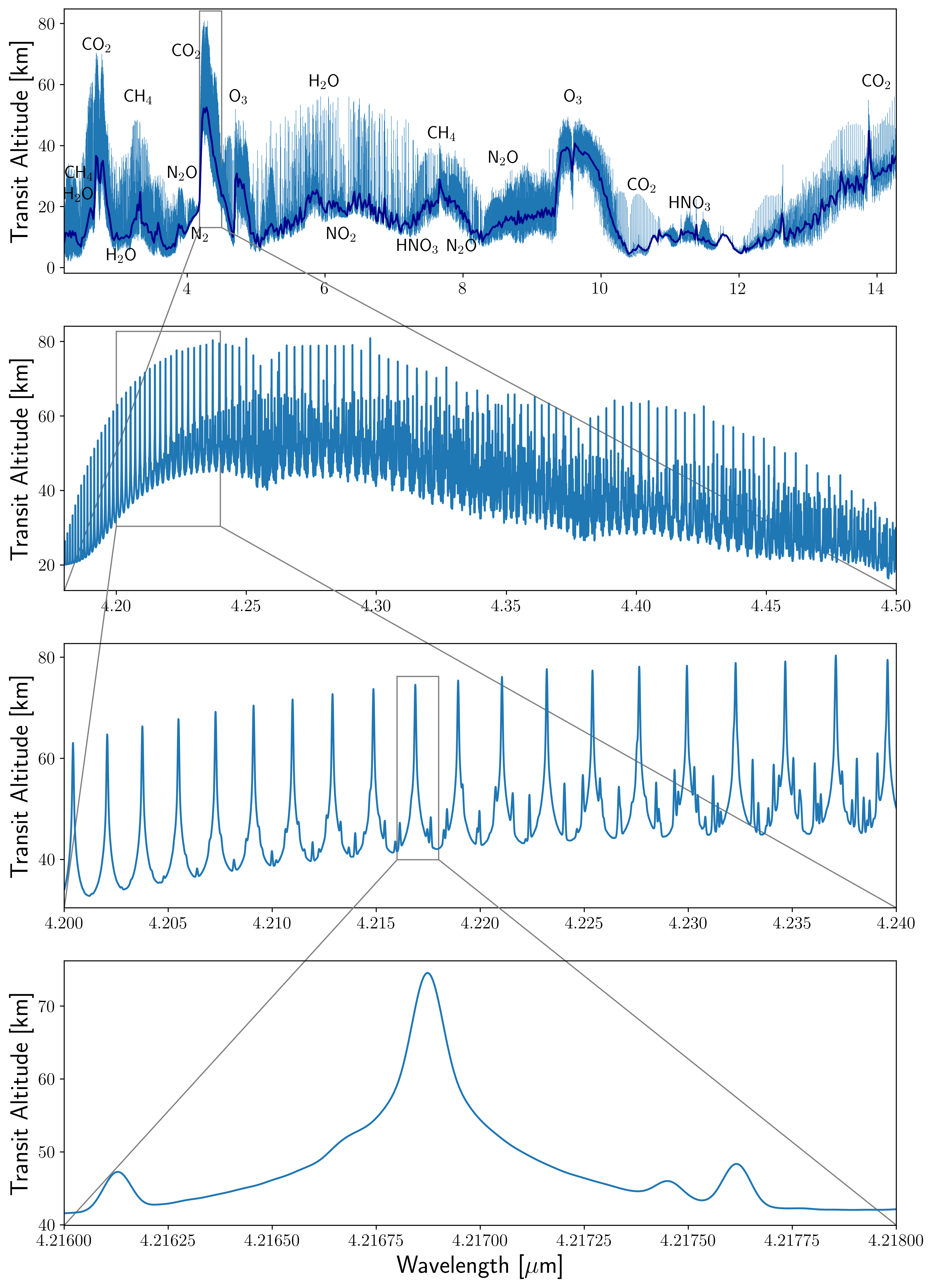}
\caption{Empirical transmission spectrum of the cloudless Earth from \citet{Macdonald2019}. The top panel shows the full wavelength range in high-resolution (light blue) and binned to a lower resolution ($\Delta \lambda = 0.02$ \um; dark blue). The lower panels zoom into the 4.2 \um \ce{CO2} band, showing progressively smaller wavelength ranges and highlighting the fully resolved spectral lines with no apparent observational noise. The data behind this figure is available at \dataset[DOI: 10.5281/zenodo.8280710]{https://doi.org/10.5281/zenodo.8280710} \citep{macdonald_evelyn_2023_8280710} which was obtained from the ACE-FTS Atmospheric Atlases \citep{Hughes2014} at \url{http://www.ace.uwaterloo.ca/atlas.php}.}  
\label{fig:earth_zoom}
\end{figure*}

% What's so great/useful about the M&C2019 spectrum
The empirical transit transmission spectrum of Earth from \citet{Macdonald2019} (hereafter \citetalias{Macdonald2019}) provides a JWST-relevant wavelength range with the highest signal-to-noise ratio (S/N) and spectral-resolution to date, with which to gauge the efficacy of transmission spectroscopy observations and atmospheric retrieval inference methods for characterizing Earth-like exoplanet atmospheres for habitability and biosignatures. \citetalias{Macdonald2019} post-processed Earth solar occultation measurements observed with the Canadian low-Earth orbit satellite, SCISAT, using the \textit{Atmospheric Chemistry Experiment Fourier Transform Spectrometer} \citep[ACE-FTS;][]{Bernath2005, Hughes2014}. The resulting $2-14$ \um transmission spectrum of Earth, shown in the upper panel of Figure \ref{fig:earth_zoom}, contains numerous absorption features from characteristic molecules in Earth's atmosphere, such as \ce{CO2}, \ce{H2O}, \ce{CH4}, and \ce{O3}, among many others \citep[see][]{Schreier2018}. 
The broad wavelength range overlaps significantly with that of JWST's primary exoplanet transmission spectroscopy instruments, including, NIRSpec, NIRISS, and MIRI, making it a particularly useful analog for upcoming observations.  
The high S/N of the Earth spectrum provides an opportunity to test best-case-scenarios for transmission spectroscopy with data that does not suffer from astrophysical (e.g. temporal, spatial, and spectral stellar variability) or astronomical (e.g. detector white and red noise) limitations of true exoplanet observations. 
The high spectral resolution affords the ability to bin the spectrum down to a variety of lower resolutions that are applicable to upcoming exoplanet observations, as well as to validate transmission spectroscopy retrievals at a high-resolution that is analogous to current and upcoming ground-based exoplanet observations \citep[e.g.,][]{Snellen2013, Brogi2014, Snellen2017, Brogi2019, Webb2020}.   

% Overview of this paper
In this paper, we describe a new terrestrial (exo)planet atmospheric retrieval model, \smarter, based on a physically-rigorous, line-by-line radiative transfer code that has been previously validated on emission and reflected spectra from Solar System terrestrials. This model has the capability to analyze UV to millimeter wave transmission, reflected light and emission data for exoplanets with a wide range of atmospheric compositions, temperatures and pressures.  Here we validate the code using Earth's empirical transmission spectrum from \citetalias{Macdonald2019}. In addition to validating our model and common modeling assumptions, we explore the achievable constraints on Earth's atmospheric composition and temperature structure at low and high spectral resolution, and using different levels of observational noise to link our findings to current and upcoming observations with JWST.  

% Outline of this section
In the next section, we describe our methods, including information about the transmission spectrum from \citetalias{Macdonald2019} (Section \ref{sec:data}) and descriptions of our atmospheric retrieval model (Section \ref{sec:methods:model}). In Section \ref{sec:results}, we describe our results. We discuss our findings in Section \ref{sec:discussion} and present our conclusions in Section \ref{sec:conclusion}.   

\section{Methods} \label{sec:methods}

\subsection{Observed Data} \label{sec:data}

% Overview of Macdonald and Cowan data and approach
\citet{Macdonald2019} processed solar occultation measurements of the Earth's atmosphere to produce an altitude-integrated and limb-averaged transmission spectrum from an observer perspective that is analogous to transiting exoplanets \citep{Seager2000, Charbonneau2002}. The original measurements were made by the Canadian low-Earth orbit satellite, SCISAT, using the \textit{Atmospheric Chemistry Experiment Fourier Transform Spectrometer} \citep[ACE-FTS;][]{Bernath2005, Hughes2014, Boone2019}. ACE-FTS is a high resolution spectrograph that observes between approximately 700--4400 wavenumbers (2.27--14.3 \um) with a full-width-at-half-maximum (FWHM) of 0.02 cm$^{-1}$ and a sampling resolution of 0.0025 cm$^{-1}$. This corresponds to a spectral resolving power of $R=100,000$ at 5 \um (2000 cm$^{-1}$). Spectra were recorded of Sunlight passing through the Earth's atmosphere as a function of time during sunrises and sunsets. The transmittance, $\mathcal{T}$, of Earth's atmosphere can be derived as a function of both wavelength and altitude by dividing out the unattenuated spectrum of the Sun acquired from observations taken just prior to atmospheric occultation. These data products support the main objective of the mission to study the vertical distribution of trace gases in the Earth's atmosphere. 

\citetalias{Macdonald2019} calculated the transmission spectrum of the Earth as if it were an exoplanet by integrating the ACE-FTS transmittance data from the bottom to the top of the atmosphere using the approach described in \citet{Robinson2014titan} for Titan and \citet{Dalba2015} for Saturn. 
\citetalias{Macdonald2019} selected for only cloud-free sightlines through the Earth's atmosphere and averaged approximately 800 datasets spanning five latitudinal regions---Arctic summer and winter ($60-85^{\circ}~\mathrm{N}$), mid-latitude summer and winter ($30-60^{\circ}~\mathrm{N}$), and the tropics ($30^{\circ}~\mathrm{S} - 30^{\circ}~\mathrm{N}$)---to mimic the limb-integrated stellar light transmitted through exoplanet atmospheres during primary transit. The final published transmission spectrum has $\mathrm{S/N} \approx 8000$ in the continuum.  
Although \citetalias{Macdonald2019} explored modifications to their resulting transmission spectrum due to refraction, their data, as well as the spectrum analyzed in this paper, does not possess the critical refraction boundary that limits the spectrum from probing below ${\sim}15$ km for an Earth-Sun exoplanet analog \citep{Misra2014a, Betremieux2014}. As a result, we are able to probe deeper into the Earth's atmosphere than a true Earth-Sun-analog exoplanet observation, where the observer can be assumed to be infinitely far away. 
While the empirical spectrum from \citetalias{Macdonald2019} and analyzed in this work does not contain cloudy sightlines, we refer readers to the recent work of \citet{Doshi2022} for a detailed analysis of the cloudy Earth spectrum derived from the same SCISAT measurements. 

% Figure/spectrum description
Figure \ref{fig:earth_zoom} shows the empirical transmission spectrum of Earth from \citet{Macdonald2019}\footnote{The published spectrum is available at \dataset[DOI: 10.5281/zenodo.8280710]{https://doi.org/10.5281/zenodo.8280710} \citep{macdonald_evelyn_2023_8280710}. }. %The transmission spectrum is of exquisite quality. 
The broad wavelength coverage spanning $2-14$ \um contains spectral features from over a dozen different molecules \citep{Bernath2005, Bernath2017, Schreier2018}. The top panel shows the spectrum at the native high resolution (light blue) and binned to a lower resolution of $\Delta \lambda = 0.02$ \um (dark blue; $R=250$ at 5 \um). The zoomed subpanels focus on the 4.3 \um \ce{CO2} band and showcase the extremely high resolution and high S/N of the spectrum. Most spectral lines appear to be fully resolved. The bottom-most panel shows that there is no visible scatter in the data, reinforcing its high S/N (${\sim}8000$) and lack of white noise.  
With its high resolution and S/N, this spectrum should be ideal for validating a spectral retrieval code designed for transmission spectroscopy. 

\subsection{Retrieval Model} \label{sec:methods:model}

%%%%%%%%%%%%%%%%%%%%%%%%%%%%%%%%%%%%%%%%%%%%%%%%%%%%%%%%%%%%%%%%%%%%%%%%%%%%%%%%%%%%%%%%

% Overview
In this work, we use a novel atmospheric retrieval code developed specifically to interpret the spectra of terrestrial exoplanets. Our retrieval model differs from other state-of-the-art retrieval codes primarily in the treatment of radiative transfer, for which we use the robust line-by-line Spectral Mapping Atmospheric Radiative Transfer (\smart) model \citep{Meadows1996, Crisp1997}. Due to the rigor of the radiative transfer, the \smart Exoplanet Retrieval (\smarter) code has been developed to use a variety of different Bayesian retrieval approaches to solve the inverse problem so that the complexity of the physical and statistical methods can be appropriately balanced to achieve a computationally tractable problem. In this section, we first describe the \smarter forward model used for exoplanet transit transmission spectroscopy (\S~\ref{sec:methods:forward}), and then describe the \smarter inverse modeling suite (\S~\ref{sec:methods:inverse}).  

% Description of state vector
\subsubsection{Forward Model} \label{sec:methods:forward} 

The \smarter forward model was designed with a modular approach capable of leveraging the many radiative transfer capabilities of \smart to perform exoplanet retrievals. As a result, \smarter can run atmospheric retrievals for both terrestrials and Jovians using transmission, thermal emission, and reflected-light spectra. However, in this work, we focus exclusively on transmission spectroscopy of the Earth and Earth-like exoplanets.  

% Radiative Transfer with SMART
\paragraph{Radiative Transfer} 

% Overview of SMART
\smarter uses the Spectral Mapping Atmospheric Radiative Transfer (\smart) model as its forward model \citep[developed by D. Crisp; ][]{Meadows1996, Crisp1997}. \smart was originally designed to accelerate radiative transfer calculations in planetary atmospheres where both gas absorption and multiple scattering by gases, clouds and aerosols contribute to extinction. In spectral regions dominated by vibration-rotation transitions of gases (i.e., spectral lines), a very fine spectral grid is needed to resolve the spectral structure of the atmospheric optical properties. If multiple scattering is also an important source of extinction in these spectral regions, a computationally-expensive multiple scatting calculation must be performed at each monochromatic spectral grid point (a single, fine wavelength point) to accurately evaluate the radiances and fluxes within the atmosphere \citep{Meadows1996, Crisp1997, Arney2016}. 

% Spectral Mapping
To accelerate these calculations, \smart exploits the fact that many spectral grid points within wide spectral regions have very similar optical properties at all points along the optical path. After generating the optical properties on high-resolution spectral grid, \smart identifies those spectral grid points with nearly identical optical properties along the entire atmospheric optical path \citep{West1990, Meadows1996}. Spectral grid points with optical properties within a specified range of values are mapped into a series of quasi-monochromatic spectral bins. For example, the binning criteria might specify that only grid points with layer-integrated optical depths, single scattering albedos and scattering phase functions with 25\% of the mean values of a bin are included in that bin. Once the optical properties are binned, radiances and their Jacobians (i.e., the first derivative of the radiances with respect to changes in the atmospheric temperatures or optical properties) are derived for the mean values of the optical properties in each bin. The Jacobians provide a linear correction to radiances for each bin when these values are mapped back to original monochromatic spectral grid points. This spectral binning and remapping approach produces high-resolution radiance spectra and typically reduces the number of radiative transfer calculations by a factor of 10 to 1000, depending on the user-specified spectral binning criteria. There are four unique binning tolerance parameters defined in terms of the fractional error in optical depth, co-single scattering albedo, aerosol asymmetry factor, and surface optical properties (e.g., surface albedo). However, only the optical depth binning parameter is relevant to this work (and only for simulations in Appendix \ref{sec:appendix:refraction}). 

% SMART Transit capability 
\smart was originally developed for plane-parallel radiative transfer calculations in planetary atmospheres \citep[e.g.,][]{Meadows1996, Robinson2011, Arney2016, Meadows2018, Lincowski2021}. For those applications, the DIScrete Ordinate Radiative Transfer code \citep[DISORT;][]{Stamnes1988, Stamnes2017DISORT} was used to perform the radiative transfer calculations for each bin. More recently, \smart was extended to perform transit calculations \citep{Misra2014a, Robinson2017a}. That is the version that is used here and it is independent of DISORT. The transmission spectrum calculations can be performed using one of three methods \citep{Robinson2017a}: (1) geometric optics using straight rays, (2) ray tracing (with the option for the refraction of stellar rays passing through the atmosphere), and (3) Monte Carlo numerical integration to account for forward-scattering by aerosol particles. For the ray-tracing option without refraction, used in most of the simulations presented here, the full-resolution monochromatic optical properties generated by \smart are used, without the binning and remapping operations described above. When ray tracing is used with refraction (See Appendix \ref{sec:appendix:refraction}), the full-resolution monochromatic optical properties are used, but the optical path length calculations assume that the refraction is constant within each bin. The Monte Carlo method uses the full spectral binning and remapping operations to improve efficiency. Here we use the ray tracing methods for all cases. 

Typically, refraction would be used for transit calculations, but in this case we turn refraction off unless otherwise noted because the solar occultation geometry of Earth's transmission spectrum from \citetalias{Macdonald2019} permits access to the near-surface atmosphere that would be restricted by the critical refraction boundary for a true Earth-analog exoplanet \citep{Betremieux2014, Misra2014a}. We explore the impacts of refraction on the optical path length using high-resolution spectral models with and without refraction in Appendix \ref{sec:appendix:refraction}, as discussed later in Section \ref{sec:results:high}, but we found that it had little impact on the results.  

% Description of LBLABC
Gas absorption cross sections are used with \smart to account for (1) vibrational–rotational transitions in the optical and infrared, (2) continuum absorption in the UV, and (3) collision-induced absorption (CIA) bands.    
\smart depends on the Line-By-Line ABsorption Coefficient (\lblabc) code to calculate molecular vibrational-rotational transition absorption coefficient input files for the radiative transfer \citep[developed by D. Crisp;][]{Meadows1996}. \lblabc takes information about the input atmospheric state---the temperature versus pressure (TP) profile, vertical gas mixing ratio, mean molecular weight of the atmosphere, and radius and surface gravity of the planet---and combines it with HITRAN line-parameter and isotope information to calculate the gas absorption coefficients as a function of pressure, temperature, and wavenumber. 
Unless otherwise stated, we use the HITRAN2016 line list \citep{Gordon2017} as an input for \lblabc and our subsequent radiative transfer calculations with \smart. 
\lblabc uses nested spectral grids to fully resolve the narrow cores of each absorption line and to resolve the far wings of the lines at every atmospheric layer over a very large range of line-center distances. Line contributions are calculated up to 1000 cm$^{-1}$ from the line center for \ce{H2O} and \ce{CO2} to capture the full wings of strong lines; all other gases have a 50 cm$^{-1}$ wing cutoff to speed up calculations where contribution from the far wings is negligible. \lblabc employs different line profiles as a function of distance from the line core. \lblabc uses a Voigt line shape \citep{Humlivcek1982} within 40 Doppler half-widths of the line center, and a van Vleck-Weisskopf profile \citep{VanVleck1945} beyond that for all gases except \ce{H2O}, \ce{CO2}, and \ce{O2}. Empirically determined $\chi$ correction factors are used in the far wings for \ce{H2O} \citep{Clough1989} and \ce{CO2} \citep{Fukabori1986, Perrin1989, Pollack1993} to reproduce the effects of quantum-mechanical line mixing, a collisional effect that occurs for overlapping lines of a given molecular species \citep{Rao2012}. For \ce{O2}, the super-Lorentzian line shape from \citet{Hirono1982} is used with an exponent of 1.958. \lblabc features the ability to simulate line absorption coefficients for fixed temperature offsets from the given TP profile, which \smart can interpolate between for added flexibility when deviating from an initial temperature structure. This can be used to generate a single set of line absorption coefficients that may be interpolated to yield estimates that are used over a broad range of atmospheric temperature structures that may be explored by the retrieval, without requiring \lblabc to be run multiple times. 

The internal wavenumber resolution of \smart is designed to completely resolve the spectral structure of the atmospheric and surface optical properties. In spectral regions where the atmospheric and surface optical properties vary slowly with wavelength, such as the near UV, the resolving power can be as low as 100.  At infrared wavelengths populated by the gas vibration-rotation transitions, the spectral resolution is set by gas absorption cross-section algorithm, \lblabc. For each spectral line, \lblabc finds the atmospheric level where the Voigt line full-width-at-half-maximum (FWHM) is the smallest. It then resolves the FWHM of this ``line core'' region with at least 8 spectral grid points. At larger distances from the line center, the distance between spectral grid points increases in a geometric series, such that the grid spacing increases 25\% for each point, such that the line wings are well resolved, with a few hundred ($<$512) points. \lblabc then maps this individual-line spectral grid onto a common spectral grid that is used for all spectral lines at all levels of the atmosphere. This grid can have a spectral resolving power as high as $10^7$. 

One important exception to the above is implemented for instances where the spectral resolution of the data is significantly lower than typical line widths. In this case, the line absorption coefficient files from \lblabc may be resampled to a lower resolution, which then sets the resolution of \smart and can significantly speed up retrieval calculations with \smarter. For our retrievals on spectra with simulated noise that use the nested sampling retrieval approach (Section \ref{sec:res:low:noise} and Section \ref{sec:results:jwst}), we resample our molecular line absorption coefficients to a resolution of 0.1 cm$^{-1}$. 

\begin{figure}[t]
\centering
\includegraphics[width=0.47\textwidth]{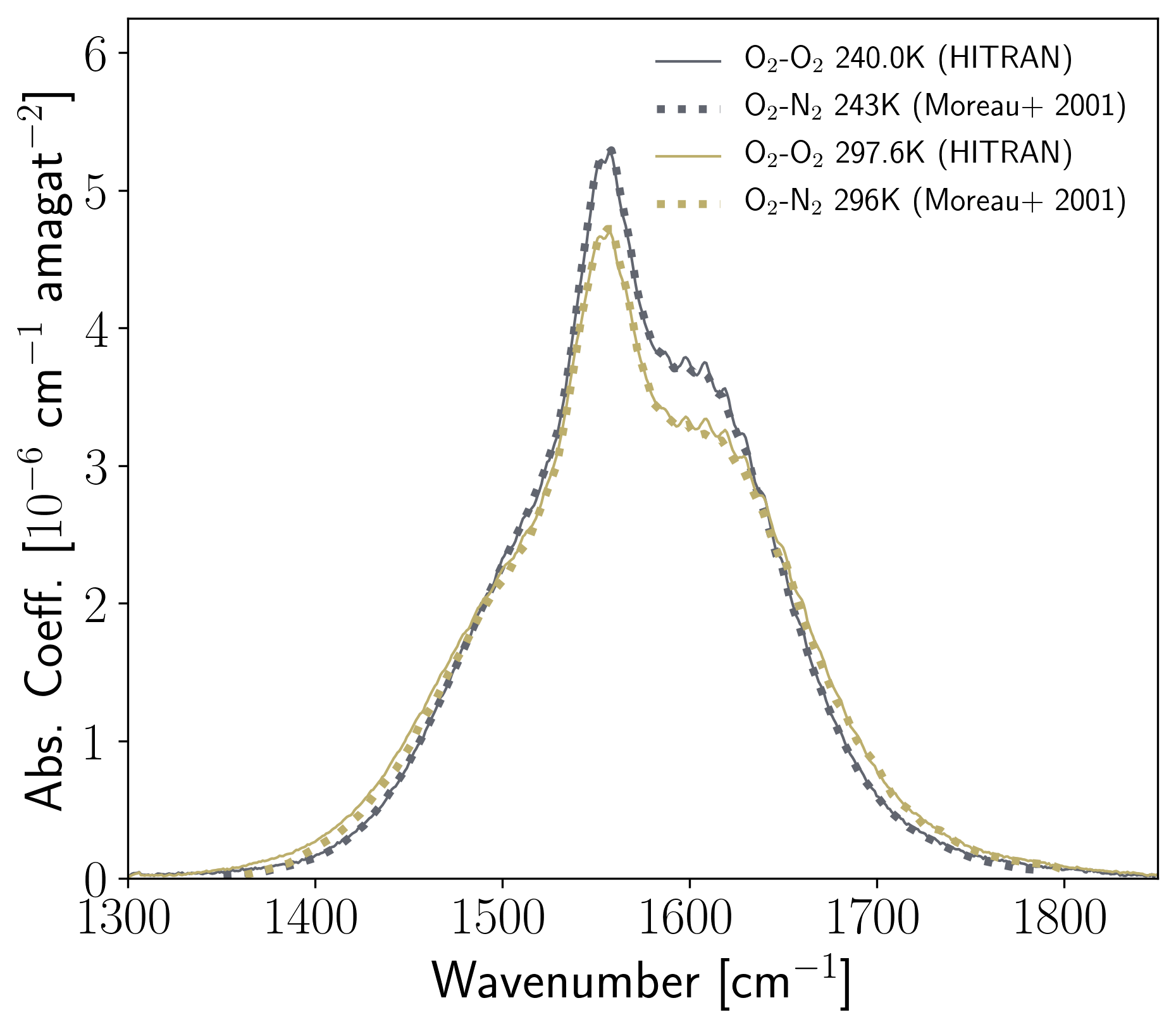}
\caption{A comparison between the 6\um \ce{O2-N2} CIA band from \citet{Moreau2001} and the 6\um \ce{O2-O2} CIA band from HITRAN 2016 \citep{Gordon2017}. At Earth-like temperatures and pressures, the \ce{O2-N2} and \ce{O2-O2} CIA bands appear to agree in intensity suggesting that the \ce{O2-O2} absorption coefficient can be accurately used for \ce{O2-N2}. } 
\label{fig:o2n2}
\end{figure}

% UV xsecs and CIA sources
We incorporate UV and visible cross sections from the MPI-Mainz UV/VIS Spectral Atlas of Gaseous Molecules \citep{Keller-Rudek2013} as described in \citet{Lincowski2018}. We use HITRAN infrared absorption cross-sections \citep{HITRAN2016_XSC} for two chlorofluorocarbons (CFCs): \ce{CCl3F} \citep[CFC-11;][]{HITRAN_CFC11} and \ce{CCl2F2} \citep[CFC-12;][]{HITRAN_CFC12}. CIA data are used for \ce{CO2-CO2} \citep{Moore1972, Kasting1984, Gruszka1997, Baranov2004, Wordsworth2010, Lee2016}, \ce{O2-O2} \citep{Greenblatt1990, Hermans1999, Mate1999, Gordon2017}, and \ce{N2-N2} \citep{Lafferty1996, Schwieterman2015b}. Following \citet{Fauchez2020}, we include \ce{O2-N2} CIA at 6\um by assuming that it is equivalent in strength to the corresponding \ce{O2-O2} CIA band at 6\um, which is incorporated from HITRAN \citep{Karman2019}. In Figure \ref{fig:o2n2} we compare the laboratory derived CIA absorption coefficients for \ce{O2-N2} at 6\um from \citet{Moreau2001} with that of \ce{O2-O2} from HITRAN \citep{Karman2019}. At Earth temperatures and pressures, using \ce{O2-O2} as a proxy for \ce{O2-N2} appears to be valid. 

% O2-N2 Figure

% SMART's past validation and uses
\smart and \lblabc were developed for Solar System terrestrial planet atmospheres and have been rigorously validated on the Earth \citep{Robinson2011} and Venus \citep{Meadows1996, Crisp1996, Arney2014, Robinson2018b} reflected and thermal observations. \smart has been used to simulate the spectra of a rich diversity of exoplanet environments, including the spectra of M dwarf habitable planets \citep{Segura2003, Segura2005, Meadows2018} and uninhabitable planets \citep{Lincowski2018, Lincowski2019}, habitable haze-enshrouded exoplanets \citep{Arney2016, Arney2017}, and high-altitude clouds in the \ce{H2}-dominated mini-Neptune exoplanet, GJ1214b \citep{Charnay2015a}. \smart also functions as the radiative transfer core for the VPL Climate Model \citep{Robinson2018b, Lincowski2018}. 

% Baseline Transit Model
\paragraph{Baseline Transmission Forward Models} 

% Basic model description
We employ two different transit transmission spectroscopy forward models in this work. Both forward models call \smart for the radiative transfer and make a number of simplifying assumptions to limit the volume of parameter space that the retrieval must traverse. However, each model differs in the treatment of vertical atmospheric profiles in order to test different modeling approaches and their underlying assumptions. 

% Model 1
The first forward model uses 1D vertical atmospheric temperature and gas mixing ratio profiles provided by the user and the retrieval fits for scaling factors that shift the entire profile to larger or smaller values without changing the shape, similar to the approach used by \citet{Barstow2016}. We refer to this as the ``scaled profiles'' model and ``Model 1''. This model makes a strong \textit{a priori} assumption about the shape of the 1D vertical profiles, but it offers a solid foundation to test the accuracy of our retrieval model on high quality data, to examine how well a 1D atmosphere can reproduce observations of a 3D planet, and to compare against our model with a homogeneous vertical structure. We note that for instances using Model 1, where the scale factor is sufficiently large so as to make any mixing ratio exceed unity, that value is capped at one. However, if the sum of all gas mixing ratios exceeds one, then the sample is rejected by the forward model. 

% Model 2 
The second forward model assumes that the 1D atmosphere is isothermal with evenly mixed gas abundance profiles---a common set of assumptions in the exoplanet atmospheric retrieval community. We refer to this as the ``evenly-mixed'' model and ``Model 2''.  Although such isothermal atmospheres are unphysical, and indeed mixing ratio biases have been reported in retrieval tests of hot Jupiter atmospheres \citep{Rocchetto2016}, transmission spectroscopy is only weakly sensitive to the atmospheric thermal structure \citep{Kempton2017}, and it remains unclear if \jwst will have the precision to justify more complex TP profile retrievals for rocky exoplanets. This model offers the opportunity to test these common exoplanet assumptions and examine how a real exoplanet-analog observation is interpreted through the lens of this simplified model. 

% Earth atm
\begin{figure}
\centering
\includegraphics[width=0.47\textwidth]{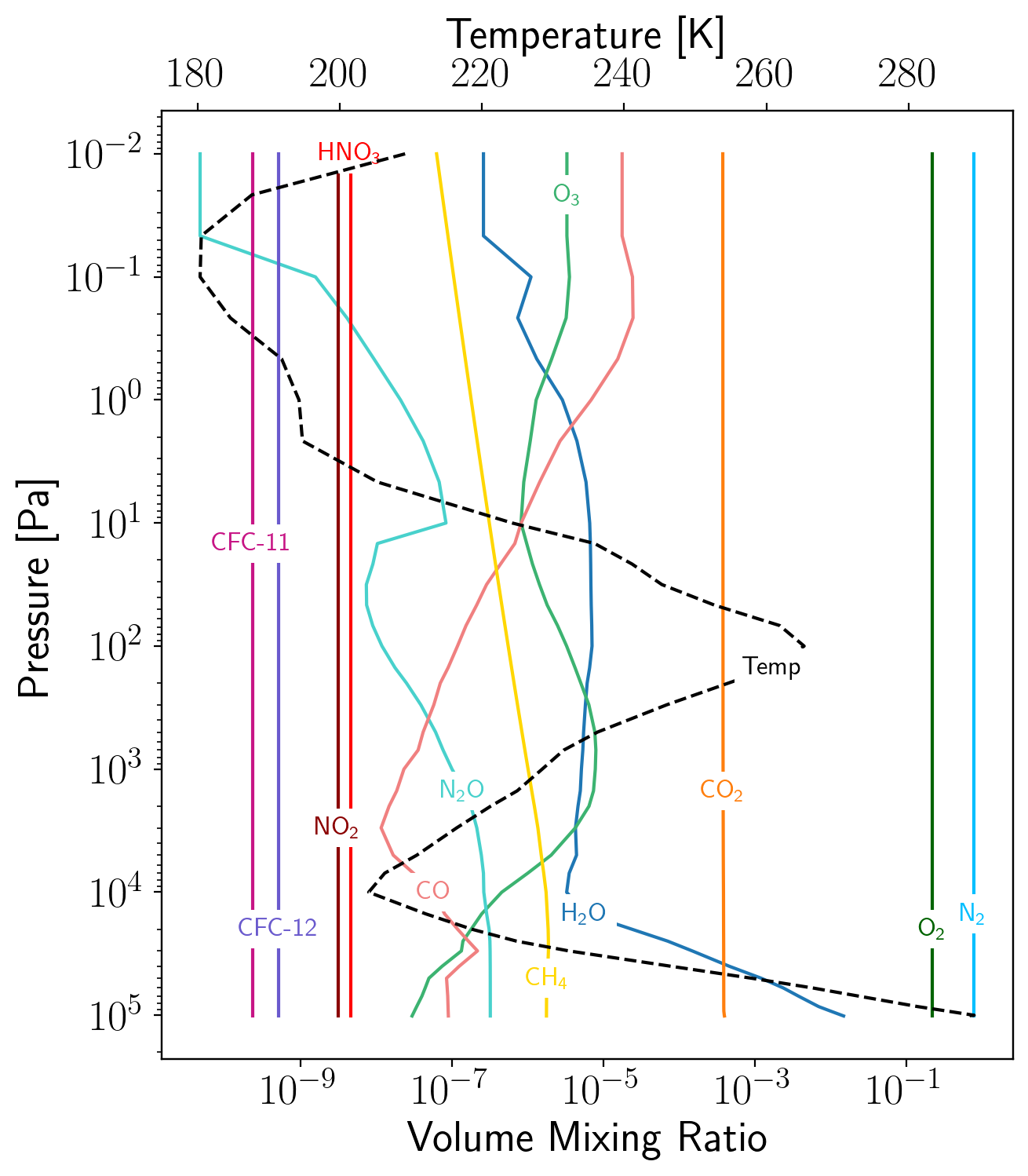}
\caption{Earth's approximate 1D atmospheric thermal structure and composition. The temperature structure (dashed black) is shown along the top x-axis and the gas mixing ratios are shown along the bottom x-axis, both as a function of pressure. This model is used to inform the initialization of our retrievals. } 
\label{fig:earth_atm}
\end{figure}

% Earth atm
We use a globally-averaged 1D atmospheric structure for Earth as a starting point for our retrievals and a basis for comparison to the known Earth atmospheric structure and composition.    
Figure \ref{fig:earth_atm} shows Earth's average 1D atmospheric vertical thermal structure and composition acquired from satellite data for the VPL Earth Model \citep{Robinson2011} to reproduce the EPOXI flyby of Earth in March and April of 2008. We use these profiles as the starting point for our retrievals. Model 1 linearly scales each of these profiles to improve the fit. Model 2 is initialized with evenly mixed atmospheric profiles that are calculated as log-pressure-weighted averages using a weighting kernel that is given by,  
\begin{equation}
    w = \frac{1}{\log_{10}(P \mathrm{~[bar]}) - \log_{10}(0.1 \mathrm{~[bar]})^2 + \epsilon}, 
\end{equation}
which favors pressures near 0.1 bar ($10^4$ Pa), approximately where transmission spectra are most sensitive, with $\epsilon = 0.1$ dex ($10^{0.1} \approx 1.26$) to control the width of averaging kernel around 0.1 bar. 
In addition to the gases included in the VPL Earth Model (\ce{H2O}, \ce{CO2}, \ce{O3}, \ce{O2}, \ce{N2}, \ce{CH4}, \ce{CO}, and \ce{N2O}), we also include \ce{HNO3}, \ce{NO2}, \ce{CCl3F} (CFC-11), and \ce{CCl2F2} (CFC-12) because they corrected notable residuals in our initial tests. For \ce{HNO3} and \ce{NO2}, we use evenly mixed atmospheric profiles as shown in Fig. \ref{fig:earth_atm}, initialized using best estimates from the literature. For \ce{HNO3}, we initialize at 10 ppb VMR based on a measurement from \citet[][]{Hanke2003}, for \ce{NO2}, we use 5 ppb VMR, which is consistent with measurements in urban areas \citep{Lamsal2013, Kopparapu2021}, and for CFC-11 and CFC-12 we used 230 ppt and 510 ppt, respectively, consistent with findings from \citet{Kellmann2012}. 

% Earth characteristic TPs
\begin{figure}
\centering
\includegraphics[width=0.49\textwidth]{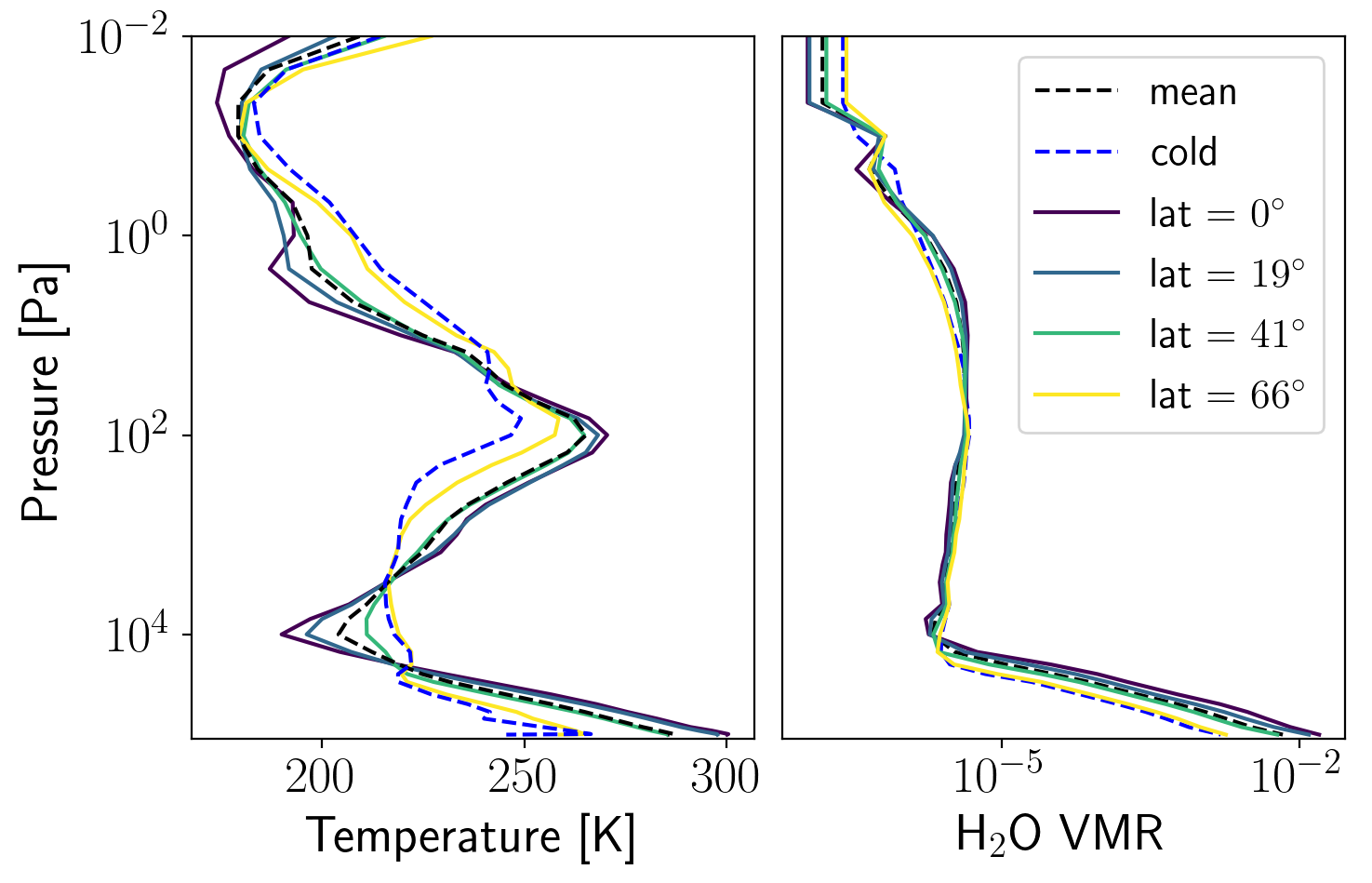}
\caption{Characteristic thermal (left) and water (right) profiles of Earth's atmosphere. The profiles are derived from satellite data and were acquired to match the dates of the EPOXI Earth flyby in 2008 \citep[see][]{Robinson2011}. Each latitude plotted represents the globally averaged vertical profile for all HEALPix bins with the corresponding north and south latitudes.} 
\label{fig:earth_TPs}
\end{figure}

% Explanation of TP figure 
Unless otherwise stated, we fix the TP profile in our forward model to the mean Earth 1D TP profile shown in Figure \ref{fig:earth_atm}. An important caveat to this is when we fit for an isothermal TP profile. Additionally, to provide an initial exploration of vertical sensitivity without performing expensive vertically-resolved retrievals, for a subset of our retrieval experiments, we run multiple cases with different globally- and latitudinally-averaged temperature and water profiles. 
Figure \ref{fig:earth_TPs} shows four characteristic TP profiles and water profiles that represent a substantial portion of the spatial variance within the data acquired for use with the VPL Earth Model \citep{Robinson2011}. These characteristic profiles correspond to averages within broad equatorially symmetric latitudinal bands (e.g. our $19^{\circ}$ profile corresponds to an average of both $19^{\circ}$ N and $19^{\circ}$ S latitudinal bins). Also shown are the average Earth profiles (black dashed line) from Figure \ref{fig:earth_atm} and a particular profile with the coldest surface in the VPL Earth Model database (blue dashed line). There is substantial variability between the different TP profiles at all relevant pressures from the surface to the upper atmosphere. The water profiles agree well above the tropopause and show the greatest difference in the troposphere, where more water vapor is seen closer to the equator. 
%We perform individual retrievals using each of these temperature and water profile pairs. 

% SMART Setup
We used a nominal \smart simulation setup for our high and low resolution Earth retrievals that, unless otherwise stated, remained fixed throughout these experiments. We fixed the mean molecular weight of the atmosphere at 29 g/mol, consistent with the Earth atmosphere, we fixed the surface pressure at 1 bar, and the solid-body planet radius and mass are fixed at 1 R$_{\oplus}$ and 1 M$_{\oplus}$, respectively.  This implicitly fixed the surface gravity at 9.81 m/s$^2$. We also placed the Earth at 1 AU from the Sun, which has a stellar radius of 1 R$_{\odot}$. These simplifying assumptions allow us to specifically focus on the retrieval of spectroscopic features while neglecting numerous parameters that are important for exoplanet atmospheric retrievals---such as the mean molecular weight and surface gravity, which are degenerate for transmission spectroscopy with the atmospheric temperature via the scale height \citep{Barstow2020}. These important considerations are beyond the scope of this study.   
We set the transit impact parameter to zero, and we turned off the stellar and thermal source functions to only simulate the transmission spectrum. Finally, for our ray tracing refraction simulations, we used an optical depth binning parameter for \smart of 0.75, which is large, but our tests have shown that negligible errors ($<1\%$) are introduced for transmission calculations even with large spectral mapping bin tolerances. For the majority of our experiments we opted to have our retrieval forward model return the transmission spectrum in units of altitude in kilometers above the surface (i.e., the effective transit altitude) to enhance physical intuition with respect to the planetary environment, rather than convert the data into units of transit depth, $(R_p/R_s)^2$. However, \jwst relevant retrievals in Section \ref{sec:results:jwst} were performed in the standard exoplanet transit depth. 

% Instrument convolution 
The final step in the \smarter forward model is to convolve the high resolution spectrum produced by \smart down to the lower resolution of the spectroscopic dataset. The exact form of the instrument convolution is flexible within \smarter and the approach should depend on the data at hand. Throughout this work, retrievals of the lower resolution transmission spectra use a tophat convolution function to bin the high resolution model spectrum (sampled at 1 cm$^{-1}$ wavenumber resolution) to the resolution of the binned data. Retrievals on the native high resolution ACE-FTS spectra use \smart's internal triangular slit convolution function with a FWHM set to match that of the instrument (0.02 cm$^{-1}$) and are oversampled by a factor of $2{\times}$ ($\Delta \nu = 0.00125$ cm$^{-1}$) to allow the spectra to be interpolated onto the observed wavelength grid. Although the ACE-FTS instrument has a spectral response function that is best described by a sinc function \citep{Bernath2017}, our initial fits using \smart's pre-existing triangular slit convolution function showed excellent agreement at high resolution using the FWHM of the ACE-FTS instrument. 

We note that although \smart has numerous features that are novel or rarely employed in exoplanet retrieval codes (e.g. ray tracing refraction, Monte Carlo scattering, aerosol phase functions, etc.), in this paper we use a baseline transmission forward model in \smarter that, unless otherwise stated, does not include these features and is roughly equivalent in complexity to other retrieval codes in the literature. This serves to aid in the extension of our results retrieving on the empirical Earth spectrum to other retrieval codes used throughout the literature that use similar complexity. It also serves the goal of validating \smarter prior to introducing additional complexity. These features will be explored in future work. 

\begin{deluxetable*}{r|ccc|ccc}
\tablewidth{0.95\linewidth}
\tablecaption{\label{tab:priors} Retrieval parameters and their respective priors.}
\tablehead{\multicolumn{1}{c}{} & \multicolumn{3}{c}{Model 1: ``scaled profiles''} & \multicolumn{3}{c}{Model 2: ``evenly-mixed''} \\
\cline{2-4} \cline{5-7}
\colhead{Parameter} & \colhead{Type} & \colhead{Initial\tablenotemark{a}} & \colhead{Prior}  & \colhead{Type} & \colhead{Initial} & \colhead{Prior}}
\startdata
Temperature, $T$ & fixed & --- & --- & linear isothermal\tablenotemark{b} & 255 K & $\mathcal{U}(100,400)$ \\
%Pressure, $P_s$ & fixed & 1 bar & $\delta_{x_0}$ & log & 1 bar & $\mathcal{U}(0.01, 10)$ \\
%Radius, $R_p$ & fixed & 1 R$_{\oplus}$ & $\delta_{x_0}$ & linear & 1 R$_{\oplus}$ & $\mathcal{U}(0.9, 1.1)$ \\
%Mass, $M_p$ & fixed & 1 M$_{\oplus}$ & $\delta_{x_0}$ & linear & 1 M$_{\oplus}$ & $\mathcal{U}(0.9, 1.1)$ \\
%MMW, $\mu$ & fixed & 28.96 g/mol & --- & self-consistent & $F(x)$ & $F(x)$ \\
\ce{H2O} & linearly scaled profile & 1.0 & $\mathcal{U}(0,100)$ & log evenly-mixed  & $-3.12$ & $\mathcal{U}(-12,0)$ \\
\ce{CO2} & linearly scaled profile & 1.0 & $\mathcal{U}(0,100)$ & log evenly-mixed  & $-3.42$ & $\mathcal{U}(-12,0)$ \\
\ce{O3} & linearly scaled profile & 1.0 & $\mathcal{U}(0,100)$ & log evenly-mixed   & $-5.88$ & $\mathcal{U}(-12,0)$ \\
\ce{O2} & linearly scaled profile & 1.0 & $\mathcal{U}(0,100)$ & log evenly-mixed   & $-0.67$ & $\mathcal{U}(-12,0)$ \\
\ce{N2} & linearly scaled profile & 1.0 & $\mathcal{U}(0,100)$ & log evenly-mixed   & $-0.11$ & $\mathcal{U}(-12,0)$ \\
\ce{CH4} & linearly scaled profile & 1.0 & $\mathcal{U}(0,100)$ & log evenly-mixed  & $-5.79$ & $\mathcal{U}(-12,0)$ \\
\ce{CO} & linearly scaled profile & 1.0 & $\mathcal{U}(0,100)$ & log evenly-mixed   & $-6.80$ & $\mathcal{U}(-12,0)$ \\
\ce{NO2} & linearly scaled profile & 1.0 & $\mathcal{U}(0,100)$ & log evenly-mixed  & $-8.50$ & $\mathcal{U}(-12,0)$ \\
\ce{N2O} & linearly scaled profile & 1.0 & $\mathcal{U}(0,100)$ & log evenly-mixed  & $-6.59$ & $\mathcal{U}(-12,0)$ \\
\ce{HNO3} & linearly scaled profile & 1.0 & $\mathcal{U}(0,100)$ & log evenly-mixed & $-8.34$ & $\mathcal{U}(-12,0)$ \\
CFC-11 & linearly scaled profile & 1.0 & $\mathcal{U}(0,100)$ & log evenly-mixed & $-9.64$ & $\mathcal{U}(-12,0)$ \\
CFC-12 & linearly scaled profile & 1.0 & $\mathcal{U}(0,100)$ & log evenly-mixed & $-9.29$ & $\mathcal{U}(-12,0)$ \\
\enddata
\tablecomments{$\mathcal{U}$ denotes a uniform prior distribution.}
\tablenotetext{a}{See Figure \ref{fig:earth_atm} for the initial profile (subject to subsequent scaling) for each of these parameters. }
\tablenotetext{b}{We also considered fixed TP profiles using Model 2.}
\end{deluxetable*}

% Inverse models
\subsubsection{Inverse Model} \label{sec:methods:inverse} 

% General SMARTER inverse modeling approach
We employ a multi-pronged inverse modeling approach similar to that of the \chimera retrieval model \citep{Line2013a, Line2014, Kreidberg2015, Feng2016, Batalha2017retrieval, Tremblay2020}. Each inverse model uses Bayes' Theorem to estimate the peak or characterize the shape of the posterior distribution. However, each method employs a different set of assumptions and numerical approaches that both affect the statistical robustness of the retrieved posteriors and the number of calls to the forward model required to converge. Three inverse models are currently available and can be used with any forward model within the \smarter framework. These retrieval models are: Optimal Estimation (OE) using \texttt{scipy.optimize.minimize} and \texttt{scipy.optimize.curve\_fit} \citep{Virtanen2019scipy, SciPy2020}, Nested Sampling (NS) using \multinest \citep{Feroz2008, Feroz2009, Feroz2013, Feroz2019} via the \pymultinest Python wrapper \citep{Buchner2014} and \dynesty \citep{Speagle2020}, and Markov Chain Monte Carlo (MCMC) using \emcee \citep{Foreman-Mackey2013}. 

% Nested Sampling
In this work we primarily use OE with \texttt{curve\_fit} when errors are small/negligible to derive maximum likelihood estimates (MLE), and NS with \dynesty to derive posterior distributions for cases with exoplanet relevant uncertainties. Nested sampling has been growing in popularity within the exoplanet atmospheric retrieval community, with most of the common retrieval codes now using it as their preferred inverse model \citep[e.g.][]{Benneke2013, Line2016, Lavie2017, Rocchetto2016, MacDonald2017, Barstow2020b}. This broad adoption stems from its efficiency---requiring ${\sim} 10 {\times}$ fewer forward model evaluations compared with traditional MCMC sampling---and its applicability in model selection problems, both of which are important for atmospheric retrievals. NS algorithms operate by iterating forward through nested shells of likelihood, progressively focusing on smaller and smaller regions of the prior volume. The procedure terminates once the marginal likelihood (evidence) is determined to some user-specified precision, for which we use the recommended value of 0.5 in log-evidence. 

% Priors 
Table \ref{tab:priors} provides the free parameters under consideration in our retrievals and their respective prior probabilities. All parameters have uninformative uniform priors. The retrievals using OE are initialized near their expected solutions based on the known vertical composition and temperature profile of the globally-averaged Earth.  

%%%%%%%%%%%%%%%%%%%%%%%%%%%%%%%%%%%%%%%%%%%%%%%%%%%%%%%%%%%%%%%%%%%%%%%%%%%%%%%%%%%%%%%%

\subsection{Retrieval Experiments} \label{sec:methods:setup}

% Overview
We considered the atmospheric retrieval analysis of the Earth spectrum shown in Fig. \ref{fig:earth_zoom} from a few different perspectives so as to (1) validate our retrieval model, (2) test common retrieval modeling assumptions, and (3) highlight different challenges and capabilities of exoplanet transmission spectroscopy. 

% Experiment 1: Low res
\paragraph{Low Resolution Data} In our first set of experiments, we used \smarter to fit the broad wavelength, low-resolution spectrum shown in the top panel of Fig. \ref{fig:earth_zoom} (dark blue line). 
This experiment was designed to test the accuracy of the forward model at a modest spectral resolution, more representative of exoplanet data, over a broad wavelength range to include as many gas absorbers as possible.

% Experiment 2
\paragraph{High Resolution Data} In our second set of experiments, we analyzed the high-resolution spectrum and performed retrievals at the native resolution of the ACE-FTS instrument. Since the high resolution spectrum resolves most individual absorption lines at these wavelengths, the spectrum provides excellent sensitivity to a large range of atmospheric altitudes (see Fig. 4 in \citetalias{Macdonald2019}), which we investigate with our retrievals.  
In this case, we started by fitting the full $2-14$ \um spectrum, and then we performed two narrow focused cases by individually fitting the 4.3 \um \ce{CO2} band ($3.7 - 5$ \um) and the 6 \um window ($5.0-7.2$ \um).  
%In this wavelength range, our forward model can be evaluated in about 30 seconds, compared to about 3 minutes for the full wavelength range. 
Fitting the $3.7 - 5$ \um wavelength range enables a more comprehensive assessment of the structure of the prominent \ce{CO2} band and the bulk atmospheric \ce{N2} abundance via \ce{N2-N2} CIA \citep{Schwieterman2015b}. Fitting the $5-7$ \um wavelength range enables a focused glimpse into one of the largest \ce{H2O} absorption bands found in the \citetalias{Macdonald2019} transmission spectrum, and a deeper dive into the \ce{O2-O2} and \ce{O2-N2} CIA bands recently highlighted by \citet{Fauchez2020}. %As a result, this wavelength range provides an opportunity to test for sensitivity to \ce{O2}, where the low resolution spectrum proved insensitive due to the lack of strong \ce{O2} bands beyond 2 \um. 
%Again, we started by retrieving gas VRM profile multiplicative constants. 
This experiment was designed to test the accuracy of our forward model at much higher spectral resolution than a typical exoplanet observation to further validate our model and to explore the limits of our 1D modeling approaches in the face of information rich transmission spectra of a known planet.

% Experiemnt 3 & 4
\paragraph{JWST Data} In our third set of experiments, we used the empirical Earth spectrum to generate synthetic \jwst data using the NIRSpec and MIRI LRS instruments, respectively. In this case, we assumed that the exoplanet TRAPPIST-1e possesses Earth's spectrum and applied a level of noise appropriate for 80 observed and stacked transits with each instrument. In this way, our mock observations with NIRSpec and MIRI LRS do not have the same level of uncertainty, but rather, they represent equivalent exposure times with two instruments that vary in wavelength range and spectral resolution.  We note that the Earth's atmosphere is not a good prediction for that of TRAPPIST-1e due primarily to the inconsistent evolutionary histories and photochemistry with the stellar UV spectrum. However, this experiment is designed simply to have noise properties consistent with expectations for a long term \jwst program. These \jwst-relevant retrievals use the \dynesty nested sampling inverse model to derive full posterior distributions for the model parameters.  
This experiment was designed to provide the context within which to interpret all of our findings in the reference frame of near-term exoplanet observations with \jwst.  

\section{Results} \label{sec:results}

Our results are organized into the following three investigations: (\ref{sec:results:low}) retrievals of the Earth's low resolution transmission spectrum, (\ref{sec:results:high}) retrievals of the Earth's high resolution transmission spectrum, and (\ref{sec:results:jwst}) retrievals of an Earth-analog TRAPPIST-1e spectrum as if it were observed with JWST.  

\subsection{Low Resolution Spectrum} \label{sec:results:low}

\subsubsection{Model 1}

% Earth spectrum low res
\begin{figure*}[t]
\centering
\includegraphics[width=0.97\textwidth]{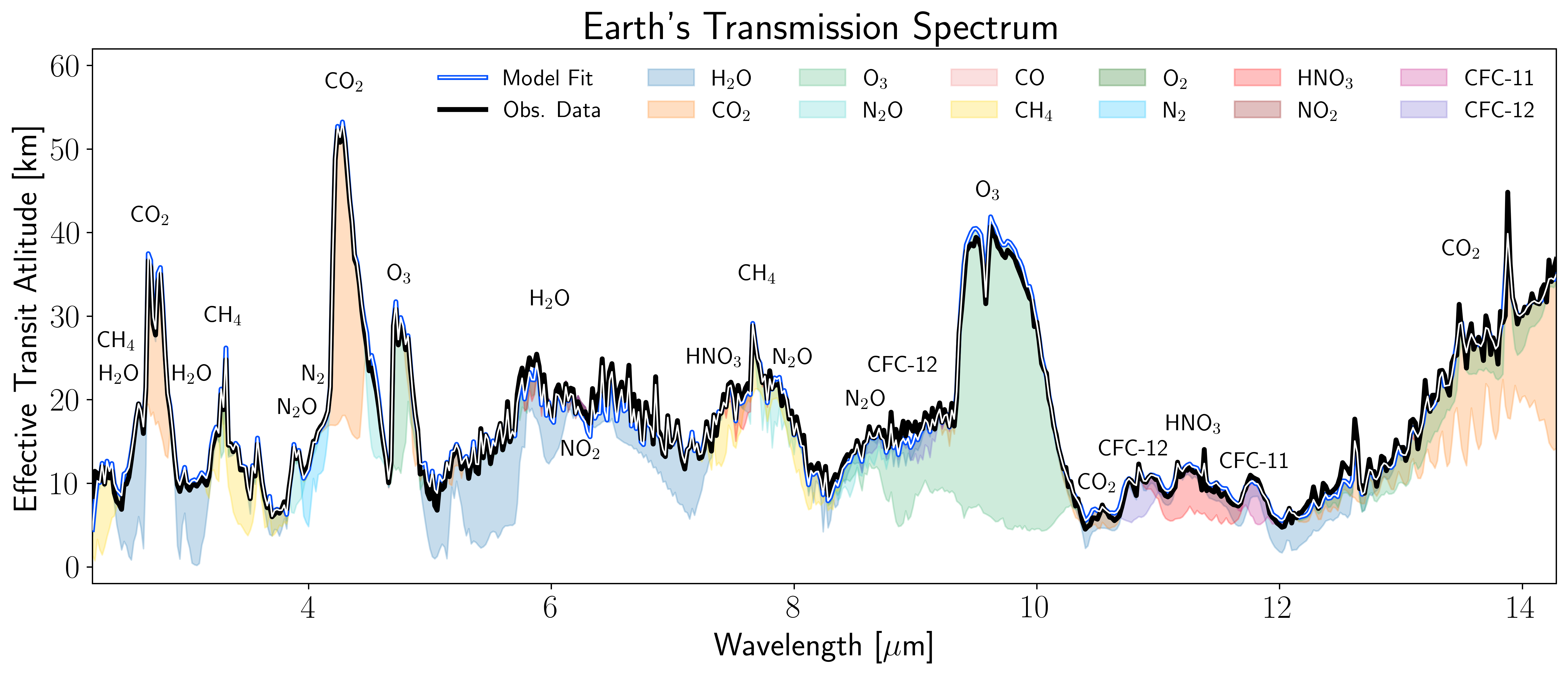}
\caption{Earth's observed cloud-free transit transmission spectrum from \citet{Macdonald2019} binned to a resolution of $\Delta \lambda = 0.02$ \um and fit using \smarter. The data are shown in black, the best-fitting model (Model 1) spectrum is shown in white and blue, and various colors are used to indicate each molecule's contribution to the model spectrum. The most prominent features are due to \ce{CO2} and \ce{O3}, but a rich diversity of habitability indicators, biosignatures, and technosignatures are present.} 
\label{fig:Earth_spectrum_fit}
\end{figure*}

% Earth low res best fit
We begin by fitting the low resolution spectrum shown in dark blue in Fig. \ref{fig:earth_zoom} using Model 1. Figure \ref{fig:Earth_spectrum_fit} shows our best-fitting solution to Earth's full $2-14$ \um transmission spectrum. 
We use semi-transparent colors to highlight the absorption features due to all molecules included in our model. Strong absorption features are clearly visible from \ce{H2O}, \ce{CO2}, and \ce{O3}, with smaller features due to \ce{CH4}, \ce{HNO3}, \ce{NO2}, \ce{N2O}, CFC-11, and CFC-12. We note that the \ce{N2-N2} CIA feature at 4.1 \um is also visible \citep[see][]{Schwieterman2015b} and provides leverage to constrain the bulk atmospheric \ce{N2} abundance. 
% Numerical agreement
There is substantial agreement between our best-fitting model and the observed data.  
Our best-fitting spectrum has a mean percent error of 5.15\% from the empirical transmission spectrum (with a maximum deviation of 5.45 km at 13.8 \um). We determined what the errors would need to be for our best fitting spectrum to have a reduced chi-squared of unity ($\chi^{2}_{red} = 1$). This yielded a characteristic error of $z_{err} = 0.94$ km. Measurement errors larger than this characteristic modeling error result in our model being an excellent fit, while measurement errors smaller than this result in our model potentially requiring additional sophistication to better fit the spectrum. By using the planet and star radius, this characteristic error (or any altitude values) can be scaled to a typical exoplanet measurement precision via  
\begin{equation}
\label{eqn:rescale}
    \Delta F_{err} = \left ( \frac{R_p + z_{err}}{R_s} \right )^2 - \left ( \frac{R_p}{R_s} \right )^2, 
\end{equation} 
which gives a mean uncertainty of about 0.024 ppm for an Earth-Sun exoplanet-analog transmission spectrum, or about 1.7 ppm for TRAPPIST-1e transiting TRAPPIST-1 \citep{VanGrootel2018}. These values are well below the noise floor of current and upcoming instruments \citep{Greene2016, Stevenson2019, Schlawin2021}. Exoplanet transmission spectra would need to be more precise than these ppm values for the deviations in our model fit with \smarter to be considered significant.  

% Table of MAP solutions
Table \ref{tab:earth_fit_low_res} provides the mixing ratio scale factors retrieved for each molecule included in our low resolution fit in the column labeled ``Model 1''. Most of the scale factors lie near unity, indicating good agreement with the expected mean composition of Earth's atmosphere. The molecules with large absorption features tend to have retrieved abundances that deviate by less than $10-30\%$ from the average Earth profiles. 
However, Earth's atmosphere naturally exhibits spatial variability for the temperature structure and many of the gas profiles, which is likely a source of systematic error in the observed and hemispherically-averaged transmission spectrum. That is, the ``true'' atmospheric composition of Earth cannot be accurately expressed using a single set of 1D vertical profiles, which on their own overconstrains the solution. To aid in our comparisons, we identify Earth's approximate spatial variability by calculating the percent deviation among the 1D profiles from the VPL Earth Model's atmospheric structure database \citep{Robinson2010, Robinson2011, Robinson2014, Schwieterman2015}. These values are reported for the majority of the molecules of interest in Table \ref{tab:earth_variability} for reference.  
% Profile variability seen across the Earth
For most molecules (e.g. \ce{H2O}, \ce{CO2}, \ce{O3}, \ce{N2O}, \ce{CH4}) our retrieved VMRs (with 1-$\sigma$ Gaussian uncertainties) are consistent within the spatial variance seen across the globe. For example, we retrieved an \ce{O3} VMR scale factor of $1.167 \pm 0.094$ times the expected mean profile, or ${\sim}17\%$ higher, however \ce{O3} profiles are known to vary spatially by about 25\%. Thus, our retrieved \ce{O3} result is consistent given both the uncertainties in the spectral fit and the systematic uncertainties that arise from fitting a 1D static atmospheric model to an observation of a time-varying 3D atmosphere. A similar result is found for \ce{H2O} where we retrieved a slightly lower than average VMR scale factor of $0.72 \pm 0.21$, which is consistent within the 30\% spatial and temporal variability exhibited across the Earth. We note that \ce{O2}, \ce{N2}, and \ce{CO} have poorly constrained mixing ratios due to the lack of substantial absorption features across the studied wavelength range, although greater sensitivity to these molecules is achieved in our subsequent high resolution retrievals. For \ce{HNO3}, CFC-11, and CFC-12, we retrieve more than double the expected Earth average mixing ratios. These findings are consistent with those of \citet{Schreier2018} who also required elevated mixing ratios for these species in their comparisons to ACE-FTS transmission spectra relative to the mid-latitude mean profiles, likely due to increased concentrations of these gases at high northern and southern latitudes \citep{Remedios2007}. 

\begin{deluxetable}{r|c|cc}
\tablewidth{0.96\textwidth}
\tabletypesize{\normalsize}
\tablecaption{Retrieved parameters for Earth's atmosphere from a fit to Earth's low-resolution transmission spectrum\label{tab:earth_fit_low_res}}
%\tablehead{   \colhead{Parameters} & \colhead{Fixed TP scaled}   &   \colhead{isothermal even} & \colhead{fixed TP even} }
\tablehead{\multicolumn{1}{c}{} & \multicolumn{1}{c}{Model 1} & \multicolumn{2}{c}{Model 2} \\
\cline{2-3} \cline{3-4}
\colhead{Parameters} & \colhead{Fixed TP}   &   \colhead{Fixed TP} & \colhead{Isothermal} }
\startdata
$T_{0}$ [K]   &       ---           &        ---           &      $192.3 \pm 3.7$ \\
    \ce{H2O}  &    $0.72 \pm 0.21$  &  $-5.177 \pm 0.093$  &     $-4.69 \pm 0.17$ \\
    \ce{CO2}  &    $1.10 \pm 0.14$  &  $-3.404 \pm 0.060$  &     $-2.58 \pm 0.14$ \\
     \ce{O3}  &  $1.167 \pm 0.094$  &  $-5.582 \pm 0.068$  &     $-5.05 \pm 0.16$ \\
    \ce{N2O}  &    $1.20 \pm 0.31$  &  $-6.507 \pm 0.027$  &   $-6.571 \pm 0.044$ \\
     \ce{CO}  &      $1.0 \pm 1.4$  &  $-6.655 \pm 0.061$  &     $-6.82 \pm 0.10$ \\
    \ce{CH4}  &    $1.18 \pm 0.21$  &    $-5.77 \pm 0.21$  &     $-5.26 \pm 0.25$ \\
     \ce{O2}  &    $0.94 \pm 0.56$  &         $-4 \pm 50$  &      $-11.9 \pm 2.1$ \\
     \ce{N2}  &      $1.0 \pm 2.1$  &    $-0.32 \pm 0.21$  &     $-0.31 \pm 0.74$ \\
   \ce{HNO3}  &    $2.06 \pm 0.27$  &  $-7.935 \pm 0.028$  &   $-7.655 \pm 0.042$ \\
    \ce{NO2}  &    $0.94 \pm 0.65$  &  $-8.409 \pm 0.041$  &   $-8.319 \pm 0.069$ \\
      CFC-11  &    $3.83 \pm 0.81$  &  $-9.150 \pm 0.037$  &  $-10.122 \pm 0.056$ \\
      CFC-12  &    $4.41 \pm 0.69$  &  $-8.820 \pm 0.031$  &   $-8.744 \pm 0.044$ \\
        \hline
         MPE  &              5.15\% &              11.74\% &               11.60\% \\
   $z_{err}$  &          $0.94$ km  &           $1.81$ km  &            $2.07$ km \\
\enddata
\tablecomments{Model 1 and Model 2 refer to the ``scaled profiles'' and ``evenly-mixed'' retrieval models, respectively. The expected Earth profiles for each molecule are displayed in Figure \ref{fig:earth_atm}, with the approximate $\log_{10}$ VMRs for Earth's atmosphere given in Table \ref{tab:priors}. The best fitting spectrum for Model 1 is shown in Figure \ref{fig:Earth_spectrum_fit}. The last two rows show the mean percentage error (MPE) and characteristic vertical uncertainty ($z_{err}$) for each spectral fit.}  
\end{deluxetable}

\begin{deluxetable}{rc}
\tablewidth{0.98\textwidth}
\tabletypesize{\normalsize}
\tablecaption{Approximate percent deviations in molecular species observed across Earth due to spatial and temporal variability \label{tab:earth_variability}}
%\tablenum{3}
\tablehead{\colhead{Molecule} & \colhead{Spatial/Temporal Deviations}} 
\startdata
\ce{H2O}   &   30\%   \\
\ce{CO2}   &   ${<}1$\%  \\
\ce{O3}    &  25\%   \\
\ce{N2O}   &  23\%   \\
\ce{CO}    &  28\%   \\
\ce{CH4}   &  34\%   \\
\ce{O2}    &  ${<}1$\%    \\
\ce{N2}    &  ${<}1$\%    \\
%\ce{HNO3}  &  2.370 
\enddata
%\tablecomments{}
\end{deluxetable}

%\subsubsection{Evenly Mixed Vertical Profiles}

\subsubsection{Model 2}

% Intro to evenly mixed results
We also fit the low resolution transmission spectrum using a forward model with evenly mixed gas profiles that exhibit no vertical structure (``Model 2''). Model 2 uses the $\log_{10}$ volume mixing ratio of each gas as free parameters and is more representative of approaches applied to exoplanet atmospheric studies. We examine two treatments for the thermal structure: (1) fixing the TP profile at Earth's characteristic spatially averaged vertical thermal structure and (2) fitting for a single-valued isothermal temperature structure. 

% Breakdown of table results
Table \ref{tab:earth_fit_low_res} presents our retrieval findings for these evenly mixed Model 2 fits. Both evenly mixed models provide good fits to the low resolution Earth spectrum and although they are quantitatively worse than the scaled profiles model (``Model 1'') as seen by the mean percent error summary statistic, they are qualitatively similar to the spectrum shown in Figure \ref{fig:Earth_spectrum_fit}. The isothermal model fit has a mean percentage error marginally lower than the fixed TP model, indicating that it fits the spectrum slightly better. This is due primarily to the additional flexibility afforded to the isothermal model, coupled to the fact that the mean Earth TP that we assumed need not be the correct nor optimal TP profile to fit this observed spectrum. However, the fixed TP model offers more precise and accurate results for most of the gases considered. For example, the molecules with the most prominent absorption features in the spectrum (see Fig. \ref{fig:Earth_spectrum_fit})---\ce{CO2}, \ce{O3}, \ce{H2O}, and \ce{CH4}---all had retrieved VMRs that were biased high when we fit for the isothermal temperature compared with our more accurate findings using a realistic fixed TP profile. Thus the slightly improved fit to the spectrum achieved with the isothermal model came at the cost of inaccurate abundance constraints. 

% Comparison with vertically resolved profiles
Our retrieved isothermal temperature of ${\sim}192$ K is slightly lower than the mean temperature of the tropopause, which occurs near 0.1 bar, but is consistent with tropopause temperature measurements in the tropics \citep{Hoinka1999}.  
For gases with substantial variability as a function of altitude, such as \ce{H2O}, \ce{O3}, \ce{N2O}, \ce{CH4}, and \ce{CO}, our retrieved evenly mixed volume mixing ratios with the fixed TP model are consistent with true values of the vertical profiles at pressures near ${\sim}0.1$ bar. Our \ce{H2O} constraint is consistent with the mean \ce{H2O} profile between $0.1-0.2$ bar. Our \ce{O3} constraint is consistent with the mean \ce{O3} profile at ${\sim}0.04$ bar. \ce{CH4} and \ce{N2O} are consistent in the lower atmosphere at pressures higher than ${\sim}0.03$ bar. \ce{CO} is consistent at pressures between $0.02-0.05$ bar. 

\subsubsection{Model 2 with Noise Added to the Spectrum} 
\label{sec:res:low:noise}

To further validate the retrieval part of \smarter, we perform one final test on the low resolution, broad wavelength transmission spectrum with 10\% random noise added to each spectral data point. We fit this noisy spectrum using Model 2 with \smarter's \dynesty nested sampling implementation with the isothermal TP profile included as a free parameter. The use of 10\% errors is motivated by the residual modeling errors for Model 2 in Table \ref{tab:earth_fit_low_res}. Despite each spectral point having 10\% uncertainties, the large number of data points results in the spectrum deviating from a flat line by $>30{\sigma}$, indicating that the spectrum far exceeds current telescope capabilities for rocky exoplanets \linktocite{Lustig-YaegerFu2023}{(e.g., Lustig-Yaeger \& Fu et al.} \citeyear{Lustig-YaegerFu2023}; \citealp{Moran2023}) Therefore this retrieval with noise enables robust validation of the model using nested sampling. 

% Earth spectrum low res with noise
\begin{figure*}[t]
\centering
\includegraphics[width=0.98\textwidth]{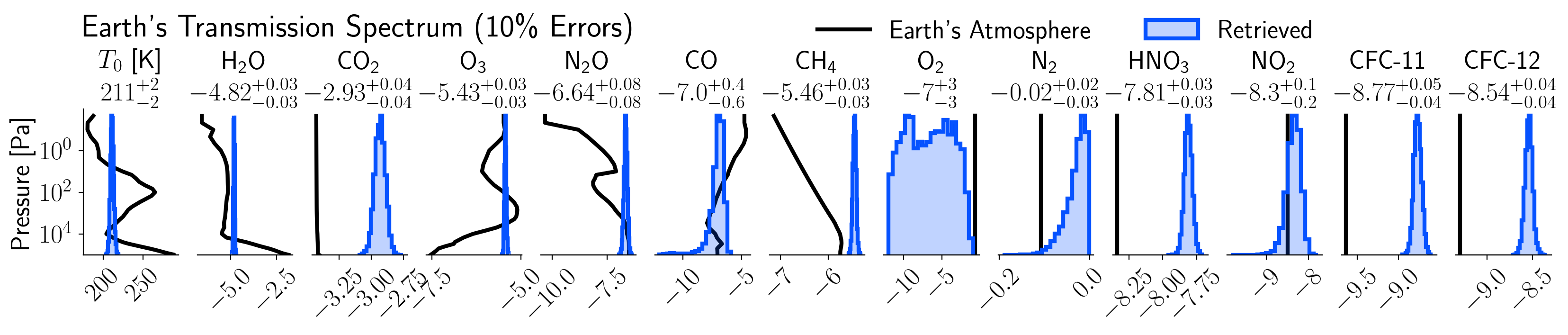}
\includegraphics[width=0.98\textwidth]{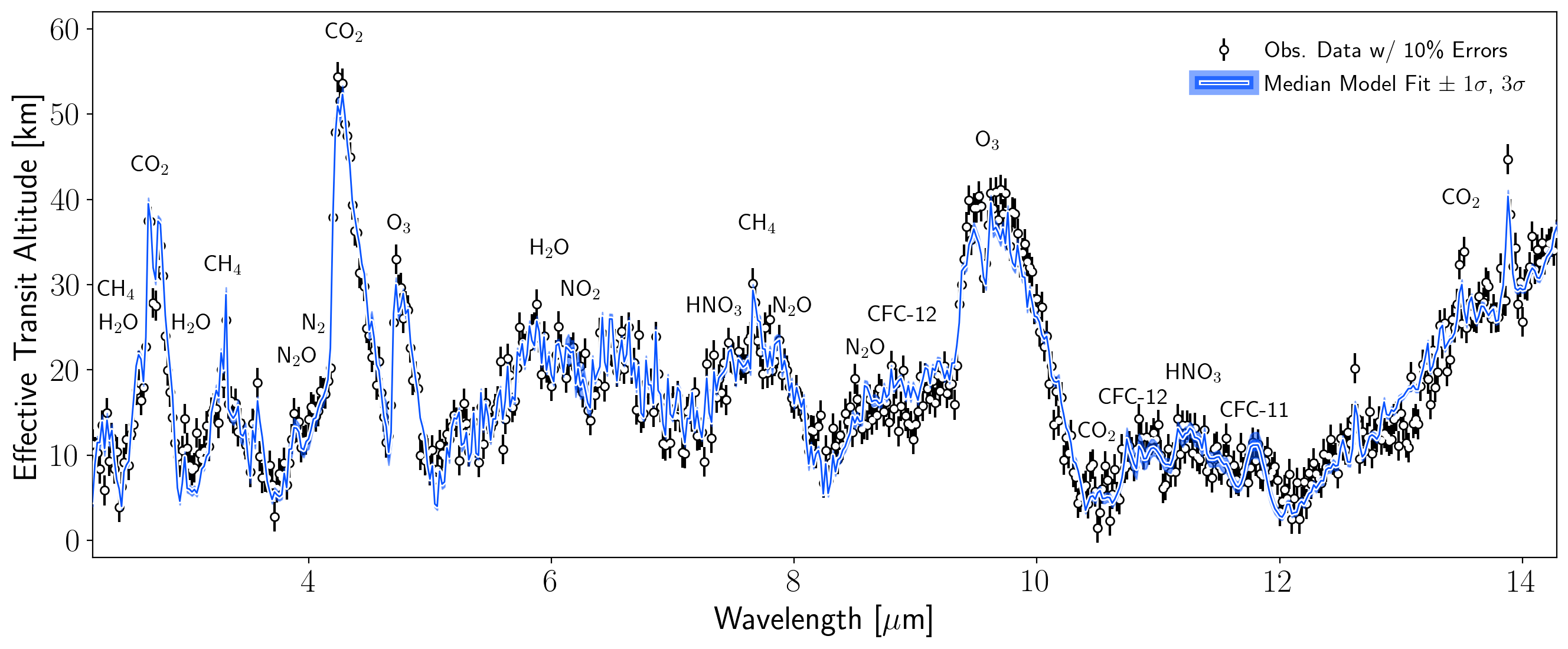}
\caption{Retrieval results on Earth's observed cloud-free transit transmission spectrum binned to a resolution of $\Delta \lambda = 0.02$ \um, with 10\% random noise added, and fit using the \dynesty nested sampling algorithm. The upper set of panels show 1D posterior histograms (blue) for each model parameter compared to the respective ``true'' atmospheric profiles from Figure \ref{fig:earth_atm}. The median and 1$\sigma$ intervals are listed above each panel. The lower panel shows the spectral data (black with uncertainties) and the median retrieved model spectrum (white and blue) surrounded by 1$\sigma$ and 3$\sigma$ credibility envelopes (often thinner than the median model line width).}  
\label{fig:Earth_spectrum_fit_noise}
\end{figure*}

Figure \ref{fig:Earth_spectrum_fit_noise} shows the results of our low resolution retrieval on the 10\% noise-added spectrum. Overall, the results are similar to the previous OE findings in that we reproduce the general characteristics of Earth's atmosphere, including major absorbing species and the bulk \ce{N2} atmospheric composition. However, the assumed uniform vertical composition introduces some biases, including overestimates of gases like \ce{CO2} and \ce{CH4}. The median retrieved spectral model has $\chi^{2}_{red} = 1.94$. Residuals between the data and fit are relatively small for the bands from evenly-mixed \ce{CO2}, but are more apparent for the bands of \ce{O3}, especially near 9.6 \um.  This disparity indicates that ozone's non-uniform vertical distribution and the resultant thermal inversion are important parameters to improve the fit for this panchromatic Earth transmission spectrum. \ce{N2} is clearly favored as the bulk atmospheric constituent, but it is overestimated, while \ce{O2} is not identified and the retrieval favors mixing ratios $< 10\%$ at 2$\sigma$. The absolute precision on each retrieved parameter is not particularly meaningful as it depends on our choice of data errors, but the relative precision between each atmospheric gas indicates their prominence in the spectrum. In this wavelength range, our most challenging gases to retrieve are \ce{O2}, \ce{CO}, and \ce{NO2}. Although \smarter began its nested sampling retrieval with broad and uniform priors (Table \ref{tab:priors}) it converged to the correct atmospheric composition of Earth (modulo the aforementioned model limitations), demonstrating the validity of the retrieval model on a high quality transmission spectrum of a habitable and inhabited planet. 

\subsection{High Resolution Spectrum} \label{sec:results:high}

% Intro to high res
We now shift to examine the Earth transmission spectrum at the high native spectral resolution ($R=100,000$ at 5 \um) of the occultation-derived measurements. 
% High res instrument convolution
To simulate the high-resolution spectrum with our forward model, we used \smart's internal triangular slit function with the appropriate FWHM for the ACE-FTS instrument (0.02 cm$^{-1}$). We output spectra on a wavenumber grid with a sampling resolution of 0.00125 cm$^{-1}$, which oversamples the data by a factor of 2. We then interpolated the model spectrum onto the observed data wavelength grid. 

% Intro to wavelength range cases
We performed three separate assessments using data covering different wavelength ranges. First, we fit the full $2 - 14$ \um wavelength range. Second, we focus on the more narrow $3.7 - 5$ \um wavelength range to specifically isolate and examine the 4.3 \um \ce{CO2} band. Third, we focus on the $5 - 7.2$ \um wavelength range to isolate the strong 6 \um \ce{H2O} band and explore the overlapping contributions from \ce{O2} and \ce{N2} CIA. 

\subsubsection{Full Spectrum: $2 - 14$ \um}

\begin{figure*}[t!]
\centering
\includegraphics[width=0.93\textwidth]{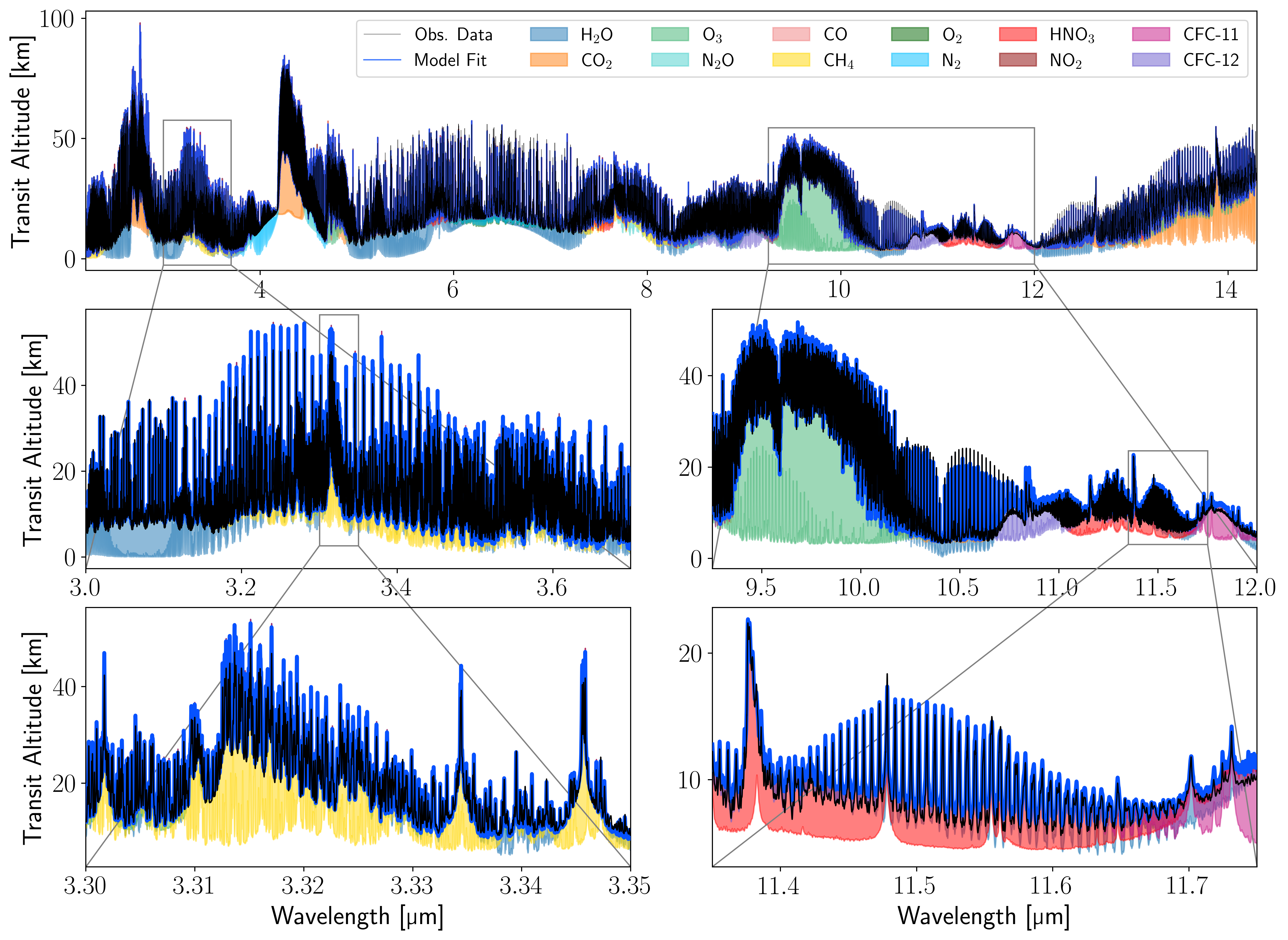}
\caption{Best fit model of Earth's high resolution transmission spectrum in the full $2 - 14$ \um wavelength range using the mean Earth TP profile. The data are shown in black, the best-fitting model spectrum is shown in blue, and various colors are used to indicate each molecule's contribution to the model spectrum. } 
\label{fig:Earth_spectrum_fit_hr_full}
\end{figure*}

Figure \ref{fig:Earth_spectrum_fit_hr_full} shows our best fit model to the high resolution Earth transmission spectrum over the full $2-14$ \um wavelength range using Model 1 and assuming the mean Earth TP profile. The zoomed panels highlight the 3.3 \um \ce{CH4} band in the NIR (left column of subplots) and the $9-12$ \um MIR range that contains significant absorption features from \ce{O3}, \ce{CO2} (${\sim}10.5$ \um), CFC-12, \ce{HNO3}, and CFC-11 (right column of subplots). 

% Earth spectrum plot

Table \ref{tab:earth_fit3_full} shows the retrieval results for our fit to the full wavelength range for the high resolution transmission spectrum using Model 1 with a variety of different Earth TP profiles. In general, the VMR scale factors reside near unity and the fits to the spectrum are good with average errors on the fits $<10\%$. The reported $1\sigma$ errors on the retrieved parameters are much smaller than those from our fits to the low resolution spectrum, which increases the significance of the deviations from the Earth averaged profiles. The deviations from unity still largely fall within the range of spatial and temporal variability reported in Table \ref{tab:earth_variability}. For each molecule, we see that the retrieved VMR scale factors exhibit greater variance with the choice of TP profile than the uncertainties on the parameters themselves. This indicates an important relationship between the TP profile and the ability to accurately retrieve gas abundances. However, an interesting extension of this is seen in Table \ref{tab:earth_fit3_full} where different TP profiles yield retrieved gas VMRs that are most consistent with the Earth averaged profiles for different molecules. For example, the $19^{\circ}$ latitude TP profile yields the most consistent \ce{H2O} retrieval, the $41^{\circ}$ latitude TP profile yields the most consistent \ce{CO2} retrieval, and the $0^{\circ}$ latitude TP profile yields the most consistent \ce{O3} retrieval. Thus, we may not have considered the optimal TP profile or there may not be an optimal profile, but rather the transmission spectrum may bear the markings of multiple TP profiles that exist simultaneously at different locations. 

\begin{deluxetable*}{r|rrrrrr}
\tablewidth{0.98\textwidth}
\tabletypesize{\footnotesize}
\tablecaption{Retrieved molecule VMR scale factors from OE fits to Earth's high-resolution transmission spectrum in the $2 - 14$ \um wavelength range\label{tab:earth_fit3_full}}
\tablehead{   \colhead{Parameters}    &   \colhead{avg}   &   \colhead{lat=$0^{\circ}$}   &   \colhead{lat=$19^{\circ}$}   &   \colhead{lat=$41^{\circ}$}   &   \colhead{lat=$66^{\circ}$}  &  \colhead{cold}}
\startdata
    \ce{H2O} & $1.3082 \pm 0.0061$ & $0.8794 \pm 0.0053$ & $1.0411 \pm 0.0053$ & $1.6548 \pm 0.0073$ & $2.2986 \pm 0.0094$ & $2.2324 \pm 0.0095$ \\
    \ce{CO2} & $1.0213 \pm 0.0038$ & $0.9373 \pm 0.0039$ & $0.9343 \pm 0.0037$ & $1.0147 \pm 0.0037$ & $1.2375 \pm 0.0045$ & $1.3753 \pm 0.0052$ \\
     \ce{O3} & $1.1562 \pm 0.0036$ & $1.0616 \pm 0.0036$ & $1.0934 \pm 0.0036$ & $1.2271 \pm 0.0038$ & $1.3155 \pm 0.0040$ & $1.5039 \pm 0.0047$ \\
    \ce{N2O} & $0.9574 \pm 0.0067$ & $0.8933 \pm 0.0072$ & $0.9403 \pm 0.0070$ & $1.1024 \pm 0.0073$ & $1.2669 \pm 0.0081$ & $1.4247 \pm 0.0094$ \\
     \ce{CO} &   $0.751 \pm 0.022$ &   $0.518 \pm 0.022$ &   $0.638 \pm 0.022$ &   $0.433 \pm 0.019$ &   $0.703 \pm 0.021$ &   $0.878 \pm 0.022$ \\
    \ce{CH4} & $1.2221 \pm 0.0039$ & $1.0400 \pm 0.0038$ & $1.0968 \pm 0.0038$ & $1.2268 \pm 0.0039$ & $1.4180 \pm 0.0044$ & $1.5784 \pm 0.0050$ \\
     \ce{O2} &   $0.692 \pm 0.012$ &   $0.746 \pm 0.014$ &   $0.637 \pm 0.013$ &   $0.527 \pm 0.012$ &   $0.774 \pm 0.013$ &   $0.947 \pm 0.014$ \\
     \ce{N2} &     $2.8 \pm 0.0$   &       $3.0 \pm 0.0$ &       $3.1 \pm 0.0$ &       $3.9 \pm 0.0$ &       $1.4 \pm 0.0$ &     $1.7 \pm 0.0$ \\
   \ce{HNO3} &   $2.329 \pm 0.016$ &   $2.412 \pm 0.018$ &   $2.338 \pm 0.016$ &   $3.127 \pm 0.020$ &   $2.683 \pm 0.018$ &   $2.800 \pm 0.019$ \\
    \ce{NO2} &   $0.738 \pm 0.015$ &   $0.730 \pm 0.016$ &   $0.632 \pm 0.015$ &   $0.810 \pm 0.015$ &   $0.774 \pm 0.016$ &   $0.914 \pm 0.017$ \\
      CFC-11 &   $6.090 \pm 0.074$ &   $3.761 \pm 0.061$ &   $4.249 \pm 0.060$ &   $4.549 \pm 0.059$ &   $7.763 \pm 0.084$ &   $6.942 \pm 0.080$ \\
      CFC-12 &   $7.222 \pm 0.053$ &   $4.900 \pm 0.044$ &   $5.455 \pm 0.045$ &   $5.323 \pm 0.043$ &   $6.640 \pm 0.049$ &   $6.964 \pm 0.053$ \\
      \hline
         MPE &              8.28\% &             10.91\% &              9.42\% &              8.03\% &              7.66\% &              8.06\% \\
   $z_{err}$ &           $1.60$ km &           $1.73$ km &           $1.66$ km &           $1.60$ km &           $1.58$ km &           $1.63$ km \\
\enddata
\tablecomments{Columns are labeled in reference to the fixed TP profiles shown in Figure \ref{fig:earth_TPs}. The fiducial Earth profiles for each molecule are displayed in Figure \ref{fig:earth_atm}. The last two rows show the mean percentage error (MPE) and characteristic vertical uncertainty ($z_{err}$) for each spectral fit.} 
\end{deluxetable*}

\begin{figure}
\centering
\includegraphics[width=0.49\textwidth]{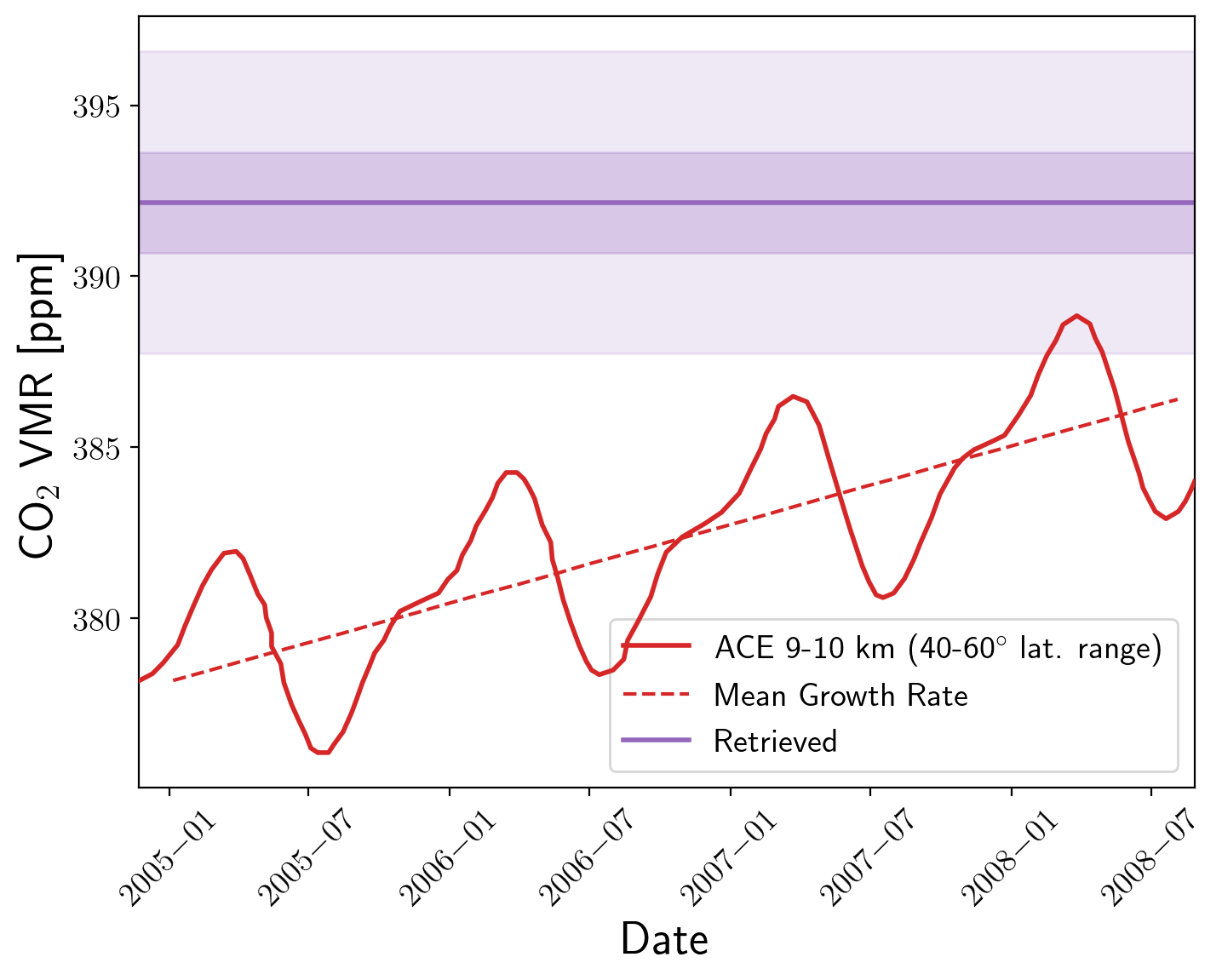}
\caption{Retrieved \ce{CO2} VMR from the high resolution full wavelength spectrum compared to the temporal evolution of the \ce{CO2} concentration retrieved from ACE as reported in \citet{Foucher2011}. The purple line shows our retrieved \ce{CO2} VMR with dark and light shading to indicate the $1 \sigma$ and $3 \sigma$ credible intervals, respectively, from our model fixed at the average Earth TP profile. The agreement is quite good considering that the transmission spectrum fit using our model is vertically integrated over the atmospheric column, spatially integrated over the entire globe, and temporally integrated over many years of SCISAT mission operation.} 
\label{fig:retrieved_co2}
\end{figure}

Figure \ref{fig:retrieved_co2} compares our retrieved \ce{CO2} concentration to the temporal evolution of \ce{CO2} that was reported using the same ACE-FTS measurements during the first few years of data collection. The nature of the empirical transmission spectrum from \citetalias{Macdonald2019} being vertically integrated, spatially integrated, and temporally integrated precludes such an in-depth temporal study as that performed by \citet{Foucher2011}. Nonetheless, our findings are statistically consistent with the results from \citet{Foucher2011} for \ce{CO2}. The exact cause of the remaining deviation from the expected value could be the result of one or more complicating effects. Since ACE-FTS continues to operate, it is possible that the \citetalias{Macdonald2019} spectrum contains measurements beyond 2008, which would agree with our slightly higher \ce{CO2} result considering the mean growth rate for \ce{CO2}, however, it is not clear from \citetalias{Macdonald2019} exactly which years were used for the construction of the transmission spectrum. We also cannot rule out non-local thermodynamic equilibrium (NLTE) effects in the upper atmosphere that could be causing minor residuals seen in the cores of the \ce{CO2} lines, since NLTE effects are neglected in our model. Finally, systematic errors from the instrument could also explain the slight discrepancy, as discussed in more detail in Section \ref{sec:discussion}.  

% Retrieved CO2 VMR

Table \ref{tab:earth_fit_high_res_even_full} lists the retrieved atmospheric parameters using the evenly mixed atmospheric composition model (Model 2) for one retrieval with a free isothermal TP profile and a set of retrievals with the TP profile fixed at the various profiles shown in Figure \ref{fig:earth_TPs}. The model with the free isothermal TP profile provides a better fit to the data ($15.21\%$ error) compared to the models with fixed TP profiles ($21.3\%$ error on average), which show minimal variation between their results. However, the isothermal model retrieves biased results that systematically overestimate the true gas mixing ratios, similar to our results at low resolution. For example, the isothermal model retrieves a \ce{CO2} VMR of 2032 ppm while the retrievals with fixed TP profiles find \ce{CO2} between 296 ppm and 500 ppm---values that are much more consistent with Earth's known \ce{CO2} (e.g., Figure \ref{fig:retrieved_co2}). Compared to the results shown using the scaled profiles retrieval model (Model 1), the evenly mixed model (Model 2) provides worse fits to the high resolution spectrum and the resulting gas mixing ratios exhibit less variance when switching between different fixed TP profiles. 

\begin{deluxetable*}{r|ccccccc}
\tablewidth{0.98\textwidth}
\tabletypesize{\scriptsize}
\tablecaption{Retrieved parameters for Earth's atmosphere from a fit to Earth's high-resolution transmission spectrum in the $2 - 14$ \um wavelength range using an evenly mixed atmospheric model (Model 2)\label{tab:earth_fit_high_res_even_full}}
%\tablehead{   \colhead{Parameters} & \colhead{Fixed TP scaled}   &   \colhead{isothermal even} & \colhead{fixed TP even} }
\tablehead{
\colhead{Parameters} & \colhead{Isothermal TP} & \colhead{avg} & \colhead{lat=$0^{\circ}$} & \colhead{lat=$19^{\circ}$} & \colhead{lat=$41^{\circ}$} & \colhead{lat=$66^{\circ}$} & \colhead{cold}}
\startdata
 $T_{0}$ [K] &    $191.912 \pm 0.066$   & ---                     & ---                     & ---                     & ---                     & ---                     & ---                     \\
    \ce{H2O} &   $-4.6893 \pm 0.0029$   &    $-5.1769 \pm 0.0026$ &    $-5.2761 \pm 0.0030$ &    $-5.0664 \pm 0.0024$ &    $-5.1771 \pm 0.0028$ &    $-5.1046 \pm 0.0026$ &    $-4.9887 \pm 0.0025$ \\
    \ce{CO2} &   $-2.6921 \pm 0.0023$   &    $-3.4953 \pm 0.0015$ &    $-3.5284 \pm 0.0015$ &    $-3.5420 \pm 0.0016$ &    $-3.5062 \pm 0.0015$ &    $-3.4084 \pm 0.0015$ &    $-3.3016 \pm 0.0014$ \\
     \ce{O3} &   $-5.0999 \pm 0.0036$   &    $-5.4944 \pm 0.0029$ &    $-5.5608 \pm 0.0030$ &    $-5.4291 \pm 0.0028$ &    $-5.5077 \pm 0.0029$ &    $-5.4465 \pm 0.0029$ &    $-5.3241 \pm 0.0027$ \\
    \ce{N2O} & $-6.05010 \pm 0.00067$   &  $-6.53024 \pm 0.00062$ &  $-6.52583 \pm 0.00062$ &  $-6.52594 \pm 0.00063$ &  $-6.52506 \pm 0.00062$ &  $-6.42177 \pm 0.00063$ &  $-6.44467 \pm 0.00063$ \\
     \ce{CO} &   $-6.5242 \pm 0.0015$   &    $-6.7230 \pm 0.0014$ &    $-6.7446 \pm 0.0013$ &    $-6.7151 \pm 0.0014$ &    $-6.7156 \pm 0.0014$ &    $-6.6341 \pm 0.0014$ &    $-6.6364 \pm 0.0014$ \\
    \ce{CH4} &   $-5.5228 \pm 0.0034$   &    $-5.8521 \pm 0.0031$ &    $-5.8930 \pm 0.0031$ &    $-5.9399 \pm 0.0032$ &    $-5.8516 \pm 0.0031$ &    $-5.8977 \pm 0.0033$ &    $-5.8493 \pm 0.0032$ \\
     \ce{O2} &     $-1.868 \pm 0.027$   &      $-1.382 \pm 0.026$ &    $-0.9128 \pm 0.0092$ &      $-1.301 \pm 0.025$ &      $-1.048 \pm 0.013$ &      $-1.410 \pm 0.027$ &      $-1.219 \pm 0.018$ \\
     \ce{N2} &   $-0.0057 \pm 0.0043$   &    $-0.1198 \pm 0.0026$ &    $-0.1934 \pm 0.0029$ &    $-0.0717 \pm 0.0023$ &    $-0.1261 \pm 0.0026$ &    $-0.0058 \pm 0.0024$ &    $-0.0168 \pm 0.0025$ \\
   \ce{HNO3} &   $-7.6634 \pm 0.0011$   &    $-8.2517 \pm 0.0012$ &    $-8.2489 \pm 0.0012$ &    $-8.2791 \pm 0.0012$ &    $-8.2519 \pm 0.0012$ &    $-8.2210 \pm 0.0012$ &    $-8.2354 \pm 0.0011$ \\
    \ce{NO2} &   $-7.9282 \pm 0.0012$   &    $-8.6241 \pm 0.0013$ &    $-8.6665 \pm 0.0014$ &    $-8.6294 \pm 0.0013$ &    $-8.6324 \pm 0.0014$ &    $-8.5310 \pm 0.0013$ &    $-8.5662 \pm 0.0013$ \\
      CFC-11 &   $-9.5848 \pm 0.0019$   &    $-9.2770 \pm 0.0018$ &    $-9.2829 \pm 0.0019$ &    $-9.2844 \pm 0.0019$ &    $-9.2671 \pm 0.0018$ &    $-9.2845 \pm 0.0018$ &    $-9.2624 \pm 0.0018$ \\
      CFC-12 &   $-8.9630 \pm 0.0017$   &    $-9.2019 \pm 0.0018$ &    $-9.2179 \pm 0.0018$ &    $-9.2155 \pm 0.0018$ &    $-9.1963 \pm 0.0018$ &    $-9.1698 \pm 0.0018$ &    $-9.1775 \pm 0.0017$ \\ 
        \hline 
         MPE &                 15.21\%  &                  21.19\% &                  21.76\% &                  21.37\% &                  21.38\% &                  21.84\% &                  20.71\% \\
   $z_{err}$ &              $2.42$ km   &               $2.90$ km &               $2.93$ km &               $2.92$ km &               $2.92$ km &               $2.95$ km &               $2.87$ km \\
\enddata
\tablecomments{Columns with fixed TP profiles are labeled in reference to those shown in Figure \ref{fig:earth_TPs}. The expected Earth profiles for each molecule are displayed in Figure \ref{fig:earth_atm}, with the approximate $\log_{10}$ VMRs for Earth's atmosphere given in Table \ref{tab:priors}. The last two rows show the mean percentage error (MPE) and characteristic vertical uncertainty ($z_{err}$) for each spectral fit.} 
\end{deluxetable*}

\subsubsection{Narrow Focus: $3.7 - 5$ \um}

We now narrow our focus to the $3.7 - 5$ \um wavelength range. This spectral interval is particularly notable for the strong 4.3 \um \ce{CO2} band, the 4.7 \um \ce{O3} band, and the 4.2 \um \ce{N2-N2} CIA band \citep{Schwieterman2015b}.  

\begin{figure*}[t!]
\centering
\includegraphics[width=0.93\textwidth]{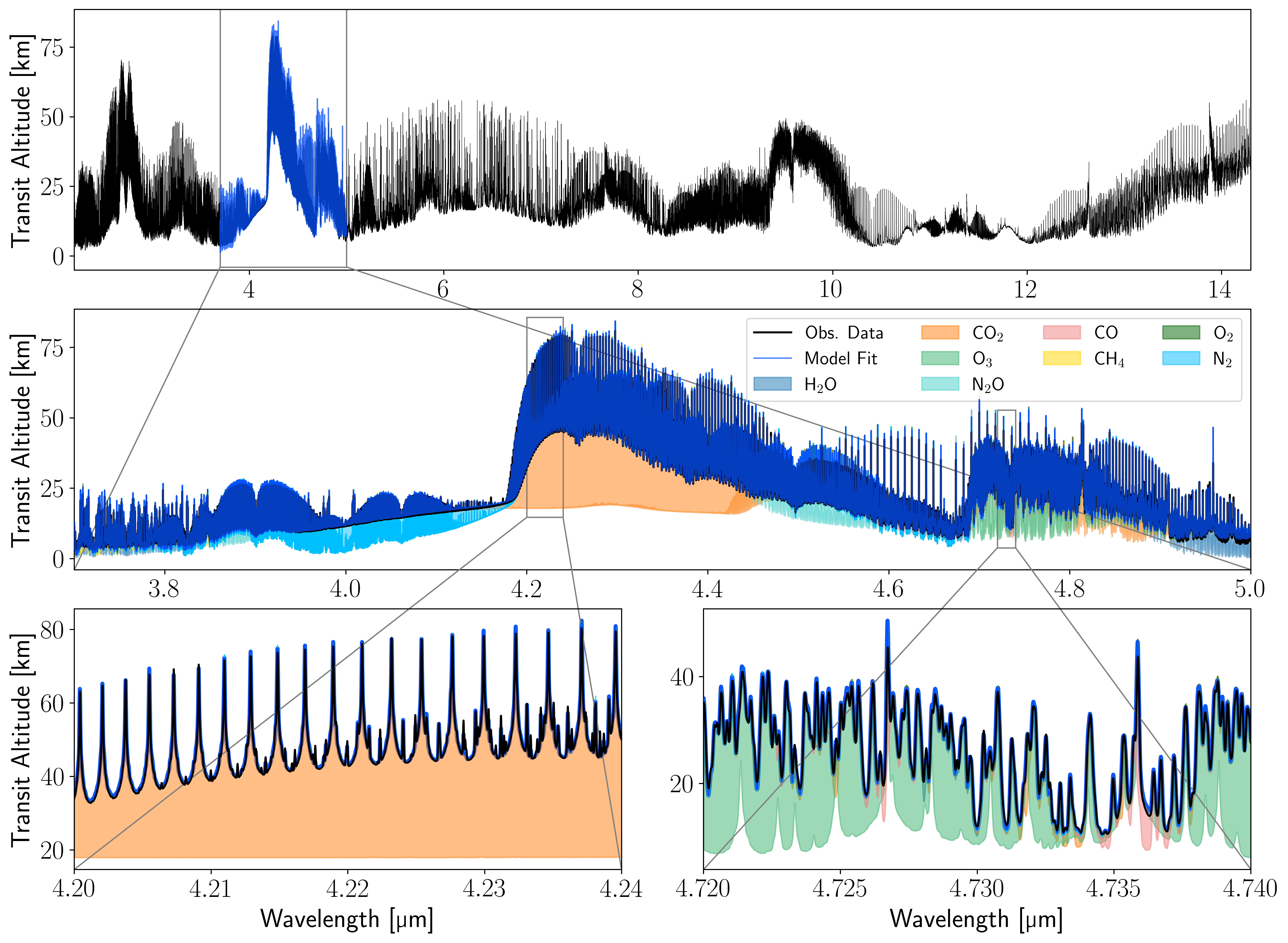}
\caption{Best fit model of Earth's high resolution transmission spectrum in the $3.7 - 5$ \um wavelength range using the mean Earth TP profile. The data are shown in black, the best-fitting model spectrum is shown in blue, and various colors are used to indicate each molecule's contribution to the model spectrum. } 
\label{fig:Earth_spectrum_fit_hr_4um}
\end{figure*}

Figure \ref{fig:Earth_spectrum_fit_hr_4um} shows our best fit spectrum using Model 1 in the $3.7 - 5$ \um wavelength range compared to the Earth observations. The middle zoomed inset reveals the molecules that are responsible for the majority of the absorption features in this spectral range: \ce{CO2}, \ce{O3}, \ce{N2}, \ce{N2O}, and \ce{H2O}. The bottom left panel highlights the individual \ce{CO2} lines within the 4.2 \um \ce{CO2} band, which are well fit by our model. The bottom right panel highlights a small region of the 4.7 \um \ce{O3} band and how it overlaps with the 4.6 \um \ce{CO} band. This \ce{CO} band is much more subtle than the \ce{O3} band, but it can be visually identified by the series of relatively strong lines seen in the middle panel of Figure \ref{fig:Earth_spectrum_fit_hr_4um} that span from around $4.55 - 4.8$ \um, and include the two strongest lines contained in the bottom right panel. Across the entire $2 - 14$ \um transmission spectrum, this is the strongest \ce{CO} band.   

% Earth spectrum plot

Table \ref{tab:earth_fit3_narrow1} displays the retrieval results for Model 1 using data covering only the $3.7 - 5$ \um wavelength range of interest.  For the aforementioned molecules with strong features in this narrow wavelength range, our inferred VMRs tend to be within 10\% of the global Earth averages. However, like our results across the full wavelength range, these results exhibit sensitivity to the assumed TP profile. Compared to the full wavelength range, focusing on this narrow range enables a more accurate retrieval of \ce{N2}. \ce{O2}, \ce{HNO3}, \ce{NO2}, CFC-11, and CFC-12 have no features in this wavelength range and are not constrained by these retrievals.     

\begin{deluxetable*}{r|rrrrrr}
\tablewidth{0.98\textwidth}
\tabletypesize{\footnotesize}
\tablecaption{Retrieved molecule VMR scale factors from OE fits to Earth's high-resolution transmission spectrum in the $3.7 - 5$ \um wavelength range\label{tab:earth_fit3_narrow1}}
\tablehead{   \colhead{Parameters}    &   \colhead{avg}   &   \colhead{lat=$0^{\circ}$}   &   \colhead{lat=$19^{\circ}$}   &   \colhead{lat=$41^{\circ}$}   &   \colhead{lat=$66^{\circ}$}  &  \colhead{cold}}
\startdata
    \ce{H2O} &    $1.064 \pm 0.018$ &    $0.573 \pm 0.014$ &    $0.712 \pm 0.015$ &    $1.332 \pm 0.023$ &    $2.827 \pm 0.046$ &    $4.186 \pm 0.066$ \\
    \ce{CO2} &  $1.1032 \pm 0.0038$ &  $1.0040 \pm 0.0035$ &  $1.0125 \pm 0.0035$ &  $1.1032 \pm 0.0038$ &  $1.3436 \pm 0.0047$ &  $1.5793 \pm 0.0057$ \\
     \ce{O3} &  $1.0744 \pm 0.0042$ &  $1.0122 \pm 0.0040$ &  $1.0368 \pm 0.0040$ &  $1.0902 \pm 0.0043$ &  $1.1900 \pm 0.0048$ &  $1.2851 \pm 0.0054$ \\
    \ce{N2O} &  $0.9010 \pm 0.0059$ &  $0.8396 \pm 0.0059$ &  $0.8543 \pm 0.0057$ &  $0.9035 \pm 0.0059$ &  $0.9930 \pm 0.0060$ &  $1.0630 \pm 0.0066$ \\
     \ce{CO} &    $0.941 \pm 0.016$ &    $0.783 \pm 0.015$ &    $0.735 \pm 0.015$ &    $0.937 \pm 0.016$ &    $0.824 \pm 0.016$ &    $0.845 \pm 0.017$ \\
    \ce{CH4} &    $1.441 \pm 0.011$ &  $1.2269 \pm 0.0096$ &  $1.3147 \pm 0.0097$ &    $1.467 \pm 0.011$ &    $1.637 \pm 0.012$ &    $1.609 \pm 0.012$ \\
     \ce{O2} &      $1.82 \pm 0.20$ &      $2.78 \pm 0.31$ &      $3.71 \pm 0.41$ &      $1.80 \pm 0.25$ &    $4.880 \pm 0.000$ &      $3.25 \pm 0.37$ \\
     \ce{N2} &    $1.172 \pm 0.083$ &    $1.088 \pm 0.047$ &    $1.104 \pm 0.050$ &    $1.176 \pm 0.082$ &    $1.439 \pm 0.000$ &    $1.311 \pm 0.000$ \\
%   \ce{HNO3} &    $1.000 \pm 0.000$ &    $1.000 \pm 0.000$ &    $1.000 \pm 0.000$ &    $1.000 \pm 0.000$ &    $1.000 \pm 0.000$ &    $1.000 \pm 0.000$ \\
%    \ce{NO2} &    $1.000 \pm 0.000$ &    $1.000 \pm 0.000$ &    $1.000 \pm 0.000$ &    $1.000 \pm 0.000$ &    $1.000 \pm 0.000$ &    $1.000 \pm 0.000$ \\
%      CFC-11 &    $1.000 \pm 0.000$ &    $1.000 \pm 0.000$ &    $1.000 \pm 0.000$ &    $1.000 \pm 0.000$ &    $1.000 \pm 0.000$ &    $1.000 \pm 0.000$ \\
%      CFC-12 &    $1.000 \pm 0.000$ &    $1.000 \pm 0.000$ &    $1.000 \pm 0.000$ &    $1.000 \pm 0.000$ &    $1.000 \pm 0.000$ &    $1.000 \pm 0.000$ \\
      \hline
         MPE &                4.50\% &                4.84\% &                4.46\% &                4.44\% &                4.45\% &                4.44\% \\
   $z_{err}$ &            $0.89$ km &            $0.90$ km &            $0.88$ km &            $0.90$ km &            $0.91$ km &            $0.93$ km \\
\enddata
\tablecomments{\ce{HNO3}, \ce{NO2}, CFC-11, and CFC-12 were not constrained and are therefore omitted from the table. Columns are labeled in reference to the fixed TP profiles shown in Figure \ref{fig:earth_TPs}. The fiducial Earth profiles for each molecule are displayed in Figure \ref{fig:earth_atm}. The last two rows show the mean percentage error (MPE) and characteristic vertical uncertainty ($z_{err}$) for each spectral fit.} 
\end{deluxetable*}

Table \ref{tab:earth_fit_high_res_even_4um} shows the retrieval results for the evenly mixed atmospheric model (Model 2) in the $3.7 - 5$ \um wavelength range. Unlike the results seen across the full $2- 14$ \um range, in this case, the isothermal model provides a marginally worse fit to the data ($10.44\%$ error) compared to the retrievals using physically-motivated fixed TP profiles ($9.92\%$ error on average). The retrieved isothermal temperature of ${\sim}246$ K is much higher than what we found using the full wavelength range. This may be a consequence of the strong \ce{CO2} band that dominates this wavelength range and causes absorption in the warm upper stratosphere. The evenly mixed VMRs also appear to be less biased when the isothermal TP profile is used compared to the full wavelength range.  

\begin{deluxetable*}{r|ccccccc}
\tablewidth{0.98\textwidth}
\tabletypesize{\scriptsize}
\tablecaption{Retrieved parameters for Earth's atmosphere from a fit to Earth's high-resolution transmission spectrum in the $3.7 - 5$ \um wavelength range using an evenly mixed atmospheric model (Model 2)\label{tab:earth_fit_high_res_even_4um}}
%\tablehead{   \colhead{Parameters} & \colhead{Fixed TP scaled}   &   \colhead{isothermal even} & \colhead{fixed TP even} }
\tablehead{
\colhead{Parameters} & \colhead{Isothermal TP} & \colhead{avg} & \colhead{lat=$0^{\circ}$} & \colhead{lat=$19^{\circ}$} & \colhead{lat=$41^{\circ}$} & \colhead{lat=$66^{\circ}$} & \colhead{cold}}
\startdata
 $T_{0}$ [K] &     $245.66 \pm 0.17$  & ---                     & ---                     & ---                     & ---                     & ---                     & ---                     \\ 
    \ce{H2O} &    $-4.406 \pm 0.015$  &      $-4.391 \pm 0.014$ &      $-5.060 \pm 0.026$ &      $-4.612 \pm 0.017$ &      $-4.382 \pm 0.014$ &      $-4.393 \pm 0.015$ &      $-4.429 \pm 0.016$ \\ 
    \ce{CO2} &  $-3.6906 \pm 0.0047$  &    $-3.4064 \pm 0.0011$ &    $-3.4567 \pm 0.0012$ &    $-3.4458 \pm 0.0011$ &    $-3.4095 \pm 0.0012$ &    $-3.3167 \pm 0.0012$ &    $-3.2327 \pm 0.0011$ \\ 
     \ce{O3} &  $-5.6016 \pm 0.0066$  &    $-5.4823 \pm 0.0047$ &    $-5.4785 \pm 0.0046$ &    $-5.4864 \pm 0.0045$ &    $-5.4666 \pm 0.0047$ &    $-5.4286 \pm 0.0048$ &    $-5.4087 \pm 0.0047$ \\ 
    \ce{N2O} &  $-6.6937 \pm 0.0013$  &  $-6.60102 \pm 0.00094$ &  $-6.63791 \pm 0.00093$ &  $-6.64024 \pm 0.00092$ &  $-6.59063 \pm 0.00094$ &  $-6.52346 \pm 0.00096$ &  $-6.50035 \pm 0.00096$ \\ 
     \ce{CO} &  $-7.1207 \pm 0.0017$  &    $-7.4039 \pm 0.0017$ &    $-7.3059 \pm 0.0016$ &    $-7.4493 \pm 0.0016$ &    $-7.4174 \pm 0.0017$ &    $-7.3946 \pm 0.0017$ &    $-7.6562 \pm 0.0018$ \\ 
    \ce{CH4} &    $-5.608 \pm 0.011$  &      $-5.736 \pm 0.011$ &      $-5.759 \pm 0.011$ &      $-5.721 \pm 0.010$ &      $-5.715 \pm 0.011$ &      $-5.688 \pm 0.011$ &      $-5.651 \pm 0.011$ \\ 
     \ce{O2} &    $-0.019 \pm 0.032$  &        $-0.68 \pm 0.10$ &      $-0.675 \pm 0.023$ &      $-0.699 \pm 0.019$ &      $-0.689 \pm 0.079$ &      $-0.691 \pm 0.019$ &      $-0.729 \pm 0.019$ \\ 
     \ce{N2} &  $-0.0564 \pm 0.0018$  &    $-0.0926 \pm 0.0019$ &    $-0.0984 \pm 0.0018$ &    $-0.0922 \pm 0.0018$ &    $-0.0877 \pm 0.0019$ &    $-0.0814 \pm 0.0021$ &    $-0.0721 \pm 0.0022$ \\ 
%   \ce{HNO3} &    $-8.338 \pm 0.000$  &      $-8.338 \pm 0.000$ &      $-8.338 \pm 0.000$ &      $-8.338 \pm 0.000$ &      $-8.338 \pm 0.000$ &      $-8.338 \pm 0.000$ &      $-8.338 \pm 0.000$ \\ 
%    \ce{NO2} &    $-8.503 \pm 0.000$  &      $-8.503 \pm 0.000$ &      $-8.503 \pm 0.000$ &      $-8.503 \pm 0.000$ &      $-8.503 \pm 0.000$ &      $-8.503 \pm 0.000$ &      $-8.503 \pm 0.000$ \\ 
%      CFC-11 &    $-9.638 \pm 0.000$  &      $-9.638 \pm 0.000$ &      $-9.638 \pm 0.000$ &      $-9.638 \pm 0.000$ &      $-9.638 \pm 0.000$ &      $-9.638 \pm 0.000$ &      $-9.638 \pm 0.000$ \\ 
%      CFC-12 &    $-9.292 \pm 0.000$  &      $-9.292 \pm 0.000$ &      $-9.292 \pm 0.000$ &      $-9.292 \pm 0.000$ &      $-9.292 \pm 0.000$ &      $-9.292 \pm 0.000$ &      $-9.292 \pm 0.000$ \\ 
      \hline
         MPE &                10.44\% &                   9.77\% &                  10.24\% &                   9.57\% &                   9.72\% &                  10.20\% &                  10.03\% \\ 
   $z_{err}$ &             $1.96$ km  &               $1.72$ km &               $1.69$ km &               $1.66$ km &               $1.73$ km &               $1.77$ km &               $1.74$ km \\ 
\enddata
\tablecomments{\ce{HNO3}, \ce{NO2}, CFC-11, and CFC-12 were not constrained and are therefore omitted from the table. Columns with fixed TP profiles are labeled in reference to those shown in Figure \ref{fig:earth_TPs}. The expected Earth profiles for each molecule are displayed in Figure \ref{fig:earth_atm}, with the approximate $\log_{10}$ VMRs for Earth's atmosphere given in Table \ref{tab:priors}. The last two rows show the mean percentage error (MPE) and characteristic vertical uncertainty ($z_{err}$) for each spectral fit.} 
\end{deluxetable*}

\subsubsection{Narrow Focus: $5 - 7.2$ \um}

\begin{deluxetable*}{r|rrrrrr}
\tablewidth{0.98\textwidth}
\tabletypesize{\footnotesize}
\tablecaption{Retrieved molecule VMR scale factors from OE fits to Earth's high-resolution transmission spectrum in the $5 - 7.2$ \um wavelength range\label{tab:earth_fit3}}
\tablehead{   \colhead{Parameters}    &   \colhead{avg}   &   \colhead{lat=$0^{\circ}$}   &   \colhead{lat=$19^{\circ}$}   &   \colhead{lat=$41^{\circ}$}   &   \colhead{lat=$66^{\circ}$}  &  \colhead{cold}}
\startdata
    \ce{H2O} & $1.5556 \pm 0.0086$ &  $1.266 \pm 0.011$ &  $1.3751 \pm 0.0099$ &  $1.5820 \pm 0.0083$ &  $1.9575 \pm 0.0095$ &   $2.246 \pm 0.011$ \\
    \ce{CO2} &   $1.034 \pm 0.014$ &  $0.737 \pm 0.020$ &    $0.734 \pm 0.016$ &    $1.085 \pm 0.013$ &    $1.782 \pm 0.017$ &   $1.926 \pm 0.019$ \\
     \ce{O3} & $0.9411 \pm 0.0090$ &  $0.654 \pm 0.013$ &    $0.703 \pm 0.011$ &  $1.1958 \pm 0.0094$ &    $1.558 \pm 0.011$ &   $1.653 \pm 0.012$ \\
    \ce{N2O} &   $0.837 \pm 0.039$ &  $0.723 \pm 0.062$ &    $0.534 \pm 0.046$ &    $0.828 \pm 0.034$ &    $1.772 \pm 0.050$ &   $1.707 \pm 0.050$ \\
     \ce{CO} &       $0.0 \pm 1.4$ &     $16.0 \pm 5.9$ &        $2.8 \pm 3.2$ &        $0.5 \pm 1.1$ &        $1.0 \pm 1.1$ &      $14.4 \pm 2.9$ \\
    \ce{CH4} &   $1.083 \pm 0.024$ &  $0.883 \pm 0.033$ &    $0.975 \pm 0.028$ &    $1.299 \pm 0.024$ &    $1.532 \pm 0.027$ &   $1.435 \pm 0.027$ \\
     \ce{O2} & $0.5775 \pm 0.0099$ &  $0.639 \pm 0.036$ &    $0.468 \pm 0.012$ &    $1.332 \pm 0.080$ &    $0.978 \pm 0.023$ &   $0.808 \pm 0.010$ \\
     \ce{N2} &   $3.848 \pm 0.000$ &    $0.77 \pm 0.37$ &    $1.653 \pm 0.000$ &      $0.28 \pm 0.17$ &      $0.97 \pm 0.26$ &   $2.412 \pm 0.000$ \\
   \ce{HNO3} &   $4.348 \pm 0.036$ &  $4.613 \pm 0.057$ &    $4.405 \pm 0.046$ &    $4.081 \pm 0.032$ &    $4.085 \pm 0.033$ &   $3.529 \pm 0.031$ \\
    \ce{NO2} &   $0.749 \pm 0.012$ &  $0.807 \pm 0.019$ &    $0.798 \pm 0.016$ &    $0.753 \pm 0.011$ &    $0.709 \pm 0.012$ &   $0.753 \pm 0.012$ \\
%      CFC-11 &   $1.000 \pm 0.000$ &  $1.000 \pm 0.000$ &    $1.000 \pm 0.000$ &    $1.000 \pm 0.000$ &    $1.000 \pm 0.000$ &   $1.000 \pm 0.000$ \\
%      CFC-12 &   $1.000 \pm 0.000$ &  $1.000 \pm 0.000$ &    $1.000 \pm 0.000$ &    $1.000 \pm 0.000$ &    $1.000 \pm 0.000$ &   $1.000 \pm 0.000$ \\
      \hline
         MPE &              4.87\% &             9.00\% &               7.01\% &               3.87\% &               3.41\% &               3.76\% \\
   $z_{err}$ &           $1.00$ km &          $1.51$ km &            $1.26$ km &            $0.92$ km &            $0.91$ km &           $0.93$ km \\
\enddata
\tablecomments{CFC-11 and CFC-12 were not constrained and are therefore omitted from the table. Columns are labeled in reference to the fixed TP profiles shown in Figure \ref{fig:earth_TPs}. The fiducial Earth profiles for each molecule are displayed in Figure \ref{fig:earth_atm}. The last two rows show the mean percentage error (MPE) and characteristic vertical uncertainty ($z_{err}$) for each spectral fit.} 
\end{deluxetable*}

We now focus exclusively on the 6 \um region ($5-7.2$ \um) to delve deeper into \ce{H2O}, \ce{O2}, and \ce{N2}. 
% Description results table for many TP profiles
Table \ref{tab:earth_fit3} displays the retrieval results from our fit to the high resolution transmission spectrum using Model 1 with various different TP and water vertical profiles. In general, the fits to the high resolution data are very good with model deviations under 10\%. Again, no single vertical profile enables an accurate VMR retrieval of all the gases; instead, some profiles provide more accurate results for specific gases. The average Earth profile yields the most accurate retrieval of the \ce{CO2} and \ce{O3} abundances, but not the \ce{O2} and \ce{N2} abundances. The high latitude/cold profiles provide a more accurate assessment of \ce{O2} and \ce{N2}, but retrieves a \ce{CO2} abundance that exceeds the current Earth value by a factor of ${\sim}2$. Furthermore, the higher latitude/cold profiles provide better fits to the entire 6 \um spectral range than the lower latitude profiles.  

% Earth spectrum high res 1
\begin{figure*}[t!]
\centering
\includegraphics[width=0.93\textwidth]{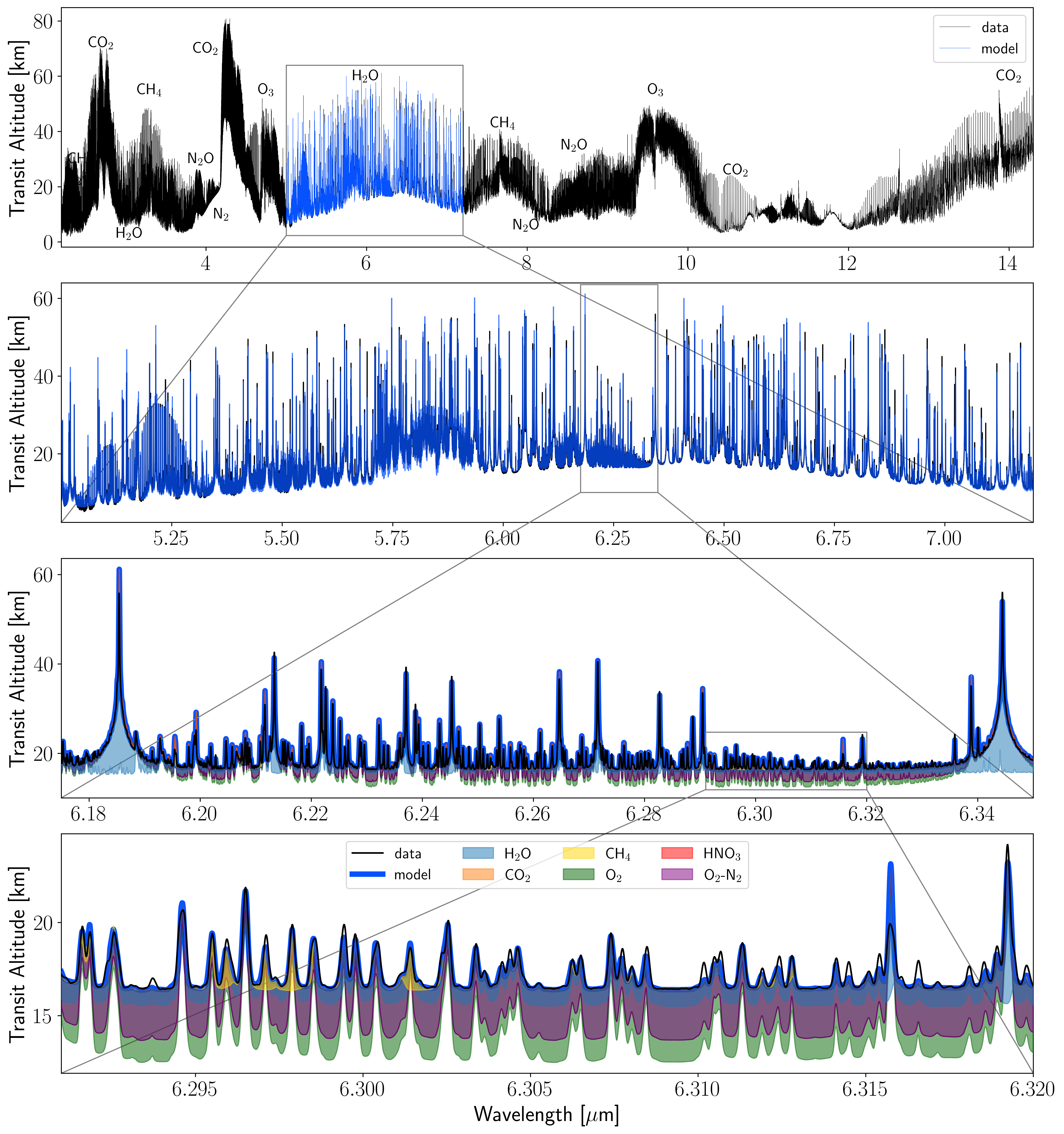}
\caption{Best fit model of Earth's high resolution transmission spectrum in the $5 - 7.2$ \um wavelength range using the $66^{\circ}$ latitude TP profile. The data are shown in black, the best-fitting model spectrum is shown in blue, and various colors are used in the bottom panels to indicate each molecule's contribution to the model spectrum. The shading for \ce{O2} indicates the contribution from both \ce{O2-O2} CIA and \ce{O2-N2} CIA. In the bottom panel, \ce{HNO3}, \ce{CH4}, and \ce{H2O} show subtle absorption features above a \ce{O2-N2} CIA, \ce{O2-O2} CIA, and \ce{H2O} continuum.} 
\label{fig:Earth_spectrum_fit_hr}
\end{figure*}

% Description of HR figure
Figure \ref{fig:Earth_spectrum_fit_hr} shows our best-fitting model (Model 1) in the 6 \um region using the $66^{\circ}$ latitude TP and water profiles. As shown in Table \ref{tab:earth_fit3}, the $66^{\circ}$ latitude profiles provided the overall best fit to the data in the spectral region and the most accurate inferred VMR for \ce{O2} and \ce{N2}, which we will examine more closely.   
In the top two panels it is difficult to discern the differences between the model (in blue) and the data (in black) because both lines plot on top of one another. 
The bottom panels of Fig. \ref{fig:Earth_spectrum_fit_hr} zoom into a narrow enough region to see small deviations. These lower two panels highlight the contribution from various line and continuum absorbers. In the third panel, there are several well-resolved \ce{H2O} lines that make up the 6 \um absorption band and span approximately 50 km in absorbing radius from the continuum to their cores. The forth panel zooms into a region between \ce{H2O} lines where the opacity is low and contributions to the continuum are visible from \ce{O2-N2} CIA, \ce{O2-O2} CIA, and \ce{H2O}. Note that the green shading for \ce{O2} shows the combined absorption due to \ce{O2-O2} CIA and \ce{O2-N2} CIA and therefore the absorption contribution from \ce{O2-O2} CIA alone is identified by the green shading that extends below the purple shading for \ce{O2-N2}. In this region, there are also subtle absorption features due to \ce{HNO3} and \ce{CH4}. 
Between $5-7.2$ \um this best-fitting spectrum has a 3.41\% percent deviation on average from the empirical transmission spectrum and a characteristic error of $z_{err} = 0.91$ km. 

% Presentation of O2-N2 degeneracies
At high spectral resolution, the continuum in the 6 \um range provides a compelling window through which to study the lower atmosphere between \ce{H2O} lines, but it appears to be home to a complex degeneracy between multiple competing continuum absorbers (Figure \ref{fig:Earth_spectrum_fit_hr}).  
The continuum between $5-7$ \um proved to be one of the most challenging spectral regions to fit in the entire $2-14$ \um Earth transmission spectrum from \citetalias{Macdonald2019}. Specifically, the residuals that caused the globally averaged, $0^{\circ}$ latitude, and $19^{\circ}$ latitude TP/water profiles to exhibit larger fitting errors, arose primarily from poor fits to the \ce{H2O} continuum between $5-5.7$ \um and beyond $6.5$ \um. Although the ``cold'' and $61^{\circ}$ latitude profiles provided the best fits to the spectral continuum, they resulted in anomalously high \ce{CO2}, \ce{H2O}, and \ce{O3} VMRs. \ce{H2O} and \ce{O3} exhibit spatial variability, but since \ce{CO2} effectively does not, these deviations from the known values further highlight a degeneracy that can bias retrievals away from the true result while still yielding an excellent fit to the observations.   
The lower two panels of Figure \ref{fig:Earth_spectrum_fit_hr} show how the spectral continuum is produced through a combination of absorption processes, including the 6 \um \ce{H2O} band continuum, \ce{O2-O2} CIA, and \ce{O2-N2} CIA. Since the choice of vertical temperature and \ce{H2O} profiles strongly affected the retrieved \ce{O2} and \ce{N2} VMR (see Table \ref{tab:earth_fit3}), we infer that there are complex degeneracies between the vertical TP structure, \ce{H2O} profile, and the VMRs of \ce{O2} and \ce{N2}. For Earth, we can hold many of these parameters fixed to break such a degeneracy, but for future exoplanet observations, degeneracies between the vertical thermal structure and mixing ratio profiles may be quite challenging to diagnose and overcome when little or no prior information is known. 

%\subsubsection{Model 2: Evenly Mixed Profiles}

\begin{deluxetable*}{r|ccccccc}
\tablewidth{0.98\textwidth}
\tabletypesize{\scriptsize}
\tablecaption{Retrieved parameters for Earth's atmosphere from a fit to Earth's high-resolution transmission spectrum in the $5 - 7.2$ \um wavelength range using an evenly mixed atmospheric model (Model 2)\label{tab:earth_fit_high_res_even}}
%\tablehead{   \colhead{Parameters} & \colhead{Fixed TP scaled}   &   \colhead{isothermal even} & \colhead{fixed TP even} }
\tablehead{
\colhead{Parameters} & \colhead{Isothermal TP} & \colhead{avg} & \colhead{lat=$0^{\circ}$} & \colhead{lat=$19^{\circ}$} & \colhead{lat=$41^{\circ}$} & \colhead{lat=$66^{\circ}$} & \colhead{cold}}
\startdata
 $T_{0}$ [K] &     $203.02 \pm 0.30$ & ---                   & ---                   & ---                   & ---                   & ---                   & ---                   \\ 
    \ce{H2O} &    $-4.728 \pm 0.012$ &  $-5.1029 \pm 0.0040$ &  $-5.1714 \pm 0.0041$ &  $-5.1685 \pm 0.0041$ &  $-5.1606 \pm 0.0048$ &  $-5.0899 \pm 0.0040$ &  $-5.0496 \pm 0.0039$ \\
    \ce{CO2} &    $-4.250 \pm 0.039$ &    $-3.412 \pm 0.010$ &  $-3.4084 \pm 0.0098$ &  $-3.4080 \pm 0.0099$ &  $-8.7712 \pm 0.0072$ &    $-3.400 \pm 0.011$ &    $-3.400 \pm 0.011$ \\
     \ce{O3} &    $-5.735 \pm 0.063$ &    $-5.872 \pm 0.046$ &    $-5.867 \pm 0.043$ &    $-5.867 \pm 0.043$ &    $-4.917 \pm 0.016$ &    $-5.863 \pm 0.047$ &    $-5.864 \pm 0.048$ \\
    \ce{N2O} &  $-6.5775 \pm 0.0026$ &  $-6.5565 \pm 0.0020$ &  $-6.4875 \pm 0.0019$ &  $-6.4798 \pm 0.0019$ &  $-6.6952 \pm 0.0036$ &  $-6.1504 \pm 0.0019$ &  $-6.1677 \pm 0.0019$ \\
     \ce{CO} &  $-6.6568 \pm 0.0085$ &  $-6.7459 \pm 0.0066$ &  $-6.7078 \pm 0.0064$ &  $-6.7070 \pm 0.0065$ &    $-6.823 \pm 0.011$ &  $-6.6712 \pm 0.0068$ &  $-6.6794 \pm 0.0069$ \\
    \ce{CH4} &    $-5.502 \pm 0.075$ &    $-5.794 \pm 0.064$ &    $-5.795 \pm 0.061$ &    $-5.795 \pm 0.061$ &    $-4.663 \pm 0.021$ &    $-5.795 \pm 0.064$ &    $-5.794 \pm 0.066$ \\
     \ce{O2} &    $-0.380 \pm 0.047$ &    $-0.724 \pm 0.073$ &    $-0.767 \pm 0.071$ &    $-0.760 \pm 0.046$ &      $-2.51 \pm 0.22$ &    $-0.733 \pm 0.069$ &    $-0.723 \pm 0.077$ \\
     \ce{N2} &      $-0.47 \pm 0.13$ &      $-0.15 \pm 0.10$ &      $-0.18 \pm 0.10$ &    $-0.179 \pm 0.066$ &           $-2 \pm 11$ &    $-0.159 \pm 0.095$ &      $-0.15 \pm 0.11$ \\
   \ce{HNO3} &  $-8.4615 \pm 0.0024$ &  $-8.4700 \pm 0.0017$ &  $-8.4733 \pm 0.0017$ &  $-8.4671 \pm 0.0017$ &  $-8.7898 \pm 0.0026$ &  $-8.3149 \pm 0.0017$ &  $-8.2996 \pm 0.0017$ \\
    \ce{NO2} &  $-9.1052 \pm 0.0034$ &  $-8.7527 \pm 0.0023$ &  $-8.8008 \pm 0.0023$ &  $-8.7969 \pm 0.0023$ &  $-9.1356 \pm 0.0028$ &  $-8.7967 \pm 0.0024$ &  $-8.7754 \pm 0.0024$ \\
%      CFC-11 &    $-9.638 \pm 0.000$ &    $-9.638 \pm 0.000$ &    $-9.638 \pm 0.000$ &    $-9.638 \pm 0.000$ &    $-9.638 \pm 0.000$ &    $-9.638 \pm 0.000$ &    $-9.638 \pm 0.000$ \\
%      CFC-12 &    $-9.292 \pm 0.000$ &    $-9.292 \pm 0.000$ &    $-9.292 \pm 0.000$ &    $-9.292 \pm 0.000$ &    $-9.292 \pm 0.000$ &    $-9.292 \pm 0.000$ &    $-9.292 \pm 0.000$ \\
      \hline
         MPE &                14.38\%&                14.68\% &                14.60\% &                14.66\% &                17.04\% &                14.54\% &                14.42\% \\
   $z_{err}$ &             $2.60$ km &             $2.40$ km &             $2.40$ km &             $2.42$ km &             $3.49$ km &             $2.43$ km &             $2.41$ km \\
\enddata
\tablecomments{CFC-11 and CFC-12 were not constrained and are therefore omitted from the table. Columns with fixed TP profiles are labeled in reference to those shown in Figure \ref{fig:earth_TPs}. The expected Earth profiles for each molecule are displayed in Figure \ref{fig:earth_atm}, with the approximate $\log_{10}$ VMRs for Earth's atmosphere given in Table \ref{tab:priors}. The last two rows show the mean percentage error (MPE) and characteristic vertical uncertainty ($z_{err}$) for each spectral fit.} 
\end{deluxetable*}

% Evenly mixed high res results
Table \ref{tab:earth_fit_high_res_even} lists the results from our evenly mixed retrievals in the $5-7$ \um range using Model 2 with an isothermal TP profile (column 1) and the assortment of fixed TP profiles (columns 2-7). Similar to our previous results, the evenly mixed models are not able to provide as good of fits to the Earth observations as the scaled profiles that assume an approximately correct 1D vertical structure (results shown in Table \ref{tab:earth_fit3}). We report mean percentage errors on the fits of 14.68\% and 14.38\% for the mean TP profile and isothermal fitted models, respectively. Again we see that the flexibility of the isothermal model provides a marginally better fit to the data than a realistic, but fixed, TP profile, however it has a propensity to retrieve biased gas VMRs. This bias is exemplified well by the \ce{CO2} VMR, which is over estimated by an order of magnitude by the isothermal model compared to all of our results with various fixed TP profiles. 
We retrieved an isothermal temperature of $203.02 \pm 0.30$ K, which agrees well with the average Earth tropopause temperature, but varies appreciably from the isothermal temperatures inferred using other portions of the spectrum. The fact that we retrieved different isothermal temperatures in each of the spectral regions considered at high resolution illuminates the sensitivity of the broad wavelength transmission spectrum to vertical temperature structure.  

% Refraction 
As an addendum to the investigations presented here, we performed a brief investigation on the impact of atmospheric refraction in our radiative transfer modeling. Appendix \ref{sec:appendix:refraction} demonstrates that a small systematic effect is present due to the increase in optical path length caused by refraction. This effect increases towards the lower atmosphere where the density is higher. While some biases in the abundance retrievals could be attributed to neglecting refraction, these effects are smaller than those exhibited by different TP profiles and vertical modeling assumptions. 

\subsection{Synthetic JWST Observations}  \label{sec:results:jwst}

% Conversion to transit depth 
We now consider the Earth's transmission spectrum as if it were observed with JWST. Until this point, we have fit the empirical Earth spectrum in units of kilometers altitude above the solid surface; now, we transform the observations into transit depths, assuming that TRAPPIST-1e exhibits this exact spectrum. We define the transit depth as, 
\begin{equation}
    \Delta F = \left ( \frac{R_p + z_{\mathrm{eff}}(\lambda)}{R_s} \right ) ^{2}
\end{equation}
where $R_p$ is TRAPPIST-1e's assumed solid body surface radius ($0.915~\mathrm{R}_{\oplus}$; \citealp{Grimm2018}), $z_{\mathrm{eff}}(\lambda)$ is the wavelength-dependent effective transit altitude from \citetalias{Macdonald2019}, and $R_s$ is the TRAPPIST-1 stellar radius ($0.121~\mathrm{R}_{\odot}$; \citealp{VanGrootel2018}).  

% JWST Noise modeling
We used the \pandexo noise model \citep[version 2.0;][]{Batalha2017b, Pandexo2018} to generate synthetic JWST observations of our TRAPPIST-1e Earth model with both the NIRSpec G395M \citep{Bagnasco2007, Ferruit2014} and MIRI LRS \citep{Bouchet2015, Kendrew2015} instruments. These two instruments were selected for this analysis because they span different, yet contiguous, spectral regions from one another, they fully overlap with the observed wavelength range of the \citetalias{Macdonald2019} spectrum, and they are ideal instruments for terrestrial exoplanet transmission spectroscopy \citep{Lustig-Yaeger2019}. Although NIRSpec PRISM is also an optimal JWST instrument mode to use for TRAPPIST-1 observations, the shortwave cutoff for PRISM is 0.6 \um and the \citetalias{Macdonald2019} spectrum only extends down to 2.2 \um. As a result, we focus primarily on the NIRSpec G395M mode, which spans 2.87 – 5.10 \um, but we show and discuss a selected PRISM simulation in Appendix \ref{sec:appendix:posteriors} for data from 2.2-5.2 \um. We simulate noise on each observation for 80 stacked transits, which is approximately the maximum number of observable transits of TRAPPIST-1e in the nominal 5-year JWST mission lifetime. This allows us to compare the atmospheric constraints retrieved from each instrument using a uniform amount of telescope time. 

% Retrieval differences
For our TRAPPIST-1e retrievals, we make a few adjustments to our model to better compare with standard exoplanet methodologies. First, we only fit for the following molecules: \ce{H2O}, \ce{CO2}, \ce{CH4}, \ce{O2}, \ce{O3}, and \ce{CO}. We use ``Model 2'' for this investigation, which assumes that these gases are evenly mixed throughout the atmospheric column and fits for their log volume mixing ratios and the isothermal temperature. Second, we also include as free parameters the solid body surface radius with uniform prior $\mathcal{U}(0.865, 0.965)$ R$_{\oplus}$ and the log-surface pressure with uniform prior $\mathcal{U}(3, 6)$ Pa. Finally, we assume \ce{N2} as the background gas in the atmosphere \citep[e.g.][]{Barstow2016, Krissansen-Totton2018, Changeat2019, Barstow2020b}, and now self-consistently determine the atmospheric mean molecular weight using the molecular composition of the forward model at each step in the retrieval. 

% JWST Spectra 
\begin{figure*}[t!]
\centering
\includegraphics[width=0.98\textwidth]{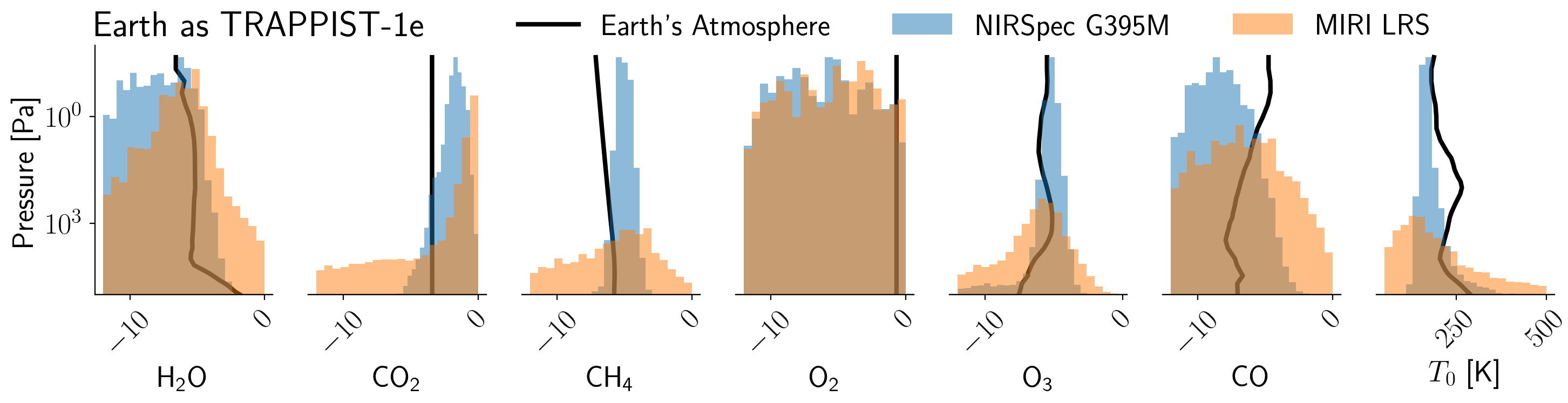}
\includegraphics[width=0.98\textwidth]{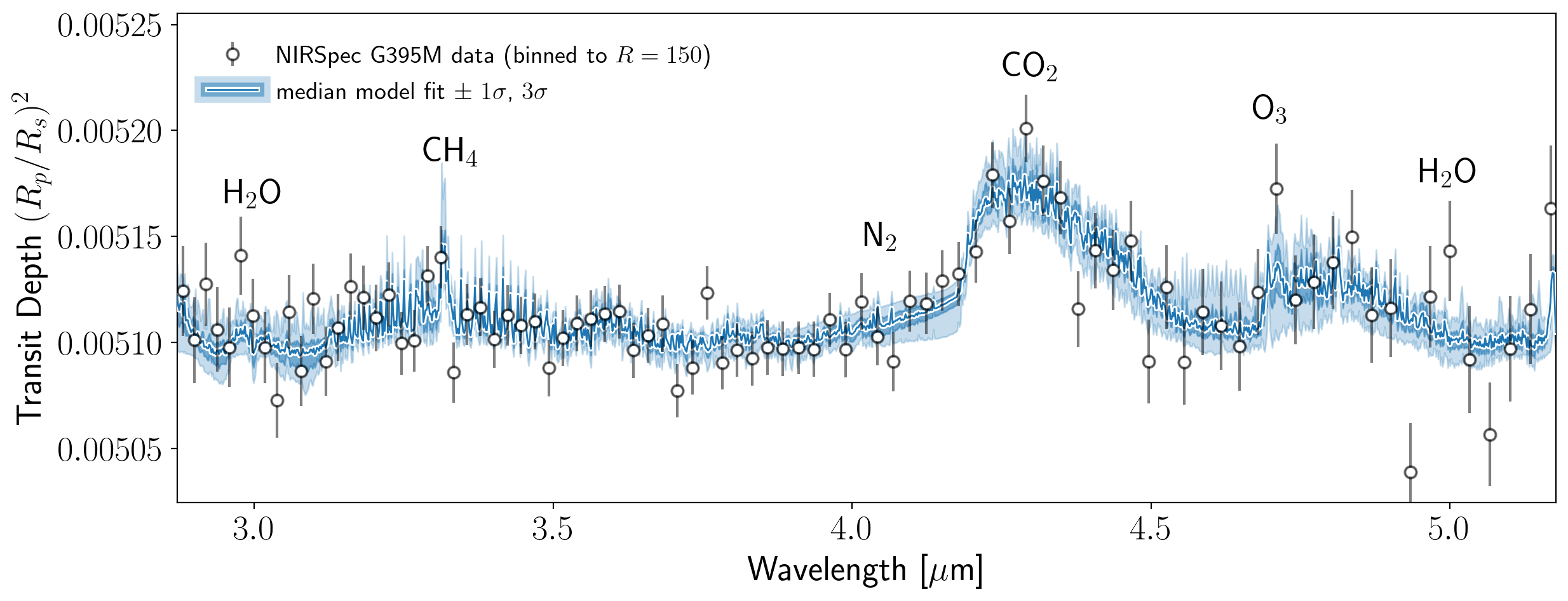}
\includegraphics[width=0.98\textwidth]{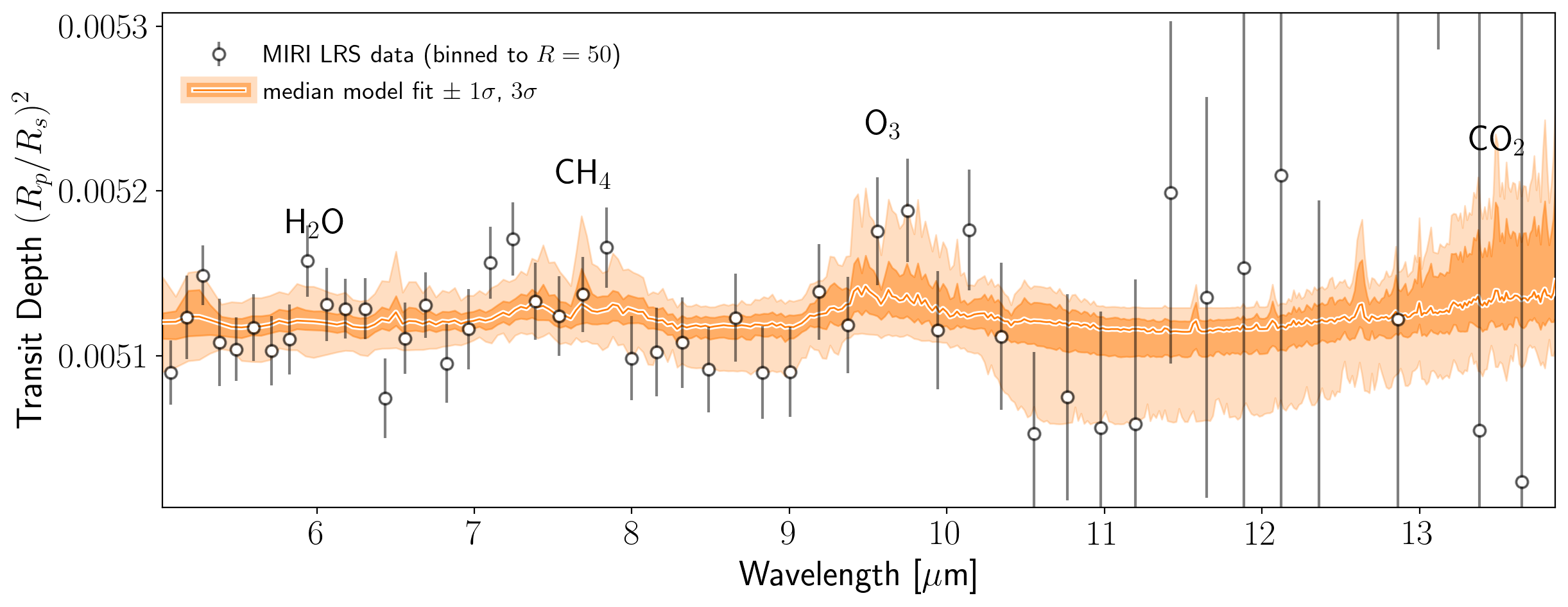}
\caption{Fits to synthetic JWST spectra and the corresponding retrieved atmospheric constraints. The upper subplots show 1D marginalized posterior distributions for retrievals using NIRSpec G395M data (blue) and MIRI LRS data (orange) compared to the true vertical profiles exhibited by Earth's atmosphere (black lines). The middle and bottom panels show the synthetic JWST spectra for 80 stacked transits of TRAPPIST-1e observed with NIRSpec G395M and MIRI LRS, respectively, along with the ${\pm}$ $1\sigma$ and $3\sigma$ uncertainty envelopes around the median model fits.  }
\label{fig:jwst_spectra}
\end{figure*}
 
% Basic description of figure
Figure \ref{fig:jwst_spectra} compares our NIRSpec G395M (blue) and MIRI LRS (orange) transmission spectrum retrieval results for TRAPPIST-1e. The top row of subplots show 1D marginalized posterior distributions for the inferred atmospheric parameters. The histograms are compared to their spatially averaged vertical profiles in Earth's atmosphere, which are shown in black as a function of pressure. The bottom two panels show our median fits to the synthetic JWST spectra with 1 and 3$\sigma$ bounding envelopes. Although we binned the NIRSpec G395M data to a resolving power of $R=150$ and the MIRI LRS data to $R=50$ for visualization purposes, the spectra were fit at the native spectral resolution of each instrument ($R \sim 1000$ for NIRSpec G395M and $R \sim 100$ for MIRI LRS). Figure \ref{fig:jwst_posteriors} in Appendix \ref{sec:appendix:posteriors} provides the full corner plot with retrieved median and 1$\sigma$ confidence intervals for each parameter and parameter covariances.  

% Detailed findings from figure
Given the same telescope time, JWST's NIRSpec G395M instrument acquires higher S/N and higher resolution spectra than MIRI LRS, and the retrieved constraints on Earth's atmosphere reflect these gains for NIRSpec. Using NIRSpec data we inferred (near) Gaussian posteriors for the abundances of \ce{CO2} and \ce{CH4}, peaked posteriors for \ce{O3} but without a lower limit, and a credible upper limit for \ce{H2O} and \ce{CO}. Each of these constraints is consistent with the true values in Earth's atmosphere. In comparison, using MIRI LRS data we achieved relatively imprecise constraints on all gases. Only \ce{O3} was accurately constrained to within three orders of magnitude with MIRI LRS ($\pm 2.4$ dex). Despite the presence of the strong 9.6 \um \ce{O3} band falling within the MIRI LRS wavelength range, the presence of the 4.7 \um \ce{O3} band within the NIRSpec G395M wavelength range enables NIRSpec to constrain \ce{O3} about twice as precisely ($\pm 0.91$ dex). Neither instrument is able to constrain \ce{O2} due to the lack of detectable bands in the 2--14 \um range, with the exception of broad CIA bands, which are well below the noise levels in the spectrum. Similarly, MIRI cannot place an upper limit on \ce{CO} due to a lack of absorption features in the wavelength range. The MIRI \ce{CO2} result has a misleading spike at high abundances due to the random white noise added to the synthetic data. This spike is eliminated in tests that omitted the randomized noise (see Appendix \ref{sec:appendix:posteriors}). 
Unlike our previous retrievals of Earth's transmission spectrum without noise, at JWST's precision the complexity of retrieved atmospheric constraints are severely limited such that homogeneous vertical profiles are sufficient to capture the general characteristics of the atmosphere. 

\section{Discussion} \label{sec:discussion}

% Concise answers to top-level research questions
We analyzed the exoplanet-analog transmission spectrum of Earth from \citet{Macdonald2019} to perform a validation of our new exoplanet retrieval model, \smarter, on the only known habitable and inhabited planet and to determine the capacity for transmission spectroscopy to retrieve signs of habitability and biosignatures. In general, we obtained excellent model fits to the high S/N observed spectra both in low resolution (Figure \ref{fig:Earth_spectrum_fit}) and in high resolution (Figures \ref{fig:Earth_spectrum_fit_hr_full}, \ref{fig:Earth_spectrum_fit_hr_4um}, \ref{fig:Earth_spectrum_fit_hr}) that demonstrate the validity of our model down to an equivalent precision of ${<}5$ ppm for the exoplanet TRAPPIST-1e.
We retrieved gas abundance estimates that are accurate within ${\sim}$25\% or better of the Earth average VMRs for the majority of the 12 molecules included in our study and demonstrated the limits to which a single set of 1D atmospheric profiles can be used to describe a dynamic 3D planet such as Earth.  
These results provide a detailed perspective on the global atmospheric environment that yields strong evidence of the planet's habitability, presence of biosignatures, and technosignatures as seen through the lens of an exoplanet-analog transmission spectrum. 
Our analyses using errors consistent with JWST observations of the TRAPPIST-1 system (Figures \ref{fig:jwst_spectra} and \ref{fig:jwst_posteriors}) reveal a simpler picture of the atmosphere wherein a large time investment is required to constrain \ce{CH4} and \ce{CO2} to better than ${\pm}1$ dex and potentially provide a weak detection of \ce{O3}, thereby illuminating practical challenges to characterizing transiting Earth-like exoplanets for habitability and biosignatures in the near term.     
We further elaborate on our model validation/limitations and the characterization of habitability and biosignatures in the following discussion. 

\subsection{Model Validation \& Limitations}

% Low resolution 
Our retrieved molecular abundances from the low resolution spectrum (Table \ref{tab:earth_fit_low_res}) are consistent with the Earth's atmospheric composition within the range of spatial and temporal variability exhibited across the globe (Table \ref{tab:earth_variability}). Specifically, in the noiseless data, we retrieved volume mixing ratios for \ce{H2O}, \ce{CO2}, \ce{O3}, \ce{N2O}, \ce{CH4}, and \ce{N2} that are within 28\% of their average Earth values.
%; \ce{CO} is within 50\% and \ce{O2} is within 70\%. 
Using TP profiles and mixing ratio profiles that vary with altitude provided better fits to the Earth spectrum (5\% errors) that yield more accurate abundance constraints than 1D isothermal TP structures with evenly-mixed gas VMRs, although vertically homogeneous 1D atmospheric models still provided a good fit to the data (12\% errors) that enabled a relatively accurate assessment of the atmospheric composition (see Table \ref{tab:earth_fit_low_res} and Figure \ref{fig:Earth_spectrum_fit_noise}), considering the simplicity of the model. 

% Effect of even mixing (LR and HR)
Our retrievals using evenly mixed molecular profiles provided $2.6{\times}$ worse fits to the observations on average than retrievals that scaled the characteristic vertical profiles of each gas. This remained consistent in both low resolution and high resolution. Thus information about the vertical distribution of gases exists in pristine transmission spectrum measurements that can in theory be used to retrieve vertically resolved molecular profiles, consistent with the modeling results of \citet{Changeat2019}. The improvements seen for the model with realistic vertical structure, however, were at the level of ${<}5$ ppm equivalent precision for observations of TRAPPIST-1e and therefore insignificant for near term studies of rocky exoplanets. While our evenly mixed retrieval on the case with 10\% noise added (Figure \ref{fig:Earth_spectrum_fit_noise}) also indicated vertical sensitivity in the 9.6 \um \ce{O3} band, our retrievals using JWST analog data further indicated that evenly mixed atmospheric models with isothermal TP profiles will suffice to accurately interpret terrestrial exoplanet transmission spectra with JWST. 

% High resolution
We investigated multiple different characteristic Earth TP profiles (Figure \ref{fig:earth_TPs}) for retrievals of the high resolution spectrum and found that each one provided a slightly different fit to the high-res data that yielded accurate retrievals for different gases (Table \ref{tab:earth_fit3}). Statistically significant differences in the retrieved gas abundances were found using different characteristic TP profiles. 
While a low latitude TP profile provided the most accurate retrieval of the \ce{CO2} abundance, \ce{N2} and \ce{O2} were most accurately retrieved using a colder TP profile representative of higher latitudes. These results indicate that such high resolution transmission spectra are sensitive to atmospheric thermal structure, but likewise illuminates a degeneracy between the TP structure and the gas abundances. These complicating factors warrant investigation in future works aimed at retrieving a characteristic altitude-dependent TP profile for Earth. However, it is possible that even such an approach may still impose biases in gas abundances because the globally-averaged exoplanet-analog transmission spectrum from \citet{Macdonald2019} contain information from more than one characteristic TP profile. These results for Earth echo the findings obtained from modeling giant exoplanet atmospheres, such as \citet{Caldas2019} \citep[see also,][]{Lacy2020, Pluriel2020, Wardenier2022} where GCM outputs were used to produce a 3D exoplanet transmission spectrum, which could be well-fit with a simple 1D atmospheric retrieval model, but doing so caused biases in gas abundances that could be challenging to diagnose in real data. Thus, even though each choice of TP profile that we used provided an excellent fit to the data, in reality none of them are correct because the data provides a sampling of all TP profiles along the terminator---an intrinsically 3D effect.    

% The continuum in HR and CIA 
Our retrievals of Earth's high resolution transmission spectrum near 6 \um revealed challenges associated with fitting the spectral continuum that, if addressed, could potentially be leveraged for more precise exoplanet atmospheric characterization. Resolving the spectral continuum between \ce{H2O} (and other gas) lines enabled the abundances of \ce{O2} and \ce{N2} to be retrieved due to the presence of broad \ce{O2-O2} and \ce{O2-N2} CIA bands (Figure \ref{fig:Earth_spectrum_fit_hr}) near 6 \um \citep[see also,][]{Fauchez2020}. However, the accuracy of this approach depended on the specific choice of TP and \ce{H2O} vertical profiles. In an exoplanet context, this assumption would not be appropriate \textit{a priori} and would therefore open up a complicated degeneracy between the TP profile, and the VMRs of \ce{H2O}, \ce{O2}, and \ce{N2}. We also note that these 6 \um CIA bands were not accessible in the low resolution spectrum due to the convolution of the strong \ce{H2O} lines with the spectral continuum. 

Furthermore, the challenges and opportunities afforded to high resolution spectroscopy may be best realized for exoplanets with the upcoming generation of ground-based extremely large telescopes \citep[ELTs;][]{Johns2012, Skidmore2015, Marconi2016}. These telescopes rely on high resolution spectroscopy to distinguish exoplanet atmospheric features from telluric lines \citep[e.g.,][]{Snellen2013, Brogi2019, Hood2020, Leung2020, Currie2023}, and are therefore well suited to investigate, and may need to properly account for, the confounding TP profile and line/continuum effects discussed here. Moreover, our validated model \smarter is an excellent tool for generating forward models and performing retrieval simulations at high spectral resolution. Future work using the high-resolution cross-correlation likelihood formulation of \citet{Brogi2019} will aim to assist groundbreaking ELT science cases. However, high-resolution ground-based observations longward of 5 \um may be prohibitively challenging due to thermal background noise. As such, our high-resolution findings in the MIR may not be observable in the near term, but rather, may serve as potential motivation for a mission concept or technique that is capable of overcoming the thermal background, likely from space.   

% Clouds
The fact that the empirical transmission spectrum from \citetalias{Macdonald2019} omitted occultation measurements with clouds has likely impacted our retrieval results and limits the scope of our study to clear sky phenomena. Since the \citetalias{Macdonald2019} transmission spectrum represents the spatially and temporally averaged Earth \textit{only along clear optical paths through the atmosphere}, this spectrum naturally contains an intrinsic bias towards atmospheric characteristics that correlate with clear skies. This could potentially explain deviations found between our retrieved atmospheric characteristics and known average values, but other factors---such as the latitudinal, seasonal, and temporal sampling of the SCISAT ACE-FTS data---cannot be ruled out.  Furthermore, our results associated with the transmission spectrum continuum, including sensitivity to CIA features (Figure \ref{fig:Earth_spectrum_fit_hr}) and refraction effects (Figure \ref{fig:refraction}), will be further complicated for exoplanets where clear skies cannot be assumed. Optically thick clouds (and/or hazes) are well known to truncate the path length of light, particularly in the transit geometry \citep{Fortney2005}, and have been responsible for the flat and featureless transmission spectra of numerous gaseous exoplanets \citep[e.g.,][]{Kreidberg2014, Knutson2014}, and are also a candidate for the flat JWST transmission spectrum of a rocky exoplanet \linktocite{Lustig-YaegerFu2023}{(Lustig-Yaeger \& Fu et al.} \citeyear{Lustig-YaegerFu2023}). In particular, partially cloudy Earth-like exoplanets will exhibit a confounding combination of clear and cloudy effects, where broad CIA bands, refraction, and clouds may all appear degenerate in retrieval analyses. Further exoplanet-analog observations and analyses of Earth and other Solar System planets in transit would be beneficial for disentangling these complicated processes \citep[e.g.,][]{Mayorga2021}. The recent work of \citet{Doshi2022} suggests that Earth's clouds are sufficiently low-altitude as to not appreciably impact transmission observations of Earth-like exoplanets transiting M dwarfs, however 3D GCM models with full atmospheric dynamics do show an impact of higher altitude clouds in the transmission spectra of similar exoplanets \citep{Komacek2020, May2021}, so the general outcome for JWST observations remains uncertain and requires both data and modeling. 

% JWST Exoplanet relevance
To further shed light on our findings in the context of upcoming exoplanet observations, we analyzed Earth's spectrum using the spectral resolution and precision calculated for a long observational campaign of TRAPPIST-1e with JWST. We found that NIR observations with NIRSpec significantly outperformed MIR observations with MIRI LRS for obtaining atmospheric constraints given equal observing times with each instrument. These results are in agreement with other studies that have compared JWST instruments for rocky exoplanet transmission spectroscopy \citep{Lustig-Yaeger2019, Fauchez2019, Pidhorodetska2020}. We retrieved an abundance constraint for \ce{O3} that was approximately 2 orders of magnitude more precise using the 4.7 \um \ce{O3} band with NIRSpec G395 compared to the much more prominent 9.6 \um \ce{O3} band with MIRI LRS, and conclude that, given current detector and telescope capabilities, \ce{O3} may be best sought at shorter NIR wavelengths rather than in the MIR.  This result is consistent with the findings of \citet{Krissansen-Totton2018} and highlights the importance of considering wavelength-dependent instrument sensitivity, and not just molecular absorption band strength, when targeting specific molecules. However, we caution that by using the true Earth transmission spectrum in our TRAPPIST-1 JWST assessment, atmospheric photochemistry for this M dwarf planet has not been considered. Photochemistry can modify the vertical \ce{O3} structure \citep[e.g.][]{Segura2005, Rugheimer2015, Kaltenegger2021, Meadows2018} and is expected to enhance the \ce{CH4} detectability, and reduce the stratospheric abundance and detectability of \ce{O3} for TRAPPIST-1e relative to the modern Earth atmosphere with solar UV forcing \citep{Meadows2023}.  

% JWST Exoplanet relevance cont'd
Our simplified atmospheric model with an isothermal and evenly mixed vertical structure provided good fits to the synthetic JWST spectra that were consistent with the true Earth values within the inferred uncertainties. 
Given the optimistic nature of our synthetic JWST observations---80 stacked transits of TRAPPIST-1e---we suggest that this common assumption may be sufficient for terrestrial exoplanet retrieval modeling in the JWST era, although more work is warranted in this area to avoid the inference biases that we identified in the high S/N Earth spectra and that have been reported in the literature for gaseous exoplanets retrieval modeling \citep[e.g.,][]{Rocchetto2016, Caldas2019, MacDonald2020}.  

% Caveats about ACE-FTS data
Thus far we have discussed how model simplifications and the 3D time-varying nature of Earth's atmosphere may explain deviations between our findings and the true Earth spectrum and atmosphere. However, systematic effects from the ACE-FTS instrument may also play a confounding role in these deviations. 
In particular, the lack of temperature coherence likely produced systematic errors on the order of a few percent in the Earth averaged spectrum \citep{Boone2019}. The metrology laser for the ACE-FTS instrument does not have a fixed wavelength because it changes with ambient temperature.  This can lead to small shifts in wavenumber, and as a result, lines in the average spectra will experience a slight artificial broadening. In an individual occultation, the relative intensities of different lines in the spectrum are well characterized using a single (ambient) temperature. In the average spectrum, the relative intensities of different lines are not perfectly characterized by a single value of temperature, and likely results in systematic errors the order of a few percent. 

% SMARTER runtime and usecase
Owing primarily to its line-by-line approach, the computational expense of \smarter exceeds the typical run time of other current exoplanet atmospheric retrieval models, which in turn limits how and when it may be optimally used. The time required to evaluate the \smarter forward model is driven primarily by the breadth of the wavelength range, since \smart operates with a high internal wavenumber resolution that is set by the density of molecular lines in the wavelength range, regardless of the resolution of the data. For example, the full $2-14$ \um spectral models in Section \ref{sec:results:low} required approximately 2 minutes per likelihood evaluation (on a cluster processor), while the narrower 4 \um simulations required approximately 30 seconds per evaluation.   The aforementioned approach of resampling the molecular absorption coefficients speeds up the forward model considerably, while incurring modeling errors at or below 1\%. Using this approach the $2-14$ \um spectrum ran in about 15 seconds, and each JWST instrument in 2-6 seconds. Since the retrieval iterates over successive calls to the forward model, computationally expensive forward models cascade into very slow retrievals. Our OE retrievals required fewer than 1000 evaluations on average to reach convergence, whereas our \dynesty retrievals required approximately 100,000 forward model evaluations. Although \smarter can be used generally to retrieve atmospheres from the transmission, emission, and reflected-light spectra of a diversity of planet types, given the computational considerations, we see \smarter as a uniquely robust, specialized model that may be optimally used on specific high priority simulated or observed targets to (1) explore the effects of physical rigor and radiative transfer complexity in exoplanet retrievals, (2) fit high resolution spectra, and (3) analyze targets where the results from \smarter are well worth the computational expense. However, science cases involving numerous retrievals or large parameter sweeps are not ideal for \smarter and should use other capable, faster retrieval models \citep[e.g., see][and references therein]{MacDonald2023}. 

\subsection{Detecting Habitability \& Biosignatures using Transmission Spectroscopy}

% Punchy opening paragraph
Assessing the habitability of an exoplanet and confirming the presence of biosignatures will be a challenging endeavor for upcoming exoplanet observations, requiring the detection of specific spectroscopic indicators \citep{Schwieterman2018, Robinson2017} synthesized together with contextual clues into the nature of the planetary environment in which these indicators reside \citep{Meadows2018b, Meadows2018c}. 
While we found that the broad wavelength, impeccably high S/N observation that we analyzed contained many these essential ingredients, transmission spectroscopy is limited in terms of information that can be gleaned about the planetary surface environment, which then limits the strength of habitability and biosignature conclusions. 

% Open with a small paragraph about the JWST NIRSpec results AND THEN wax on about the noise-free data and all of the beautiful details
Through the eyes of JWST quality measurements, assessing the habitability and inhabited nature of Earth is challenging and leaves much unknown about the atmosphere. Although we obtained compelling abundance constraints for \ce{CH4}, \ce{CO2}, \ce{O3}, and \ce{H2O} using simulated NIRSpec observations for an exhaustive 80 transit campaign of TRAPPIST-1e, the habitability of the planet is not readily apparent in our findings. Our retrieved \ce{H2O} abundance is not consistent with that at the Earth's surface, but instead it is most consistent with the upper troposphere. Similarly, the characteristic isothermal TP profile that we retrieved is most consistent with the upper atmosphere. Thus, the insensitivity of the transmission spectrum to the surface limits habitability assessments even using relatively high quality observations with JWST. The simultaneous presence of \ce{CO2}, \ce{CH4}, and \ce{O3} are indeed a biosignature that JWST could potentially detect for an Earth-like planet, but may prove challenging to interpret without contextual knowledge about the habitable environment. 

% Bulk composition, N2
Our inferences about Earth's atmosphere from the \textit{noiseless} exoplanet-analog transmission spectrum provide compelling evidence in support of the planet's habitable surface conditions and presence of a global biosphere that has and continues to significantly modify the atmospheric composition. We will now walk through the lines of evidence that can be used to support this claim by considering the bulk atmospheric composition, the atmospheric indicators of habitability, and the detection of specific biosignature and technosignature gases. 

Evidence of Earth's primarily \ce{N2} and \ce{O2} bulk atmospheric composition proved to be one of the most subtle and difficult signals to interpret from the transmission spectrum observations. We were able to roughly infer the volume mixing ratio of \ce{N2} (${\sim} 80 \%$), particularly in high resolution spectra, and thereby identify \ce{N2} as the bulk atmospheric constituent. This finding depended almost exclusively on the 4.1 \um \ce{N2-N2} CIA feature that resides in the wings of the 4.3 \um \ce{CO2} band (see Figure \ref{fig:Earth_spectrum_fit}). This underscores the previous findings of \citet{Schwieterman2015b} by demonstrating that exoplanet transmission spectroscopy observations can target this \ce{N2} CIA band in the future as a means to detect and constrain the abundance of \ce{N2} in terrestrial atmospheres. This also agrees with similar findings for \ce{N2-N2} CIA that were obtained by \citet{Kaltenegger2020} in the context of an Earth-like planet transiting a white dwarf, where high S/N transmission spectra may be more readily achieved. 
At MIR wavelengths near 6 \um, the spectral continuum appears to be dominated by partial contributions from \ce{N2-O2} CIA, \ce{O2-O2} CIA, and \ce{H2O} line absorption wings (see Figure \ref{fig:Earth_spectrum_fit_hr}). These three absorbing species were degenerate in our model fits and proved difficult to accurately constrain without imposing a specific choice of TP profile. Constraining the atmospheric mean molecular weight can provide another line of evidence for the bulk atmospheric composition, however we did not vary it in our retrievals.  

% Habitability via H2O
We were able to infer the presence and abundance of \ce{H2O} vapor in Earth's atmosphere from a combination of absorption features in the near- and mid-IR. Our VMR constraints are consistent with upper tropospheric water abundances, which supports the interpretation of Earth as a habitable planet. Crucially, the small and subtle nature of water features in Earth's transmission spectrum---while difficult to remotely sense---is a good sign for habitability because substantial water features would indicate abundant stratospheric water vapor, which may indicate that the planet is in a runaway greenhouse state and is therefore uninhabitable \citep{Goldblatt2012}. 
Furthermore, our assessment of atmospheric \ce{CO2} provides important contextual clues into the climatic state of the planet that would greatly aid in the indirect assessment of habitability. 
However, assessing exoplanet habitability from transmission spectroscopy alone may not provide sufficient evidence due to difficulty sensing the surface, including surface temperature and pressure estimates, the near-surface \ce{H2O} abundance, and abundances of other greenhouse gases. In the absence of more direct methods for habitability assessment \citep[e.g.][]{Cowan2009, Robinson2010, Lustig-Yaeger2018}, these quantities would all be required for climate models to accurately simulate the likelihood of unobserved liquid water existing on the planetary surface \citep{Meadows2018c}. 
Our high resolution investigations yielded sensitivity to the vertical thermal structure and \ce{H2O} profile, which could be used to detect the atmospheric cold trap that keeps water vapor concentrated in Earth's lower atmosphere. Since we found that retrievals using realistic vertical structure provided better fits to the spectra than assuming even mixing, future studies should further investigate the potential for vertically resolved water vapor retrievals to enhance the robustness of indirect exoplanet habitability assessments.  

% Biosignatures
We retrieved robust constraints on the presence and abundance of numerous classical biosignature gases and disequilibrium biosignature pairs. 
In particular, we retrieved robust constraints on \ce{O3} from the prominent 4.7 \um and 9.65 \um absorption bands. \ce{O3} forms from the photolysis of photosyntheically produced \ce{O2}, which we were notably not able to detect due to \ce{O2} lacking prominent absorption features in the $2-14$ \um wavelength range. \ce{O3} has been proposed as a proxy for \ce{O2} for MIR exoplanet observations \citep{Leger1993, Leger2011, DesMarais2002}, which is strongly supported by our analyses of \ce{O2} and \ce{O3} in the observed transmission spectrum. 
We retrieved the \ce{CH4} abundance from the 2.3, 3.3, and 7.7 \um absorption bands, which exists in Earth's atmosphere as a result of geological sources \citep{Etiope2013}, methanogenesis from anaerobic single-celled microbes, and anthropogenic sources \citep{Schwieterman2018}. The simultaneous presence of \ce{CH4} alongside \ce{O3} is a compelling biosignature pair because \ce{CH4} is thermodynamically disfavored in an oxidized atmosphere and must therefore originate from a strong flux source. 
We also retrieved the abundance of \ce{N2O}, which is a biogenic product of incomplete denitrification of \ce{NO3-} to \ce{N2} gas and has minimal abiotic sources \citep{Schwieterman2018}. \ce{N2O} has been proposed as a strong biosignature \citep{Segura2005, Rugheimer2015}. Our sensitivity to \ce{N2O} derives primarily from features at 4.5, 7.8, and 8.6 \um. 
We also detected and retrieved the abundance of \ce{NO2} primarily using the absorption feature at 6.2 \um. \ce{NO2} was recently proposed as a possible technosignature due to its anthropogenic production from combustion processes \citep{Kopparapu2021}. 
Finally, we constrained the abundance of the chlorofluorocarbons (CFCs) CFC-11 and CFC-12 at 11.8 and 10.8 \um, respectively. CFCs have also been proposed as technosignatures due to their many industrial applications and infamy catalyzing the destruction of \ce{O3} \citep{Marinova2005, Schneider2010, Lin2014, Haqq-Misra2020}. 

Taken together, these climate and habitability indicators (e.g., \ce{H2O}, \ce{CO2}, \ce{CO}), biosignatures (e.g., \ce{O3}, \ce{CH4}, \ce{N2O}, \ce{O2}), and technosignatures (e.g., \ce{NO2}, CFC-11, CFC-12) paint a striking picture of the modern Earth as a living planet. However, despite the robust detection of individual touchstone molecules and their respective abundances in the noiseless \citetalias{Macdonald2019} data, the claim that we have detected life, while perhaps evident, remains unquantified. Likewise, the confidence of life detection from our results using simulated exoplanet observations with JWST, while surely much lower, is not readily quantified. Such gaps in the statistical astrobiology used for life detection have recently been illuminated as requiring significant focus and broad interdisciplinary engagement  \citep[e.g.,][]{Catling2018, Green2021, Meadows2022AGU}. The data analyses undertaken here as well as the empirical Earth spectrum from \citetalias{Macdonald2019} may serve as an important resource to test and validate life detection frameworks as they are put forth.

\section{Conclusion} \label{sec:conclusion} 

In this paper, we validated our terrestrial exoplanet atmospheric retrieval model \smarter against a high resolution, high S/N empirical transmission spectrum of Earth and were able to retrieve abundances for the majority of spectrally active molecules to better than 25\% while maintaining residuals below 5 ppm equivalent precision for TRAPPIST-1 observations. The presence and abundance of retrieved gases provides compelling evidence of the planet's habitable surface conditions, global biosphere, and emerging technosphere as viewed exclusively through the analysis of an exoplanet-analog transmission spectrum. 
Although the data wavelength range did not contain ro-vibrational or electronic transitions for the Earth's bulk atmospheric gases, \ce{N2} and \ce{O2}, nonetheless we were able to roughly constrain their abundances using collisionally-induced absorption features at high spectral resolution where they could be distinguished from overlying absorption lines. 

When we degraded the empirical spectrum down to the spectral resolution and precision of expected JWST/NIRSpec observations, even after coadding all of the transits available in the nominal 5-year mission, we were only able to obtain robust abundance constraints for \ce{CO2} and \ce{CH4}, with a tentative indication of \ce{O3}. 

Just as observations and retrievals of extrasolar giant planets are leading the way towards future analyses of Earth-like exoplanets with higher-sensitivity instruments, the study of Earth provides high-precision exoplanet-analog data that can prepare our exoplanet models to interpret vertically complex and spatially dynamic 3D atmospheres, beginning with near-term studies of extrasolar giant planets.  The Earth transmission spectrum from \citetalias{Macdonald2019} is a treasure trove of information on the physics and chemistry of exoplanet transmission spectroscopy in the limit of high precision and resolution. In this work we have only begun to scratch the surface of what is possible with this benchmark dataset, and we encourage community engagement with these data for continued model validations and upgrades as the field progresses towards the precise characterization of Earth-like exoplanets.   

 %% Acknowledgements %%
\acknowledgments

We would like to sincerely thank E. Macdonald for providing the spectrum that made our investigation possible, as well as N. Cowan and C. Boone for useful discussions. We also thank two anonymous reviewers for their thorough reviews of our manuscript and thoughtful comments that helped to improve the quality, clarity, and reproducibility of our work. This work was performed by the Virtual Planetary Laboratory Team, which is a member of the NASA Nexus for Exoplanet System Science, and funded via NASA Astrobiology Program Grant 80NSSC18K0829. J.L.Y acknowledges support from JHU APL’s Independent Research And Development program. This work made use of the advanced computational, storage, and networking infrastructure provided by the Hyak supercomputer system at the University of Washington. Some of the results in this paper were derived using the healpy and HEALPix package. 

\software{Astropy \citep{Astropy2013, Astropy2018}, Dynesty \citep{Speagle2020}, HEALPix \citep{Gorski2005}, healpy \citep{Zonca2019}, LBLABC \citep{Meadows1996}, Matplotlib \citep{Hunter2007}, MultiNest \citep{Feroz2008, Feroz2009}, NumPy \citep{Walt2011, NumPy2020}, SciPy \citep{Virtanen2019scipy, SciPy2020}, SMART \citep{Meadows1996}, SMARTER (this paper; \citealp{Lustig-Yaeger2022}), Pandas \citep{reback2020pandas}, Pandeia \citep{Pontoppidan2016}, PandExo \citep{Batalha2017b, Pandexo2018}, pysynphot \citep{STScI2013}}

%% Bibliography %%
\bibliography{ms}

\newpage

\appendix

\section{Refraction Path Length Effects} \label{sec:appendix:refraction}

% Refraction intro
Refraction is widely appreciated as a mechanism that can set the continuum of exoplanet transmission spectra, much like a cloud deck \citep{GarciaMunoz2012, Betremieux2014, Misra2014a}. However, in solar occultation geometry with an orbiting observer, this limiting exoplanet-analog refraction effect is not immediately realized, despite the unavoidable presence of refraction in the observed transmittance data. To investigate the role of refraction in our transmission spectrum retrievals, we activated the ray tracing refraction module within \smart \citep{Robinson2017a} for use in our retrieval forward model, and performed a series of spectroscopic and inference tests. To accurately simulate the increase in optical path length through the atmosphere due to refraction without modeling the critical refraction boundary for an exoplanet observer at infinity, we increased the stellar radius in our radiative transfer model to the arbitrarily large value 100 $R_{\odot}$. Although this change would strongly scale the transit depth $(R_p/R_s)^2$, it does not modify the resultant effective transit altitude, with the exception of the increase in optical path length that is of interest.  

\begin{figure*}[b!]
\centering
\includegraphics[width=0.97\textwidth]{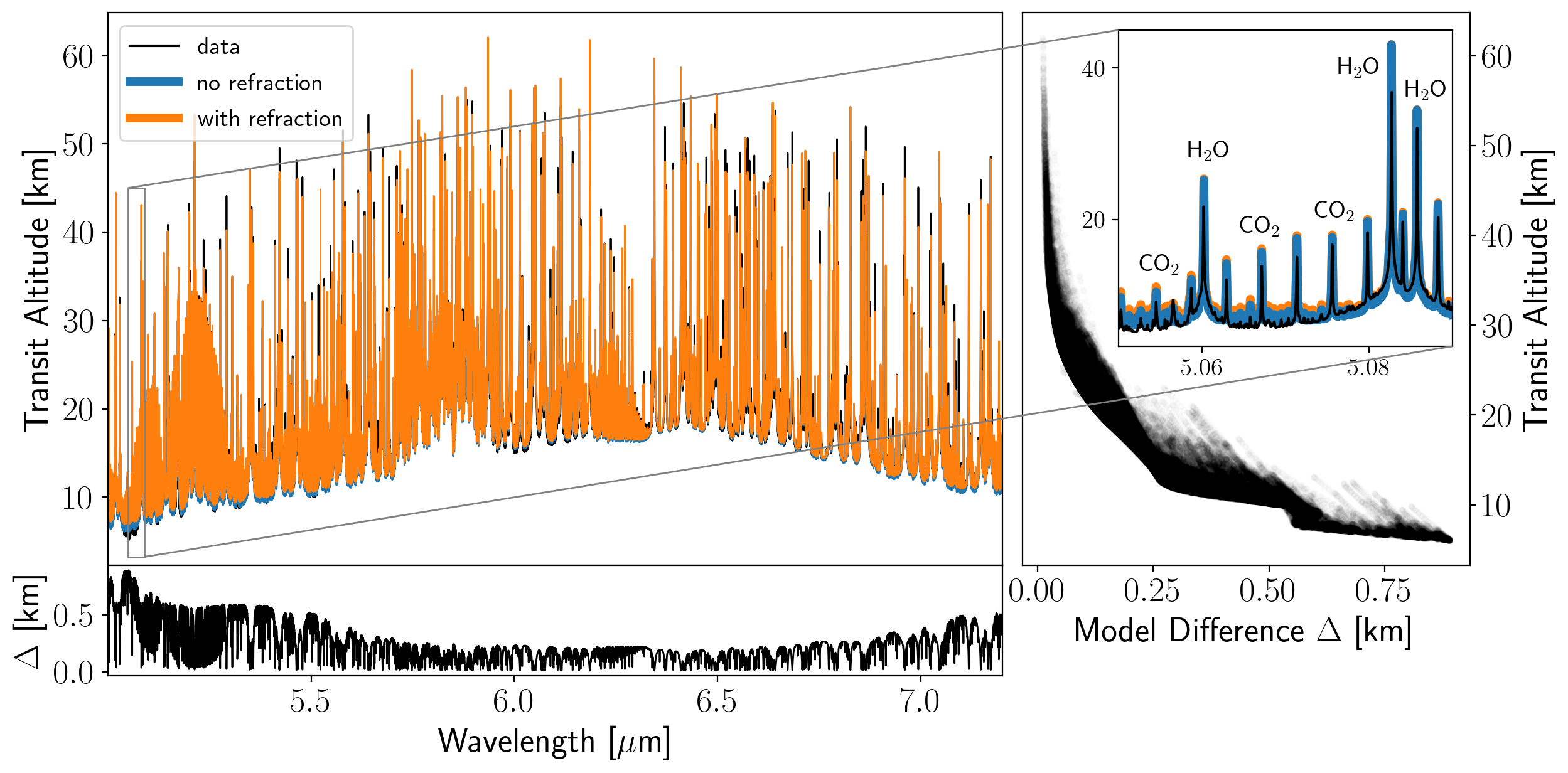}
\caption{Comparison of transmission spectrum models with and without refraction. The upper left panel shows the two spectral models atop the observed data in the 6 \um band of interest. The lower left panel shows the residuals between the model with and without refraction, which appears anti-correlated with the spectral continuum. This relationship is borne out more clearly in the right panel, which shows the model residuals as a function of transit altitude where larger model difference due to refraction are observed at lower altitudes. Finally, the zoomed insert visually emphasizes this point by showing a comparison of individual \ce{CO2} and \ce{H2O} lines. The cores of optically thick lines that probe higher altitudes are not affected by refraction, while the continuum, and optically thin line and cores are slightly affected, with larger refraction effects seen at lower altitudes.} 
\label{fig:refraction}
\end{figure*}

% Refraction figure
We took the best-fitting ``cold'' spectral model from our high-resolution retrieval in the 6 \um window (see Table \ref{tab:earth_fit3}) and simulated the spectrum again with refraction turned on. Figure \ref{fig:refraction} compares the spectral models with and without refraction and provides an assessment of their differences as a function of wavelength and altitude. At all wavelengths, the model with refraction results in slightly larger effective transit altitudes. Shortward of 5.5 \um, where the continuum altitude within the spectral window is the lowest, the difference between the two models is largest. This distinction is most evident in the right panel of Figure \ref{fig:refraction}, which shows that the model differences due to refraction are systematically larger at the lowest altitudes probed by the transmission spectrum, while the cores of the strongest lines that extend up to high altitudes are essentially unaffected by refraction. The zoomed inset provides a visual aid for this effect as the differences due to refraction increase stepping down from the strongest \ce{H2O} and \ce{CO2} lines---where there is no discernible refraction effects---to the weakest lines and the continuum---where the refraction effects are maximal. Although this demonstrates the existence of a systematic modeling artifact due to the omission of refraction physics, and the resultant increase in optical path length, the maximum model deviations seen near the continuum still remain quite small ($< 1$ km) and are exceeded by the average deviations of our best fitting models.  

\section{Simulated JWST Retrieval Posteriors \& Tests} \label{sec:appendix:posteriors}

\begin{figure*}[b!]
\centering
\includegraphics[width=0.98\textwidth]{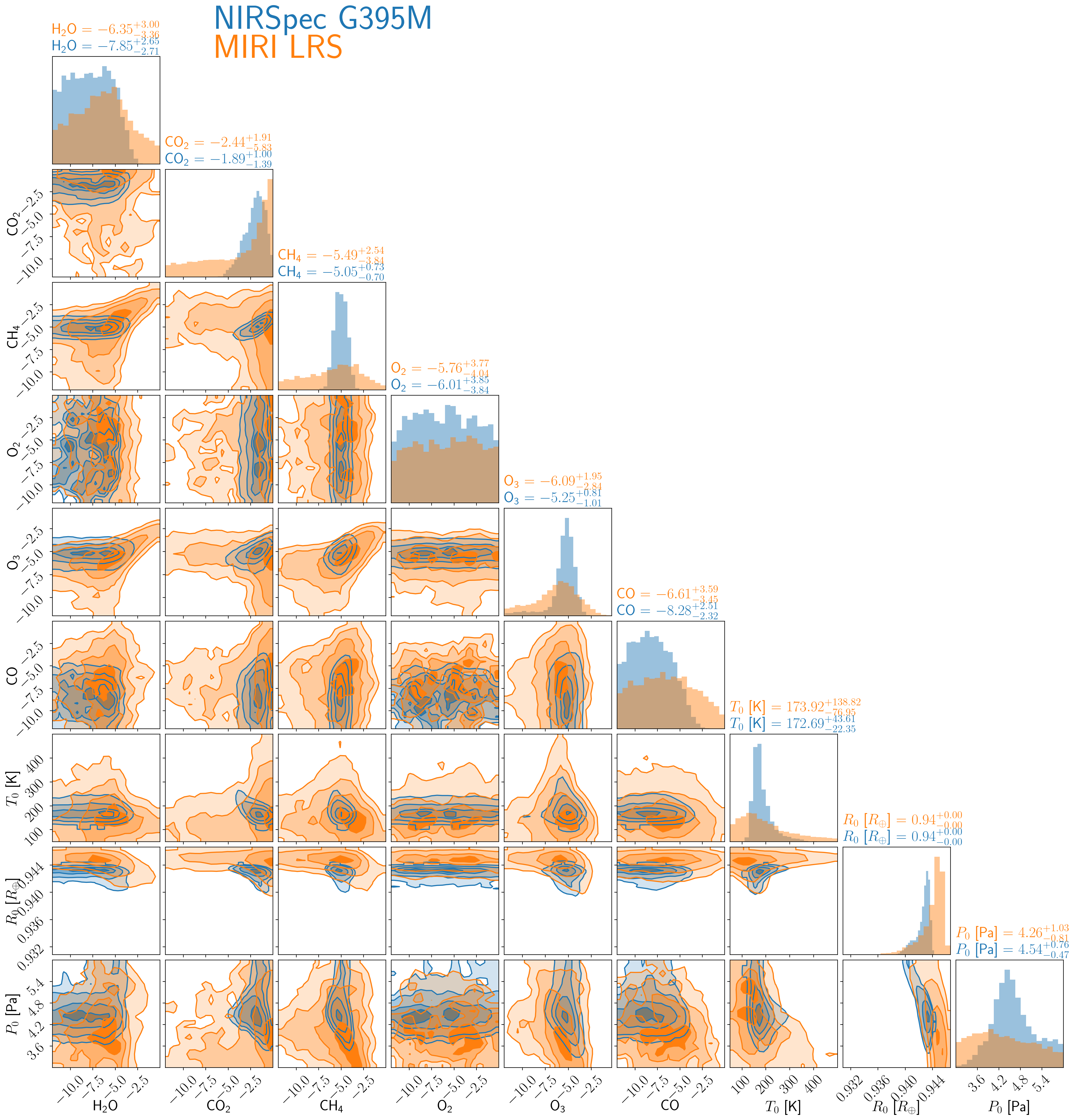}
\caption{Marginalized posterior distributions for the NIRSpec G395M (blue) and MIRI LRS (orange) retrievals discussed in Section \ref{sec:results:jwst} and Figure \ref{fig:jwst_spectra}. These constraints correspond to 80 stacked transits of TRAPPIST-1e using each instrument.} 
\label{fig:jwst_posteriors}
\end{figure*}

Figure \ref{fig:jwst_posteriors} shows a corner plot of the posterior distributions retrieved for TRAPPIST-1e from simulated NIRSpec G395M and MIRI LRS observations, as discussed in Section \ref{sec:results:jwst} and Figure \ref{fig:jwst_spectra}. 

% JWST Spectra 

\begin{figure*}[t!]
\centering
\includegraphics[width=0.98\textwidth]{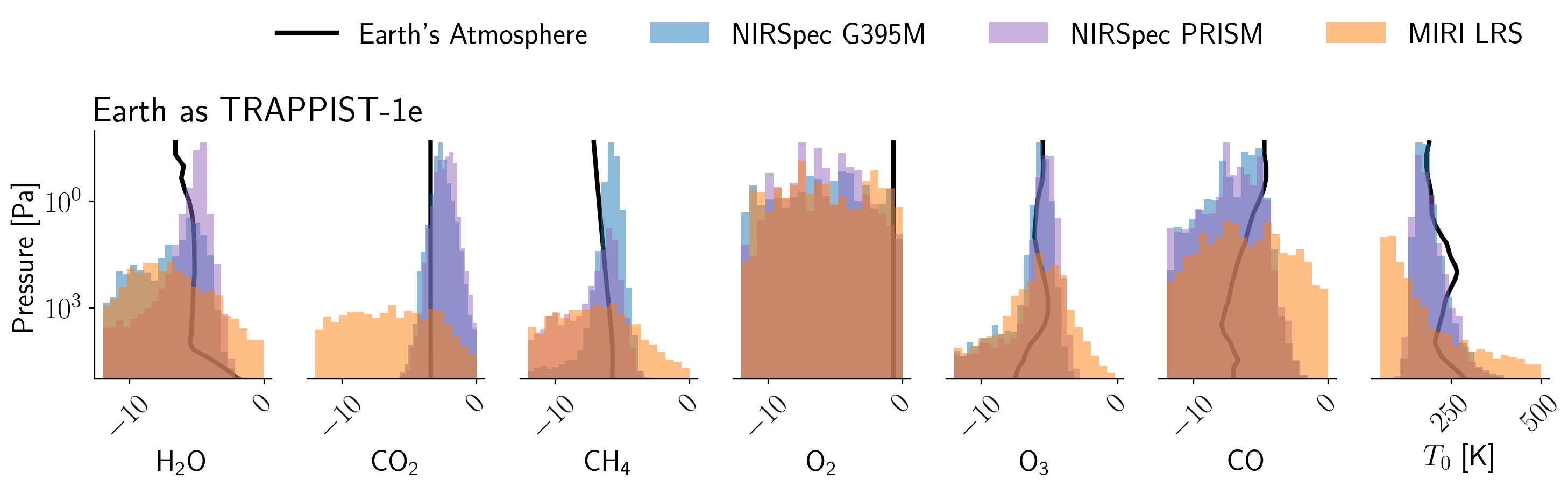}
\caption{1-D marginalized posterior distributions for NIRSpec G395M (blue), NIRSpec PRISM (purple), and MIRI LRS (orange) retrievals. These constraints correspond to 80 stacked transits of TRAPPIST-1e using each instrument, but in this case without randomized noise added to the observations.} 
\label{fig:jwst_prism}
\end{figure*}

We also investigated a limited case using the NIRSpec PRISM instrument mode. Figure \ref{fig:jwst_prism} is similar to the upper panel of Figure \ref{fig:jwst_spectra} with the addition of retrieval results from a simulated spectrum using the NIRSpec PRISM mode with a shortwave cutoff of 2.2 \um. All results are shown for retrievals where the data points were centered on their true model value (similar to, for example, \citet{Feng2018}, but unlike our Figure \ref{fig:jwst_spectra})  to also investigate the sensitivity of our results to the specific random noise instance that was previously used. The G395M and PRISM results are very similar, likely due to the fact that we could not include Earth data for the PRISM case shortward of 2.2 \um so the two modes cover a similar wavelength range, but differ in spectral resolution (PRISM has $R\sim100$ and G395M has $R\sim1000$). However, given this caveat, the PRISM data yields slightly higher sensitivity to \ce{H2O} and the G395M data yields slightly higher sensitivity to \ce{CH4}. The high abundance \ce{CO2} posterior spike seen for MIRI LRS in Figure \ref{fig:jwst_spectra} is absent in this test, thereby demonstrating that it was a manifestation of random noise. 

% JWST PRISM test 

%\section{Table of Contents \textit{(to appear only in draft)}}
%\tableofcontents

%% End Doc
\end{document}